 \documentclass[final,5p,times,sort&compress,fleqn]{elsarticle}

\usepackage{amssymb}
\usepackage{amsthm}

\usepackage{lineno}

\usepackage{graphicx}
\usepackage{amsmath}
\usepackage[version=4]{mhchem}
\usepackage{siunitx}
\usepackage{upgreek}  
\usepackage{longtable,tabularx}
\usepackage{xcolor}
\usepackage{float}  
\usepackage{caption}
\usepackage{subcaption}
\usepackage[many]{tcolorbox}  

\usepackage{framed}  
\usepackage{multicol}  

\usepackage{soul}  
\usepackage{ulem}  

\setstcolor{red}  
\newcommand\redsout{\bgroup\markoverwith{\textcolor{red}{\rule[0.5ex]{2pt}{0.4pt}}}\ULon}  

\tcbset{highlight math style={enhanced,%
    colframe=black, colback=white, boxsep=0pt, 
    left=1mm, right=1mm, top=1mm, bottom=1mm, 
    leftrule=0.20mm, rightrule=0.20mm, toprule=0.20mm, bottomrule=0.20mm}}

\usepackage{nomencl}  
\makenomenclature
\usepackage{etoolbox}
\renewcommand\nomgroup[1]{%
  \item[\bfseries
  \ifstrequal{#1}{L}{Latin Symbols}{%
  \ifstrequal{#1}{M}{Greek Symbols}{%
  \ifstrequal{#1}{N}{Abbreviations}{}}}%
]}

\journal{Acta Astronautica}

\begin{document}

\normalem  

\begin{frontmatter}



\title{Structural Stability of a Lightsail for Laser-Driven Interstellar Flight}


\author[inst1]{Dan-Cornelius Savu \texorpdfstring{\fnref{DCS}}{}}
\author[inst1]{Andrew J. Higgins \texorpdfstring{\fnref{AJH}}{}}

\affiliation[inst1]{organization={Department of Mechanical Engineering, McGill University},  
            addressline={817 Sherbrooke St. W.},
            city={Montreal},
            state={Quebec},
            postcode={H3A 0C3},
            country={Canada}}

\fntext[DCS]{Undergraduate Research Assistant, Department of Mechanical Engineering, 817 Sherbrooke St. W. Email address: \href{mailto:x@x.com}{\underline{dan-cornelius.savu@mail.mcgill.ca}}}
\fntext[AJH]{Professor, Department of Mechanical Engineering, 817 Sherbrooke St. W. Email address: \href{mailto:x@x.com}{\underline{andrew.higgins@mcgill.ca}}}

\begin{abstract}
The structural stability of a lightsail under the intense laser flux necessary for interstellar flight is studied analytically and numerically. A sinusoidal perturbation is introduced into a two-dimensional thin-film sail to determine if the sail remains stable or if the perturbations grow in amplitude. A perfectly reflective sail material that gives specular reflection of the laser illumination is assumed in determining the resulting loading on the sail, although other reflection models can be incorporated as well. The quasi-static solution of the critical point between shape stability and instability is found by equating the bending moments induced on the sail due to radiation pressure with the restoring moments caused by the strength of the sail material and the tension applied at the edges of the sail. From this quasi-static solution, analytical expressions for the critical value of elastic modulus and boundary tension magnitude are found as a function of sail properties (e.g., thickness) and the amplitude and wave number of the initial sinusoidal perturbation. These same expressions are also derived from a more formal variational energy (virtual work) approach. A numerical model of the complete lightsail dynamics is developed by discretizing the lightsail into rectangular finite elements. By introducing torsional and rectilinear springs between the elements into the numerical model, a hierarchy of models is produced that can incorporate the effects of bending and applied tension. The numerical models permit the transient dynamics of a perturbed lightsail to be compared to the analytic results of the quasi-static analysis, visualized as stability maps that show the rate of perturbation growth as a function of sail thickness, elastic modulus, and applied tension. The analytic theory is able to correctly predict the stability boundary found in the numerical simulations. The stiffness required to make a thin lightsail stable against uncontrolled perturbation growth appears to be unfeasible for known materials, however, a relatively modest tensioning of the sail (e.g., via an inflatable structure or spinning of the sail) is able to maintain the sail shape under all wavelengths and amplitudes of perturbations.
\end{abstract}

\onecolumn 
\begin{highlights}
    \item The shape of a flat lightsail was perturbed to test its resistance to deformation.
    \item Quasi-static analytical expressions were developed for the critical stability point.
    \item Dynamics of the perturbed lightsail was studied using numerical models.
    \item Analytical expressions predict the stability boundary seen in numerical simulations.
    \item Sail tensioning, \emph{not} increasing material stiffness, is a method to obtain stability.
\end{highlights}
\twocolumn


\end{frontmatter}



\nomenclature[L#00]{$a$}{absorption coefficient}
\nomenclature[L#02]{$\mathrm{amp_{max}}$}{maximum lightsail perturbation amplitude}
\nomenclature[L#03]{$A$}{discrete element in-plane area}
\nomenclature[L#04]{$A_\mathrm{c}$}{cross-sectional area}
\nomenclature[L#05]{$B_\mathrm{b}$}{non-Lambertian coefficient of the lightsail back (non-reflecting) surface}
\nomenclature[L#06]{$B_\mathrm{f}$}{non-Lambertian coefficient of the lightsail front (reflecting) surface}
\nomenclature[L#07]{$c$}{speed of light in vacuum}
\nomenclature[L#08]{$d$}{diameter}
\nomenclature[L#09]{$E$}{material Young's modulus}
\nomenclature[L#10]{$E_\mathrm{cr}$}{critical Young's modulus predicted by the direct approach}
\nomenclature[L#11]{$E_\mathrm{cr_{max}}$}{maximum critical Young's modulus value predicted by the direct approach}
\nomenclature[L#12]{$E_\mathrm{cr_{\delta \mathcal{W}}}$}{critical Young's modulus predicted by the energy approach}
\nomenclature[L#13]{$E_\mathrm{cr_{\mathrm{max}_{\delta \mathcal{W}}}}$}{maximum critical Young's modulus value predicted by the energy approach}
\nomenclature[L#14]{$f_0$}{flat element radiation force magnitude}
\nomenclature[L#15]{$\mathbf{f}_i$}{$i\mathrm{th}$ element radiation force}
\nomenclature[L#16]{$\mathbf{f}_\mathrm{n}$}{radiation force component directed along the element normal direction}
\nomenclature[L#17]{$\mathbf{f}_\mathrm{t}$}{radiation force component directed along the element tangential direction}
\nomenclature[L#18]{$\mathbf{f}$}{forcing vector}
\nomenclature[L#19]{$\hat{\mathbf{f}}$}{modified forcing vector}
\nomenclature[L#20]{$g_0$}{non-inertial D'Alembert vertical acceleration}
\nomenclature[L#21]{$h$}{thickness}
\nomenclature[L#22]{$I$}{second moment of area}
\nomenclature[L#23]{$I_\mathrm{G}$}{moment of inertia about the center of mass}
\nomenclature[L#24]{$I_0$}{laser beam intensity}
\nomenclature[L#25]{$k_\mathrm{s}$}{rectilinear spring constant}
\nomenclature[L#26]{$k_\mathrm{t}$}{torsional spring constant}
\nomenclature[L#27]{$l$}{element length}
\nomenclature[L#28]{$l_\mathrm{s}$}{rectilinear spring elongation at rest (always set to $0$)}
\nomenclature[L#29]{$L$}{working length}
\nomenclature[L#30]{$\mathcal{L}$}{the Lagrangian}
\nomenclature[L#31]{$m$}{discrete element mass}
\nomenclature[L#32]{$\mathbf{m}$}{general radiated element force direction}
\nomenclature[L#33]{$M$}{bulk mass}
\nomenclature[L#34]{$\mathcal{M}$}{moment/torque}
\nomenclature[L#35]{$\mathbf{M}$}{mass matrix}
\nomenclature[L#36]{$\hat{\mathbf{M}}$}{modified mass matrix}
\nomenclature[L#37]{$n$}{number of rigid sail elements}
\nomenclature[L#38]{$\mathbf{n}$}{normal vector}
\nomenclature[L#39]{$p_\mathrm{r}$}{radiation pressure}
\nomenclature[L#42]{$\mathbf{q}$}{vector of generalised coordinates with components $q_i$}
\nomenclature[L#43]{$r$}{reflection coefficient}
\nomenclature[L#44]{$\mathbf{r}$}{position vector}
\nomenclature[L#45]{$s$}{fraction of light that is specularly reflected}
\nomenclature[L#46]{$t$}{time}
\nomenclature[L#47]{$t_\mathrm{final}$}{final simulation runtime}
\nomenclature[L#49]{$T$}{boundary tension magnitude}
\nomenclature[L#50]{$T_\mathrm{cr}$}{critical boundary tension magnitude}
\nomenclature[L#51]{$T_\mathrm{cr_{max}}$}{maximum critical boundary tension value}
\nomenclature[L#52]{$\mathbf{T}$}{boundary tension vector}
\nomenclature[L#53]{$w$}{initial perturbations vertical displacement}
\nomenclature[L#54]{$W$}{width}
\nomenclature[L#55]{$\mathcal{W}$}{work done}
\nomenclature[L#56]{$x_i$}{$x$-position of the $i$th element}
\nomenclature[L#57]{$\mathbf{x}$}{modified vector of generalised coordinates}
\nomenclature[L#58]{$y_i$}{$y$-position of the $i$th element}
\nomenclature[L#59]{$z_i$}{elongation of the $i$th element}

\nomenclature[M#00]{$\varepsilon_{\mathrm{b}}$}{emissivity of the lightsail back (non-reflecting) surface}
\nomenclature[M#01]{$\varepsilon_{\mathrm{f}}$}{emissivity of the lightsail front (reflecting) surface}
\nomenclature[M#02]{$\theta_i$}{angular position of the $i$th element with respect to the positive $x$-axis}
\nomenclature[M#03]{$\kappa$}{curvature}
\nomenclature[M#04]{$\nu$}{mode number of initial perturbations}
\nomenclature[M#05]{$\rho$}{density}
\nomenclature[M#06]{$\tau$}{time to doubling of initial perturbation amplitude}
\nomenclature[M#07]{$\Tilde{\tau}$}{transmission coefficient}

\printnomenclature[1.5cm]  


\section{Introduction}
\label{sec:intro}
Concentrating laser-light energy onto a reflective foil to permit fast transportation within the solar system and beyond has been actively considered since the 1980s \cite{forward1984roundtrip,landis1989optics} with serious academic discussions dating as far back as the 1960s \cite{marx1966interstellar,redding1967interstellar}. Because of the exponential rate of development of fiber-optic-based lasers within the telecommunication and laser machining industries, the laser-driven spacecraft is now steadily turning from concept to a present-day reality. Arbitrarily large laser beams can now be made by constructing phased arrays of lasers using inexpensive optical components \cite{lubin2016roadmap,parkin2018breakthrough}. With modern initiatives such as the Breakthrough Starshot project, which intends to use a ${\sim}1$~gram lightsail laser-accelerated to 20\% the speed of light to reach the nearest star Proxima Centauri within 20 human years, interstellar flight in the 21st century may become an actuality, but for this to happen various interdisciplinary scientific and engineering challenges still need to be overcome \cite{lubin2016roadmap,parkin2018breakthrough,atwater2018materials}. One such challenge that needs to be solved, given the large laser intensities involved (${\sim} \SI{10}{\giga \watt \per \meter^2}$) and the ideally low sail inertia, is the problem of the dynamic and structural stability of the lightsail.

The directional or \emph{beam-riding} stability of laser-driven lightsails has already been studied for a variety of rigid sail shapes. The beam-riding stability of conical and spherically curved lightsails attached to the spacecraft via a rigid boom have been investigated \cite{benford2003experimental,popova2016stability}, and Srinivansan et al. have shown that a hyperboloid shaped lightsail impinged upon its convex surface by the laser beam has passive directional stability \cite{srinivasan2016stability}. Manchester and Loeb also demonstrated that a spherically shaped lightsail---a lightsail akin to a balloon---is inherently directionally stable provided that the laser beam intensity profile was Gaussian multi-modal or donut-shaped \cite{manchester2017stability}. The influence of dynamic dampening upon directional lightsail stability has also been considered recently \cite{shirin2021dampening,rafat2022selfstabilization}. The inclusion of dissipation effects appears to effectively dampen the lightsail motions lateral to the beam axis thereby further increasing directional stability. Shirin et al. have numerically shown, for example, that dynamically dampened conical-shaped lightsails become exponentially stable---as opposed to their marginally stable undampened counterparts \cite{shirin2021dampening}.

\begin{figure*}[t]
    \centering
    \includegraphics[width=0.90\textwidth]{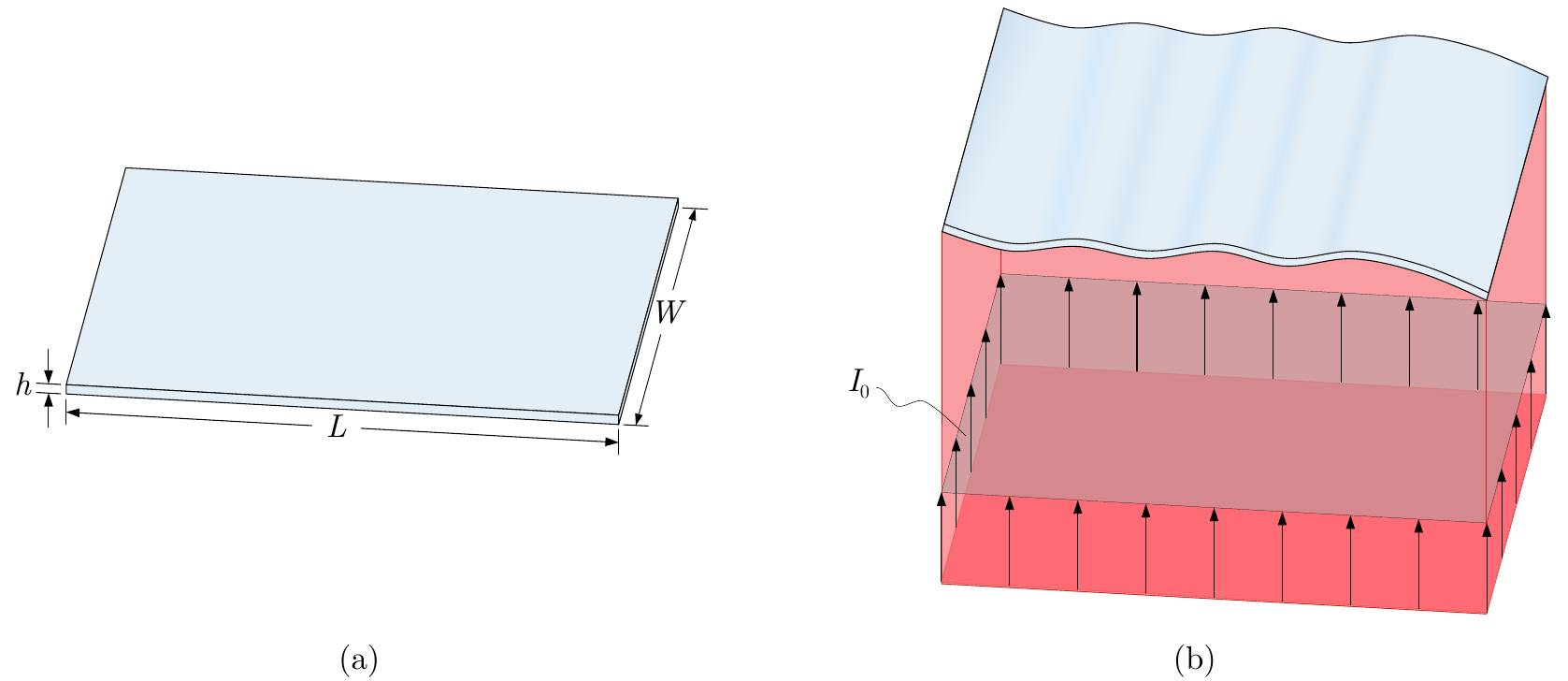}
    \caption{Three-dimensional analytical (plate) model of the lightsail; (a)~the lightsail, flat; (b)~the lightsail, smoothly perturbed with an incident uniform laser beam. The problem here considered is whether the perturbations will grow in amplitude or not under the large laser loads.}
    \label{fig:TheoConcept}
\end{figure*}

Directional stability has also been shown to be possible without recourse to curved sail shapes by means of engineered optical lightsail surfaces and materials. This has been achieved through the addition of nanoscale (dielectric) structures to the surface of a flat lightsail, a procedure enabled through the recent advances in optical design and nanofabrication. These so-called photonic metasurfaces allow for control over the magnitude and direction of the reflected and transmitted incident laser light while adding relatively little mass to the lightsail because of their nanoscopic nature. Swartzlander et al., in a series of theoretical, computational, and experimental studies have shown that a flat lightsail whose reflecting surface is equipped with diffractive gratings is directionally stable \cite{swartzlander2017diffractive,chu2018measurements,srivastava2019stable,chu2019experimental,srivastava2020optomech}. Myilswamy et al. have also shown that a nonlinear photonic crystal lightsail can help minimize the dynamic asymmetry caused by the atmospherically distorted laser beam \cite{myilswamy2020photonic}. Lightsails employing Bloch-wave type scatterers and other engineered optical metasurfaces have also been reported to be directionally stable \cite{atwater2019opto,siegel2019selfstabilizing,salary2020photonic}. Santi et al. have also suggested the use of thin-film multilayered optical structures (Bragg mirrors) for the actualization of curved lightsails with enhanced passive beam-riding stability \cite{santi2022multilayers}.

The studies mentioned so far have assumed the lightsail to be rigid and either ideally flat or perfectly smooth, that is, absent of deformation of the ideal sail shape. Deformation of the lightsail has been built into several other models in an attempt to quantify their influence upon the orientation and trajectory of solar-driven sails \cite{sakamoto2006FE01,sakamoto2007FE02,huang2021cloud}. Huang et al., for example, inquired into the deviation of the resultant solar radiation pressure force due to sail deformation effects caused by wrinkling and billowing via the use of point cloud and triangular mesh methods \cite{huang2021cloud}. Structural analysis of the sail, considering beams/booms and membranes, has also been undertaken \cite{wong2003wrinkle,liu2014highlyflexible,liu2014flexibleSail,boni2017solarsail,Zhang2020stripped}. Liu et al. studied the attitude dynamics and the vibrations of a square solar sail supported by four beams, the presence of which allowed them to neglect the detailed sail membrane vibrations and wrinkle effects \cite{liu2014highlyflexible,liu2014flexibleSail}. Wong and Pellegrino inquired theoretically, numerically, and experimentally into the visible membrane wrinkling amplitude and wavelength growth when tension is gradually applied to the corners of an initially flat, square solar sail membrane \cite{wong2003wrinkle}. Other studies concerning the problem of sail structural response have also been undertaken for particular sail shapes \cite{cassenti1996structural,genta2017preliminary}. Cassenti and Cassenti have also proposed the tensioning of the lightsail via the use of a boundary ring---much akin to a drumskin---as a potential solution to the structural vibration problem caused by the presence of a non-uniform laser beam \cite{cassenticassenti2020}. A more recent study inquired into the thermal and mechanical stresses that the lightsail experiences during laser-driven acceleration and has concluded that spherically curved lightsails of appreciable curvature are better suited to sustain the large laser loads \cite{campbell2022billow}. Of note are also the experimental efforts of Myrabo et al. who conducted investigations on the problem of sail stability by subjecting lightsail prototypes to laser loads in vacuum \cite{myrabo2000experimental}.

Recently, the application of the radiation pressure regime for laser-driven acceleration of thin foils under extremely intense fluxes (exceeding $\SI{e21}{\watt / \centi \meter^2}$) has been considered as a technique for heavy ion acceleration \cite{qiao2009ionbeam,macchi2009lightsail,macchi2017radiation}. Such radiation pressure acceleration technology could overcome the velocity limitations of more traditional pulsed-laser acceleration technologies, such as ablation, and has potential application to fast ignition in inertial confinement fusion. Under such intense fluxes, the structural stability of the thin-foil lightsail under radiation pressure can be treated using the Rayleigh-Taylor formalism of interface instability \cite{palmer2012rayleigh}. The regime of laser flux in this application is fifteen orders of magnitude greater than that proposed for the laser-driven interstellar lightsail and thus is not likely of direct relevance to the problem under consideration in the present study.

Altogether, while some studies have inquired into the structural stresses and strains supported by solar sails, it should be noted that solar-driven sails are generally not designed to tolerate high photon pressures and resultingly most solar sails envisioned so far---like the IKAROS, NanoSail-D2, and LightSail 2 solar sails \citep{mori2010IKAROS,tsuda2013achievement,mansell2020lightsail2,davoyan2021photonic}---sustain accelerations orders of magnitude less than the accelerations that, for example, the Breakthrough Starshot laser-driven lightsail would need to withstand \citep{parkin2018breakthrough,atwater2018materials}. Consequently, the structural analysis of solar-driven sails has been chiefly concerned with the influence of lightsail deformations upon the spacecraft trajectory whereas, by contrast, the comparatively large accelerations sustained by laser-driven lightsails could cause a deformed lightsail to crumple beyond functional use. Thus far, no inquiry into the detailed vibrations and deformations of a lightsail under high photon radiation loading has been completed. In particular, no study inquired into whether an ideally thin lightsail is capable of sustaining large laser-driven accelerations despite the inevitable presence of perturbations (see Fig.~\ref{fig:TheoConcept}). These perturbations---arising from multiple possible sources such as atmospheric disturbances, beam non-uniformities, shape distortion of the sail, etc.---prevent perfectly uniform loading of the sail, and this complication of the laser-lightsail dynamics may cause the distortions in sail shape to grow in amplitude. Even if the perturbations are, through careful lightsail deployment, kept small in magnitude, the following question remains: Under the large photonic pressure loads required for feasible interstellar flight, will the structural perturbations of the lightsail grow out of bound or will the lightsail remain flat?

The present study addresses the question of lightsail shape stability with perturbations by considering a first-principles approach to the problem. First, a continuous model of a sinusoidally perturbed lightsail under radiation load allows an analytical, quasi-static analysis of the critical point between lightsail shape stability and instability in Section~\ref{sec:theory}. A Lagrangian-based finite element (FE) numerical model of the perturbed lightsail is then constructed using rigid rectangular slices to simulate the full dynamics of the lightsail while undergoing acceleration in Section~\ref{sec:numerical}. The rigid-element numerical model was further generalized to include torsion and rectilinear springs to investigate the influence upon lightsail structural stability of material bending stiffness and applied tension, respectively. In Section~\ref{sec:RandD}, the quasi-static derived analytical expressions are compared to the numerical results, which consisted of multiple lightsail dynamics simulations with each simulation varying lightsail geometry, material modulus, and/or applied tension magnitude. The engineering implications of the analysis are then explored.


\section{Theoretical Considerations}
\label{sec:theory}

\begin{figure*}[!htp]
    \centering
    \includegraphics[scale=1.0]{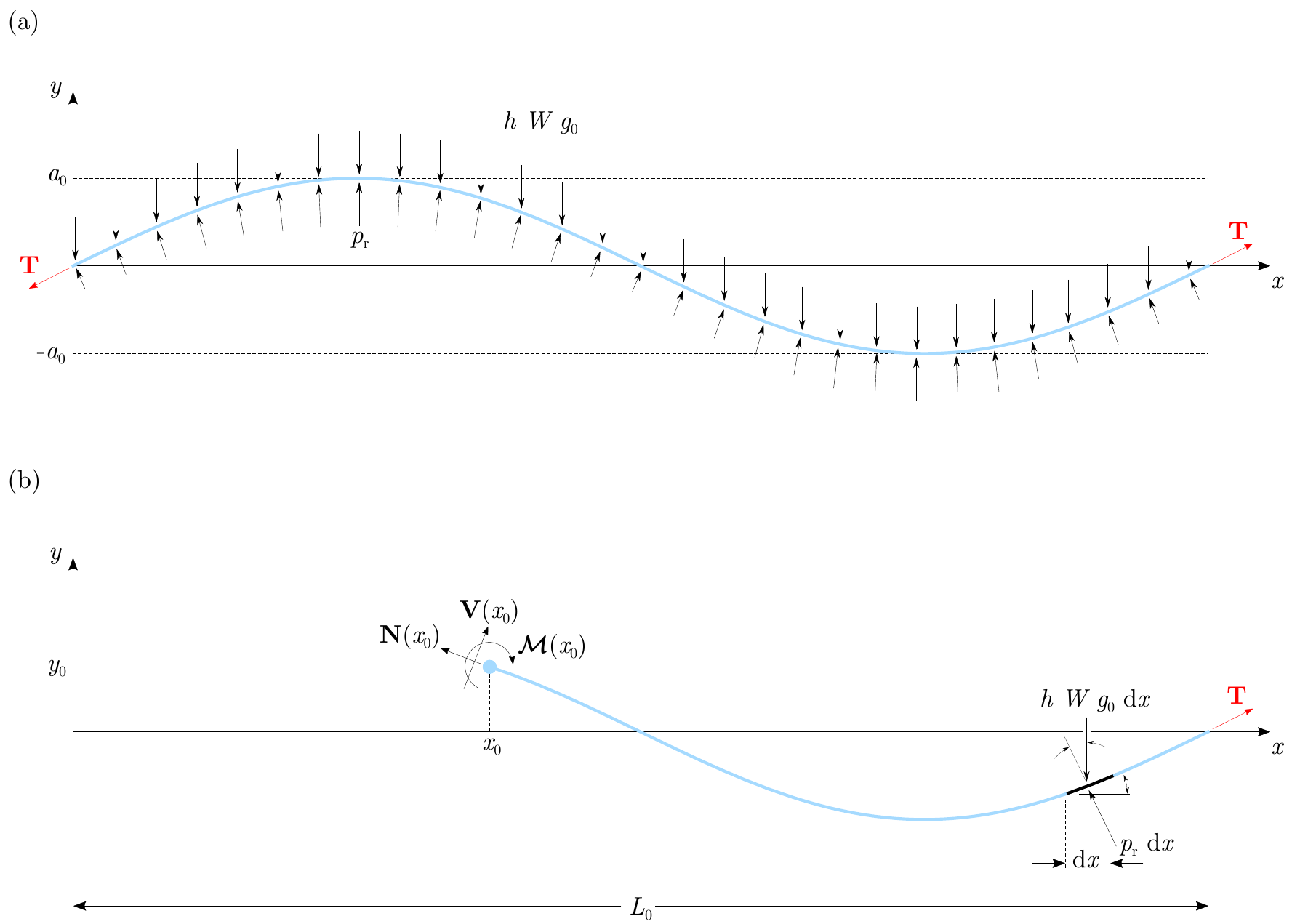}
    \caption{Deformed two-dimensional (unit width) beam lightsail stability model; (a)~force analysis on entire lightsail; (b)~moment analysis about a point on the lightsail located at $x_0$.}
    \label{fig:theoretika}
\end{figure*}

As a preliminary consideration of the problem of a lightsail under radiation loading, an ${L\times W\times h}$ continuous elastic plate model was first constructed that would allow for a static analysis of the criticality between a structurally stable and unstable lightsail (see Fig.~\ref{fig:TheoConcept}). The 3-dimensional plate model was then simplified to a 2-dimensional beam model by setting the lightsail width, $W$, equal to unit length (in meters). The continuous elastic beam model is shown in Fig.~\ref{fig:theoretika}. The following analysis considers introducing a sinusoidal perturbation into the lightsail shape. The restoring bending moment caused by this deformation is then compared to the moment induced by the incident radiation interacting with the curve surface of the lightsail. If the radiation-induced moment exceeds the bending moment associated with the imposed perturbation and acts in the same direction as the perturbation, then presumably the lightsail continues to further deform. If the radiation-induced moment is less than the restoring bending moment of the perturbation, then the sail would be expected to return toward its original, flat configuration. By equating the bending moment of the sinusoidal perturbation with the radiation-induced moment, a critical condition for lightsail stability can be defined in terms of the radiation intensity, the elastic modulus and dimensions of the lightsail, and the amplitude and wave number of the perturbation. The formalism of this approach follows here.

\begin{figure*}[!htp]
    \centering
    \includegraphics[scale=1.0]{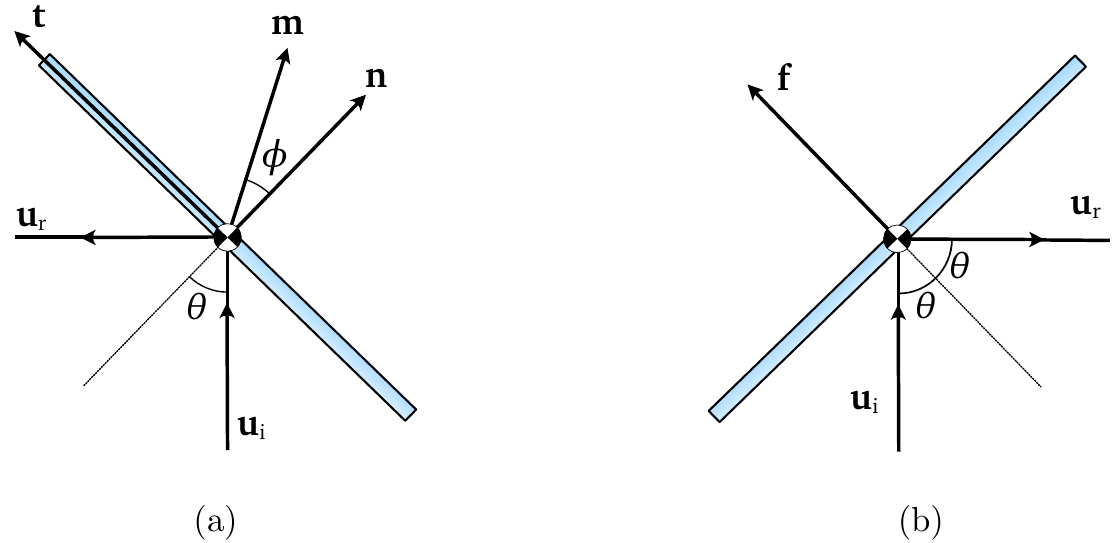}
    \caption{FWM flat lightsail element model; (a)~single element kinematics; (b)~single, specular element dynamics.}
    \label{fig:element}
\end{figure*}

The beam model analysis is here conducted in a non-inertial reference frame accelerating at $g_0$, the lightsail's vertical acceleration, with a body force term ${\rho \, h \, W  \, g_0}$ acting in the $y$-direction on the mass elements of the lightsail in accordance with D’Alembert’s principle, thereby reducing the dynamic problem to a quasi-static problem.\footnote{Given the large magnitude of the vertical accelerations, it is here assumed that ${g_0=\ddot{y}\gg \ddot{x},\ddot{\theta}(s)}$, and thus the horizontal and rotational inertia of the lightsail were deemed negligible and their d'Alembert equivalent was not included into the lightsail beam model.} The presence of radiation pressure is modeled by the distributed loading, $p_\mathrm{r}$. The expression of $p_\mathrm{r}$ can be derived by first considering the kinematics and dynamics of an infinitesimal flat sail element of length $\mathrm{d}s$ as shown in Fig.~\ref{fig:element} for reference. For a lightsail element made of a material with reflection coefficient $r$, absorption coefficient $a$, and transmission coefficient $\Tilde{\tau}$, the total force imparted by radiation of oblique incidence angle $\theta$ can be resolved into the normal-tangential components
\begin{equation}
    \label{eqn:genForce01}
    \begin{aligned}
        \mathbf{f}_\mathrm{n} = \frac{I_0}{c} &\left[ \vphantom{\frac{\varepsilon_\mathrm{f}\,B_\mathrm{f}+\varepsilon_\mathrm{b}\,B_\mathrm{b}}
            {\varepsilon_\mathrm{f}+\varepsilon_\mathrm{b}}} 
        (1+rs) \cos^2{\theta} + B_\mathrm{f} (1-s)r \cos{\theta} \right. \\
        &+ \left. (1-r-\Tilde{\tau}) \frac{\varepsilon_\mathrm{f}\,B_\mathrm{f}+\varepsilon_\mathrm{b}\,B_\mathrm{b}}{\varepsilon_\mathrm{f}
        +\varepsilon_\mathrm{b}} \cos{\theta} \right] \, \mathbf{n},
    \end{aligned}
\end{equation}
\begin{equation}
    \label{eqn:genForce02}
    \mathbf{f}_\mathrm{t} = \frac{I_0}{c} \left( 1-rs \right) \cos{\theta} \sin{\theta} \, \, \mathbf{t},
\end{equation}
where $s$ stands for the fraction of light that is specularly reflected; $\varepsilon_\mathrm{f}$ and $\varepsilon_\mathrm{b}$ stand for the emissivity of the front (reflecting) and back (non-reflecting) surfaces of the sail, respectively; and $B_\mathrm{f}$ and $B_\mathrm{b}$ are the coefficients accounting for the potentially non-Lambertian nature of the sail element surfaces. Together, these two forces generate a resultant whose direction, $\mathbf{m}$, is skewed away from the element's normal:
\begin{equation}
    \label{eqn:genForce03}
    \mathbf{f} = \sqrt{f^2_\mathrm{n}+f^2_\mathrm{t}} \, \, \mathbf{m}.
\end{equation}
This optical lightsail model was first proposed by Forward \cite{forward1989grey} and then was further discussed by Wright and McInnes \cite{wright1992space,mcinnes2004solar} now termed the Forward-Wright-McInnes or FWM model.\footnote{The authors are aware that more accurate lightsail optical models using vector theories such as the Rayleigh-Rice theory have been proposed \cite{vulpetti2012fast,vulpetti2014scattering}, but the FWM model was used here to ensure a straightforward simplification to the case of an ideal reflector. The implementation of physically more accurate optical models can be incorporated into future studies.} For the purposes of this paper, the lightsail element will be assumed to be perfectly reflective (${r=1 \implies a=\Tilde{\tau}=0}$) with all beam reflections being specular (${s=1}$). The resulting force per element then becomes
\begin{equation}
    \label{eqn:genForce04}
    \mathbf{f} = \frac{2I_0}{c} A \cos^2{\theta} \, \, \mathbf{n},
\end{equation}
a force that is entirely normal to the sail element. Normalizing (\ref{eqn:genForce04}) with respect to the element area generates the radiation pressure:
\begin{equation}
\label{eqn:genForce05}
    p_\mathrm{r} = \frac{2I_0}{c} \cos{\theta}.
\end{equation}

To model the presence of surface defects, the lightsail was initially sinusoidally deformed with a deformation profile given by
\begin{equation}
    \label{eqn:deform}
    w = a_0 \sin{\left( \frac{2 \, \pi \, \nu \, x}{L} \right)}
\end{equation}
where $\nu$ is the mode number of the perturbation, which was taken as whole and halved positive integers, i.e., ${\nu = \, \frac{1}{2}, \, 1, \, \frac{3}{2}, \,2,\,}$ \ldots The acceleration of the sinusoidal lightsail can be found by integrating the forces acting on the sail in the $y$-direction:
\begin{equation}
    \sum F_y = M \, g_0
\end{equation}
\begin{equation}
    \label{eqn:vertNewt}
    \int_0^{L} \frac{2 I_0}{c} \, W \, \cos^2{\theta} \, \mathrm{d}x = g_0 \int_0^{L} \rho \, h \, W \frac{\mathrm{d}x}{\cos{\theta}}
\end{equation}
where $\theta$ is the local angle of the lightsail.

Using the small angle approximations $\sin(\theta) \approx \theta \approx \frac{\mathrm{d}w}{\mathrm{d}x}$ and $\cos(\theta) \approx 1 - \frac{\theta^2}{2}$,\footnote{The second order small angle approximation for the secant function here used is $\sec(\theta)\approx 1 + \frac{\theta^2}{2}$.} the acceleration of the lightsail can be solved for
\begin{equation}
    g_0 = \frac{2 I_0}{c \, \rho \, h} \left[ \frac{3L^2}{L^2+a_0^2 \pi^2\nu^2} - 2 \right]
\end{equation}
The term in the brackets represents the second order correction or decrement of lightsail acceleration due to the momentum of photons being scattered off the perturbed reflective surface.  Of note here is that as the amplitude of the perturbations, $a_0$, goes to zero, that is, as the lightsail flattens, the vertical acceleration becomes the acceleration of a flat lightsail (${g_0 \approx \frac{2 I_0}{c \, \rho \, h} \approx \frac{2 I_0}{c \, m}}$) \cite{lubin2016roadmap}. This same result for $g_0$ can be obtained if only a first order small angle approximation is employed by letting ${\cos(\theta) \approx 1}$ instead, making equation (\ref{eqn:vertNewt}) then yield ${g_0 = \frac{2 I_0}{c \, \rho \, h}}$. To keep all subsequent analytical expressions tractable and conceptually insightful, for the rest of this section, only a first-order approximation will be employed.

To investigate the material strength and/or boundary tension that is required to keep the deformed lightsail in equilibrium, a standard method of sections was performed by conceptually cutting the lightsail at a given point \emph{O} (see Fig.~\ref{fig:theoretika}b). Summing the moments acting at that point then gives:
\begin{equation}
    \label{eqn:anamom}
    \begin{aligned}
        \sum &\mathcal{M}_0 - \mathcal{M}^* = - \mathcal{M}(x_0) \\
        &+ \int_{x_0}^{L} \frac{2 I_0}{c} \, W \, \left( x-x_0 \right)  \cos^2{\theta} \mathrm{d}x \\
        &- \int_{x_0}^{L} \frac{2 I_0}{c} \, W \, \left( w_0-w \right) \cos{\theta} \sin{\theta} \, \mathrm{d}x \\
        &+ T \, w(x_0) \cos{\theta}
        - \int_{x_0}^{L} \rho \, h \, g_0 \, W \, \left( x-x_0 \right ) \frac{\mathrm{d}x}{\cos{\theta}}.
    \end{aligned}
\end{equation}
The tension $T$ is tangent to the $L$ end of the lightsail, however, given the small amplitude of the sinusoidal deformation, the dynamic contribution of its vertical component is neglected. The first two integral terms in equation (\ref{eqn:anamom}) represent the moment contributions of the vertical and horizontal components of the radiation pressure loads, respectively. The third integral term containing $\rho \, h \, g_0$ in the integrand represents the moment arising from the d'Alembert body force, $\mathcal{M}^*$. Again using the small angle approximations and retaining only first order terms, equation (\ref{eqn:anamom}) yields
\begin{equation}
    \label{eqn:anamom02}
    \begin{aligned}
        - \mathcal{M}(x_0) &+ \int_{x_0}^{L} \frac{2 I_0}{c} \, W \, \left( x-x_0 \right) \mathrm{d}x \\
        &- \int_{x_0}^{L} \frac{2 I_0}{c} \, W \, \left( w_0-w \right) \frac{\mathrm{d}w}{\mathrm{d}x} \, \mathrm{d}x
        + T \, w(x_0) \\
        &- \int_{x_0}^{L} \rho \, h \, g_0 \, W \, \left( x-x_0 \right) \mathrm{d}x = 0.
    \end{aligned}
\end{equation}
Substituting the first order result for the lightsail acceleration, $g_0 = \frac{2 I_0}{c \, \rho \, h}$ into equation (\ref{eqn:anamom02}) causes the moments generated by the vertical component of the radiation pressure loading and by the d'Alembert body force to cancel each other to first order. Equation (\ref{eqn:anamom02}) thus reduces to
\begin{equation}
    \label{eqn:anamom03}
    - \mathcal{M}(x_0) - \int_{x_0}^{L} \frac{2 I_0}{c} \, W \, \left( w_0-w \right)
    \frac{\mathrm{d}w}{\mathrm{d}x} \, \mathrm{d}x
    + T \, w(x_0) = 0.
\end{equation}
To carry out the remaining integral, recall that the lightsail deformation is defined by equation (\ref{eqn:deform}). Substituting the appropriate expressions for $w$, $w_0$, and $\frac{\mathrm{d}w}{\mathrm{d}x}$, the integration can be carried out to yield
\begin{equation}
    \label{eqn:anamom04}
    - \mathcal{M}(x_0)
    - \frac{ I_0}{c} \, W \, a_0^2 \sin^2{\left( \frac{2 \, \pi \, \nu \, x_0}{L} \right)}
    + T \, a_0 \sin{\left( \frac{2 \, \pi \, \nu \, x_0}{L} \right)} = 0.
\end{equation}
Equation (\ref{eqn:anamom04}) defines a state of static equilibrium of the lightsail: The net moment applied to the sail by the acceleration body-force term and the radiation pressure is balanced by a material moment resulting from the bending of the lightsail. The elastic modulus necessary to provide the moment $\mathcal{M}_0(x_0)$ can be found from the well known moment-curvature relation
\begin{equation}
    \label{eqn:MomCurv}
    \mathcal{M}_0 = E \, I \, \kappa(x_0)
\end{equation}
where $E$ is the material elastic modulus (Young's modulus), $I$ is the second moment of area about the $z$-axis of the sail cross-section,\footnote{For the current lightsail model, $I = \frac{1}{12} W h^3$.} and $\kappa$ is the curvature of the deformed lightsail.  Using the small angle approximation for the curvature, ${\kappa \approx \frac{\mathrm{d}^2 w}{\mathrm{d}x^2}}$, the material strength required for equilibrium can be solved for:
\begin{equation}
    \label{eqn:anamom05}
    E = - \frac{3}{\pi^2} \, \frac{L^2}{\nu^2 h^3}
    \left[ \frac{ I_0}{c} \, a_0 \sin{\left( \frac{2 \, \pi \, \nu \, x_0}{L} \right)}
    + \frac{T}{W} \right].
\end{equation}

By comparing the applied moments due to radiation and acceleration (in the non-inertial frame) and the moments due to the bending of the stiff sail and due to the applied tension, criteria for the structural stability of the lightsail can be obtained.  For example, if the moments due to the lightsail's bending stiffness and tension exceed the moments applied by radiation and acceleration, it is plausible that the sail remains stable. If the moments caused by radiation pressure exceed the bending moment of the lightsail, however, the lightsail is likely to undergo further--potentially catastrophic--deformation. From these considerations, it is possible to define a critical value of elastic modulus in the absence of tension, that is, by letting ${T = 0}$:
\begin{equation}
    \label{eqn:ECR}
    E_\mathrm{cr} = - \frac{3}{2} \left( \frac{2 I_0}{c} \right) \frac{a_0 L^2}{\pi^2 \, \nu^2 \, h^3}
    \sin{\left( \frac{2 \, \pi \, \nu \, x_0}{L} \right)} \text{, with}
\end{equation}
\begin{equation}
    \label{eqn:ECRMAX}
    E_\mathrm{cr_{max}} = \frac{3}{2} \left( \frac{2 I_0}{c} \right) \frac{a_0 L^2}{\pi^2 \, \nu^2 \, h^3},
\end{equation}
where the extrema values, $E_\mathrm{cr_{max}}$, occur at the peaks and troughs of the sinusoid.  For the case of tension (in the absence of bending stiffness), the critical value of tension (per unit width of the lightsail) can be found
\begin{equation}
    \label{eqn:TCR}
    T_\mathrm{cr} = - \frac{1}{2} \left( \frac{2 I_0}{c} \right) a_0
    \sin{\left( \frac{2 \, \pi \, \nu \, x_0}{L} \right)} \text{, with}
\end{equation}
\begin{equation}
    \label{eqn:TCRMAX}
    T_\mathrm{cr_{max}} = \frac{1}{2} \left ( \frac{2 I_0}{c} \right ) a_0.
\end{equation}
The results of this quasi-static analysis provide a candidate for the functional form of the relation between material properties (i.e., elastic modulus), perturbation amplitude and mode number, and the radiation intensity. The results of equation (\ref{eqn:TCR}) also suggest a critical value of the tension necessary to prevent the development of instability. \ref{sec:appendix A} contains a derivation of the functional expression of the critical material modulus based on a different, energy-based approach. These findings will help guide the computational simulations of the full lightsail dynamics found in the next section of this paper.


\section{Numerical Considerations}
\label{sec:numerical}


\subsection{Rigid-Element Lightsail Model}
\label{subsec:Rigid}
\begin{figure*}[t]
    \centering
    \includegraphics[width=0.90\textwidth]{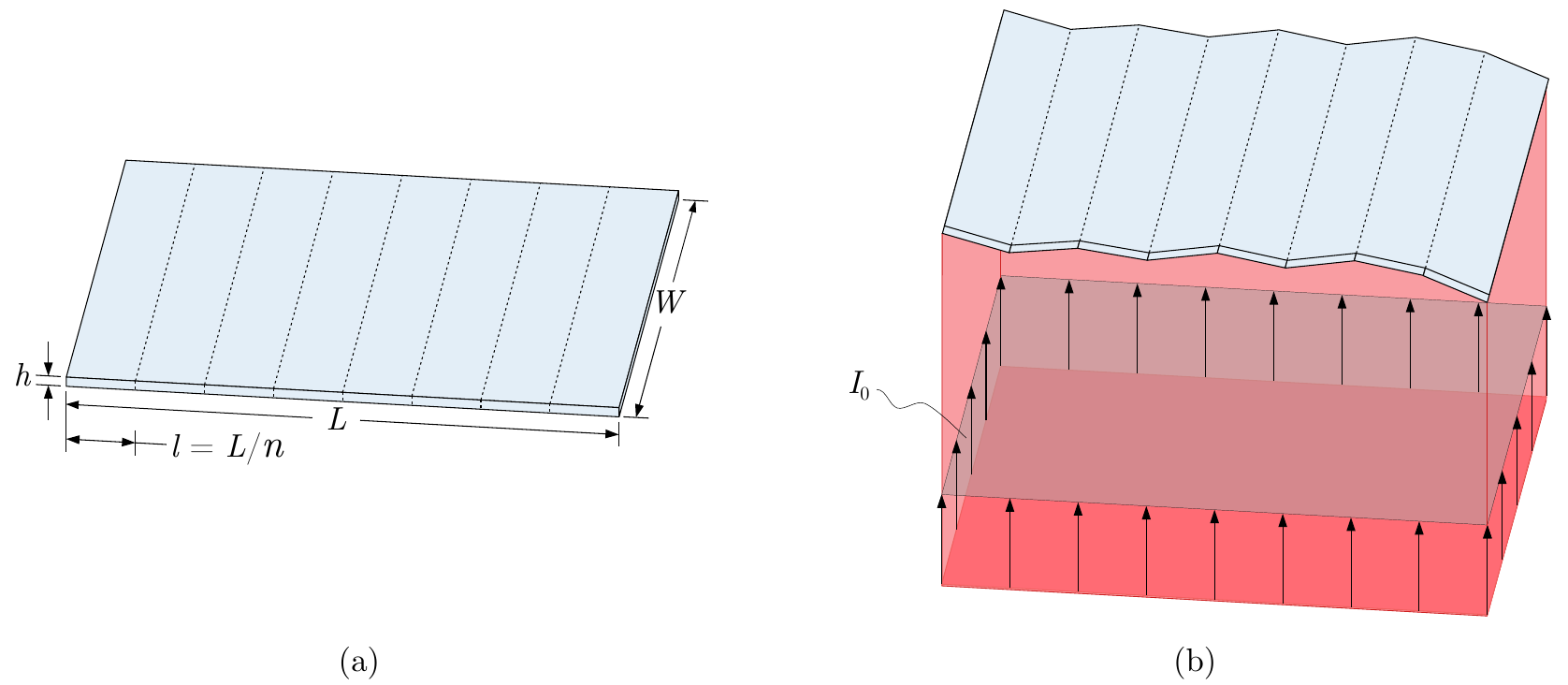}
    \caption{Three-dimensional numerical model of the lightsail; (a)~the lightsail, sliced; (b)~the lightsail, perturbed with an incident uniform laser beam.}
    \label{fig:NumConcept}
\end{figure*}

\begin{figure*}[!htp]
    \centering
    \includegraphics[width=0.90\textwidth]{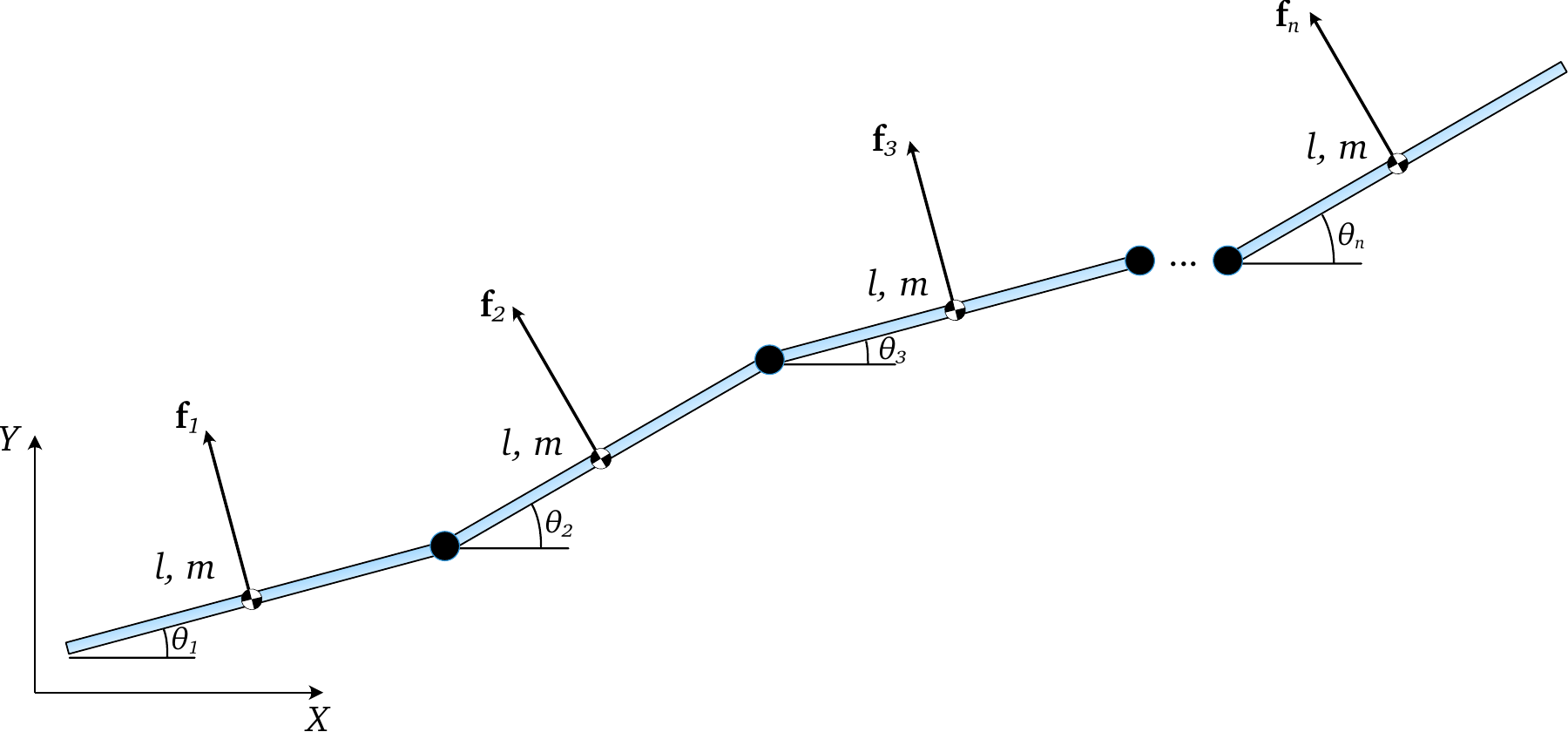}
    \caption{Two-dimensional (unit width) finite-element model of the lightsail using rigid slices and frictionless connections.}
    \label{fig:RigidModel}
\end{figure*}

With the intent of achieving the rapid transit of a small-scale spacecraft equipped with a 1~meter sail to Mars (1~AU) in 30~minutes or to 0.3~$c$ for interstellar flight within three minutes, the mission at launch would require the use of a ${0.1-\SI{100}{\giga \watt}}$ laser load \cite{lubin2016roadmap,parkin2018breakthrough}. The dynamic stability of the lightsail at launch was analyzed by first considering a rectangular sail with dimensions ${L \times W \times h}$ where ${L = W = \SI{1}{\meter}}$ and $h$ is the sail's thickness. The sail was then divided into $n$ equal slices of length ${l=L/n}$ and mass $m$, each attached together via frictionless hinges (see Figs.~\ref{fig:NumConcept} and~\ref{fig:RigidModel}). The system was further subjected to a laser beam of uniform intensity distribution. Given the idealized flat surface of each slice, specular reflection was assumed to occur between an individual photon and a sail element. Because the reflection was specular and complete, the force resulting from the interaction between the laser beam and the lightsail was normal to the sail element at the point of photon incidence. This uniform pressure thus results in the presence of concentrated forces along the center of mass (CoM) of each individual slice (recall Fig.~\ref{fig:element}b). In the limit where the number of rigid slice elements is very large (i.e., when ${n \rightarrow \infty}$) this assumption holds true even for non-uniform laser intensities because of the infinitesimal length of each element. The force imparted to the $i$th element by the photon pressure is thus
\begin{equation}
    \label{eqn:perEle}
    \mathbf{f}_i = \frac{2I_0}{c} A \cos^2{\theta_i} \, \, \mathbf{n}
\end{equation}
where ${A = W\times l}$ is the slice element area and $\theta_i$ is the $i\mathrm{th}$ element's inclination with respect to the horizontal.

To investigate the dynamical behavior of the system, the Lagrangian formalism of mechanics was used. First, the Lagrangian of the system was constructed alongside its constraints (note that ${I_\mathrm{G} = \frac{1}{12}ml^2}$ is the moment of inertia of the sail element about its center of mass).
\begin{equation}
\label{eqn:rigidLagrangian}
    \mathcal{L}_\mathrm{rigid} = \frac{1}{2} m \sum_{j=1}^n \left( \Dot{x}^2_j + \Dot{y}^2_j \right) 
    + \frac{1}{2} I_\mathrm{G} \sum_{j=1}^n \Dot{\theta}_j^2 \text{, with}
\end{equation}
\vspace{-20pt}
\begin{align}
    \label{eqn:consX01}
    x_j &= x_1(t) + \frac{l}{2} \left( \cos{\theta_1} + 2 \sum_{i=2}^{j-1} \cos{\theta_i} + \cos{\theta_j} \right)
    \text{, and} \\
    \label{eqn:consY01}
    y_j &= y_1(t) + \frac{l}{2} \left( \sin{\theta_1} + 2 \sum_{i=2}^{j-1} \sin{\theta_i} + \sin{\theta_j} \right).
\end{align}
Further, the non-conservative generalized forces that arise due to the radiation forces, $\mathbf{f}_i$, are
\begin{equation}
    \label{eqn:genF}
    Q_{j} = \sum_{i=1}^{n} \mathbf{f}_{i} \cdot \frac{\partial \mathbf{r}_{i}}{\partial q_{j}} \text{, with}
\end{equation}
\begin{equation*}
    \mathbf{f}_{i} = f_{0} \cos^2{\theta_i} \left( -\sin{\theta_i}, \, \cos{\theta_i}, \, 0 \right) \text{, and}  
\end{equation*}
\begin{equation*}
    f_0 = \frac{2I_0}{c}Wl \text{,\quad}
    \mathbf{r}_i = \left( x_i, \, y_i, \, 0 \right).
\end{equation*}
Using the above terms, the equations of motion (EOMs) of the system were derived using the Euler-Lagrange equations
\begin{equation}
    \label{eqn:EulerLagrange}
    \frac{\mathrm{d}}{\mathrm{d} t} \left( \frac{\partial \mathcal{L}}{\partial \dot{q}_j} \right)
    - \frac{\partial \mathcal{L}}{\partial q_j} = Q_j
\end{equation}
where ${q_i = x_1, \, y_1, \, \theta_1, \, \dots, \, \theta_n}$ are the generalized coordinates of the system encapsulated in the vector of unknowns $\mathbf{q}_{\mathrm{rigid}} = [x_1, \, y_1, \, \boldsymbol{\theta}]^{\mathrm{T}}$. After applying the derivative operators, the following equations were obtained:
\\
--- \quad $x_1$ equation of motion for the rigid model
\begin{equation}
    \label{eqn:x1Rigid}
    \begin{aligned}
        - 2 n \Ddot{x}_1 + &\sum_{j=1}^n a^{xy}_j  \sin{\theta_j} \Ddot{\theta}_j \, = \\
        &- \sum_{j=1}^n a^{xy}_j \cos{\theta_j} \dot{\theta}_j^2
        + \frac{2}{m}f_0 \sum_{j=1}^n \sin{\theta_j} \cos^2{\theta_j};
    \end{aligned}
\end{equation}
--- \quad $y_1$ equation of motion for the rigid model
\begin{equation}
    \label{eqn:y1Rigid}
    \begin{aligned}
        2 n \Ddot{y}_1 + \sum_{j=1}^n &a^{xy}_j \cos{\theta_j} \Ddot{\theta}_j \, = \\
        &\sum_{j=1}^n a^{xy}_j \sin{\theta_j} \dot{\theta}_j^2 + \frac{2}{m}f_0 \sum_{j=1}^n \cos^3{\theta_j};
    \end{aligned}
\end{equation}

\vspace{0.5cm}

\vspace{0.6cm}

\noindent --- \quad $\theta_1$ equation of motion  for the rigid model
\begin{equation}
    \label{eqn:th1rigid}
    \begin{aligned}
        - 6 (n&-1) l \sin{\theta_1} \Ddot{x}_1 + 6 \left( n-1 \right) l \cos{\theta_1} \Ddot{y}_1 \\
        &+ \sum_{j=1}^n a^{\theta_1}_{j} \cos{\left( \theta_1 - \theta_j \right)} \Ddot{\theta}_j \, = \\
        &- \sum_{j=1}^n a^{\theta_1}_{j} \sin{\left( \theta_1 - \theta _j \right)} \dot{\theta}_j^2 \\
        &+ \frac{12}{m} \frac{l}{2} f_0 \sum_{j=2}^n \left( \sin{\theta_1} \sin{\theta_j} \cos^2{\theta_j} + \cos{\theta_1}
        \cos^3{\theta_j} \right);
    \end{aligned}
\end{equation}
\noindent --- \quad $\theta_k$ equations of motion  for the rigid model where $k = 2, \, 3, \, 4, \, \ldots, \, n-1$
\begin{equation}
    \label{eqn:thkRigid}
    \begin{aligned}
        - 6 (2n&-2k+1) l \sin{\theta_k} \Ddot{x}_1 + 6 (2n-2k+1) l \cos{\theta_k} \Ddot{y}_1 \\
        &+ \sum_{j=1}^n a^{\theta_k}_{kj} \cos{\left( \theta_k - \theta_j \right)} \Ddot{\theta}_j \, = \\
        &- \sum_{j=1}^n a^{\theta_k}_{kj} \sin{\left( \theta_k - \theta_j \right)} \dot{\theta}_j^2 \\
        &+ \frac{12}{m} \frac{l}{2} f_0 \left( \sin^2{\theta_k} \cos^2{\theta_k} + \cos^4{\theta_k} \right) \\
        &+ \frac{12}{m} l f_0 \sum_{j=k+1}^n \left( \sin{\theta_k} \sin{\theta_j} \cos^2{\theta_j} + \cos{\theta_k}
        \cos^3{\theta_j} \right);
    \end{aligned}
\end{equation}
--- \quad $\theta_n$ equation of motion  for the rigid model
\begin{equation}
    \label{eqn:thnRigid}
    \begin{aligned}
        - 6 l \sin{\theta_n} \Ddot{x}_1 &+ 6 l \cos{\theta_n} \Ddot{y}_1
        + \sum_{j=1}^n a^{\theta_n}_{nj} \cos{\left( \theta_n - \theta_j \right)} \Ddot{\theta}_j \, = \\
        &- \sum_{j=1}^n a^{\theta_k}_{nj} \sin{\left( \theta_n - \theta_j \right)} \dot{\theta}_j^2 \\
        &+ \frac{12}{m} \frac{l}{2} f_0 \left( \sin^2{\theta_n} \cos^2{\theta_n} + \cos^4{\theta_n} \right);
    \end{aligned}
\end{equation}
with the auxiliary coefficients
\begin{align*}
     a^{xy}_j = & 
    \begin{cases} 
          \left( n-1 \right) l &\text{for } j=1 \\
          (2n-2j+1)l &\text{for } j>1 
    \end{cases}
    \\
    a^{\theta_1}_{j} = & 
    \begin{cases} 
        3 (n-1)l^2 + l^2 &\text{for } j=1 \\
        3 (2n-2j+1) l^2 &\text{for } j>1
    \end{cases}
    \\
    a^{\theta_k}_{kj} = & 
    \begin{cases} 
        3 (2n-2k+1) l^2 &\text{for } j=1 \\
        6 (2n-2k+1) l^2 &\text{for } 1<j<k \\
        12 (n-k) l^2 + 4 l^2 &\text{for } j=k \\
        6 (2n-2j+1) l^2 &\text{for } j>k. 
    \end{cases}
\end{align*}

\subsection{Torsion Lightsail Model}
\label{subsec:Torsion}
\begin{figure*}[!htp]
    \centering
    \includegraphics[width=0.9\textwidth]{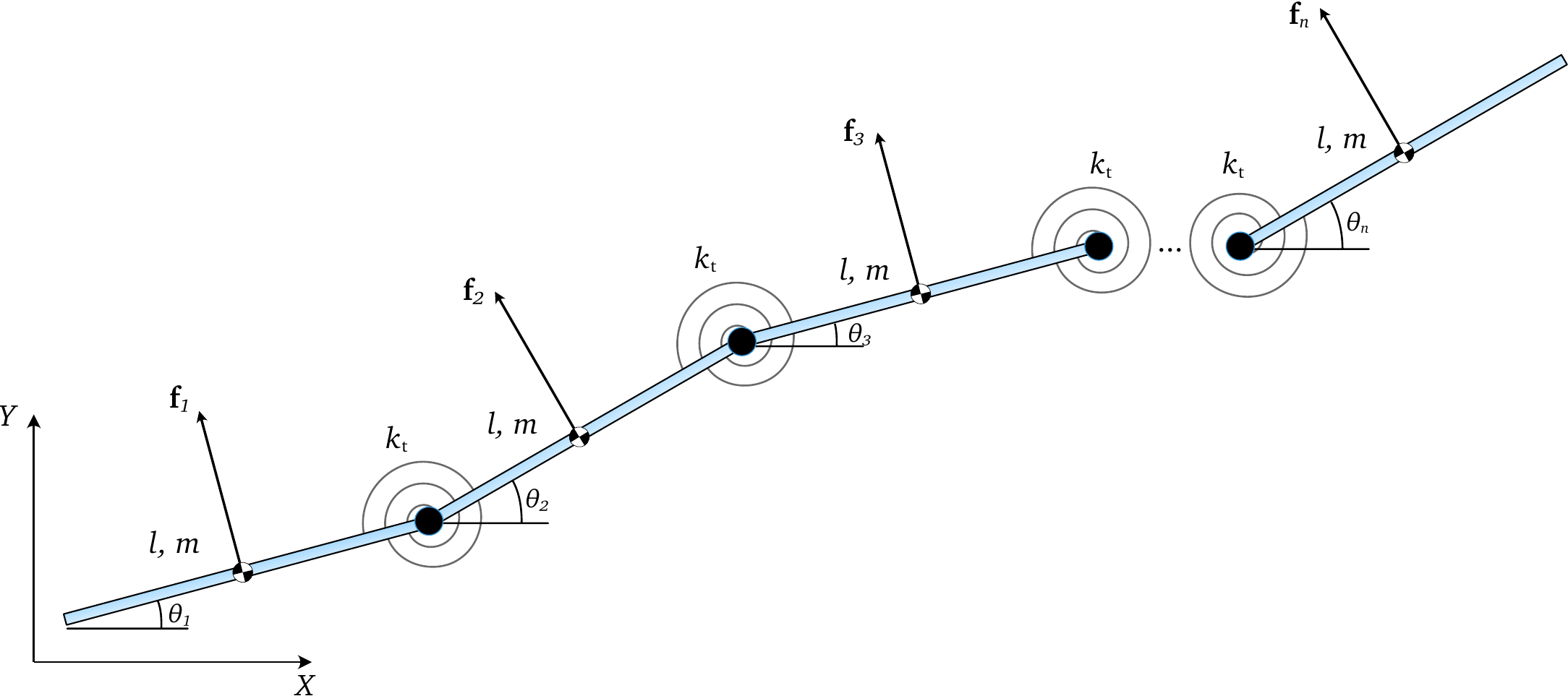}
    \caption{Finite-element model generalization using torsional spring elements to simulate material bending stiffness.}
    \label{fig:torsion01}
\end{figure*}

 To include the material bending stiffness of the lightsail within the study, the finite-element model was further generalized by attaching torsional spring elements with stiffness constant $k_\mathrm{t} = EI/l$ to the hinges that connect the rigid slices together (see Fig.~\ref{fig:torsion01}). To see a full derivation of this discrete torsional spring constant from the moment-curvature relation, the reader is referred to \cite{fritzkowski2011tetherTorsion}. The quantity $EI$ stands for the flexural rigidity of the individual slices with $E$ being the sail material elastic modulus and $I$ being the second moment of area of each rigid element:
\begin{equation}
    \label{eqn:flexRigidty}
    EI = E \left( \frac{1}{12} L h^3 \right).
\end{equation}

Employing the same variational approach as for the rigid lightsail model, the Lagrangian of the system is modified with the appearance of a potential energy term arising from the presence of the linear torsion springs:
\begin{equation}
    \label{eqn:torsionLagrangian}
    \mathcal{L}_\mathrm{torsion} = \frac{1}{2} m \sum_{j=1}^n \left( \Dot{x}_j^2 + \Dot{y}_j^2 \right) 
    + \frac{1}{2} I_\mathrm{G} \sum_{j=1}^n \Dot{\theta}_j^2
    - \frac{1}{2} k_\mathrm{t} \sum_{j=2}^n \left( \theta_j - \theta_{j-1} \right)^2
\end{equation}
Coupling equation (\ref{eqn:torsionLagrangian}) with equations (\ref{eqn:consX01}), (\ref{eqn:consY01}), and (\ref{eqn:genF}) and applying the Euler-Lagrange equation, a new set of ordinary differential equations (ODEs) describing this extended lightsail model was obtained:
\\
--- \quad $x_1$ equation of motion for the torsion model
\begin{equation}
    \label{eqn:x1Torsion}
    \begin{aligned}
        - 2 n \Ddot{x}_1 + &\sum_{j=1}^n a^{xy}_j  \sin{\theta_j} \Ddot{\theta}_j \, = \\
        &- \sum_{j=1}^n a^{xy}_j \cos{\theta_j} \dot{\theta}_j^2
        + \frac{2}{m}f_0 \sum_{j=1}^n \sin{\theta_j} \cos^2{\theta_j};
    \end{aligned}
\end{equation}
--- \quad $y_1$ equation of motion for the torsion model
\begin{equation}
    \label{eqn:y1Torsion}
    \begin{aligned}
        2 n \Ddot{y}_1 + \sum_{j=1}^n &a^{xy}_j  \cos{\theta_j} \Ddot{\theta}_j \, = \\
        &\sum_{j=1}^n a^{xy}_j \sin{\theta_j} \dot{\theta}_j^2 + \frac{2}{m}f_0 \sum_{j=1}^n \cos^3{\theta_j};
    \end{aligned}
\end{equation}
--- \quad $\theta_1$ equation of motion  for the torsion model
\begin{equation}
    \label{eqn:th1Torsion}
    \begin{aligned}
        -6 (n&-1) l \sin{\theta_1} \Ddot{x}_1 + 6 \left( n-1 \right) l \cos{\theta}_1 \Ddot{y}_1 \\
        &+ \sum_{j=1}^n a^{\theta_1}_{j} \cos{\left( \theta_1 - \theta_j \right)} \Ddot{\theta}_j \, = \\ 
        &- \sum_{j=1}^n a^{\theta_1}_{j} \sin{\left( \theta_1 - \theta_j \right)} \dot{\theta}_j^2 \;
        \tcbhighmath{+ \frac{12}{m} k_\mathrm{t} \left( \theta_{2} - \theta_1 \right)} \\
        &+\frac{12}{m} \frac{l}{2} f_0 \sum_{j=2}^n \left( \sin{\theta_1} \sin{\theta_j} \cos^2{\theta_j} + \cos{\theta_1} \cos^3{\theta_j} \right);
    \end{aligned}
\end{equation}
--- \quad $\theta_k$ equations of motion  for the torsion model where $k=2, \, 3, \, 4, \, \ldots, \, n-1$
\begin{equation}
    \label{eqn:thkTorsion}
    \begin{aligned}
        - 6 (2n&-2k+1) l \sin{\theta_k} \Ddot{x}_1 + 6 (2n-2k+1) l \cos{\theta_k} \Ddot{y}_1 \\
        &+ \sum_{j=1}^n a^{\theta_k}_{kj} \cos{\left( \theta_k - \theta_j \right)} \Ddot{\theta}_j \, = \\
        &- \sum_{j=1}^n a^{\theta_k}_{kj} \sin{\left( \theta_k - \theta_j \right)} \dot{\theta}_j^2 \\
        &\tcbhighmath{- \frac{12}{m} k_\mathrm{t} \left( \theta_k - \theta_{k-1} \right)
            +\frac{12}{m} k_\mathrm{t} \left( \theta_{k+1} - \theta_k \right)} \\
        &+ \frac{12}{m}\frac{l}{2} f_0 \left( \sin^2{\theta_k} \cos^2{\theta_k} + \cos^4{\theta_k} \right) \\
        &+ \frac{12}{m} l f_0 \sum_{j=k+1}^n \left( \sin{\theta_k} \sin{\theta_j} \cos^2{\theta_j}
        + \cos{\theta_k} \cos^3{\theta_j} \right);
    \end{aligned}
\end{equation}
--- \quad $\theta_n$ equation of motion  for the torsion model
\begin{equation}
    \label{eqn:thnTorsion}
    \begin{aligned}
        -6 l \sin{\theta_n} \Ddot{x}_1 &+ 6 l \cos{\theta_n} \Ddot{y}_1
        + \sum_{j=1}^n a^{\theta_n}_{nj} \cos{\left( \theta_n - \theta_j \right)} \Ddot{\theta}_j \, = \\
        &- \sum_{j=1}^n a^{\theta_k}_{nj} \sin{\left( \theta_n - \theta_j \right)} \dot{\theta}_j^2 \;
        \tcbhighmath{- \frac{12}{m} k_\mathrm{t} \left( \theta_n - \theta_{n-1} \right)} \\
        &+ \frac{12}{m} \frac{l}{2} f_0 \left( \sin^2{\theta_n} \cos^2{\theta_n} + \cos^4{\theta_n} \right);
    \end{aligned}
\end{equation}
with the same auxiliary coefficients $a^{xy}_j$, $a^{\theta_1}_{j}$, and $a^{\theta_k}_{kj}$ as in the rigid case. Note that the only difference between the equations of motion of the torsion and rigid models are the addition of the torsional spring restoring moments boxed in the RHS of equations (\ref{eqn:th1Torsion}) to (\ref{eqn:thnTorsion}). For the sake of completion and clarity, however, the entire set of ODEs is given. The Supplementary Material accompanying this paper contains a section describing how the torsion lightsail model was validated against classical Euler--Bernoulli beam theory.

\subsection{Tension and Torsion (TnT) Lightsail Model}
\label{subsec:TnT}
\begin{figure*}[!htp]
    \centering
    \includegraphics[width=0.9\textwidth]{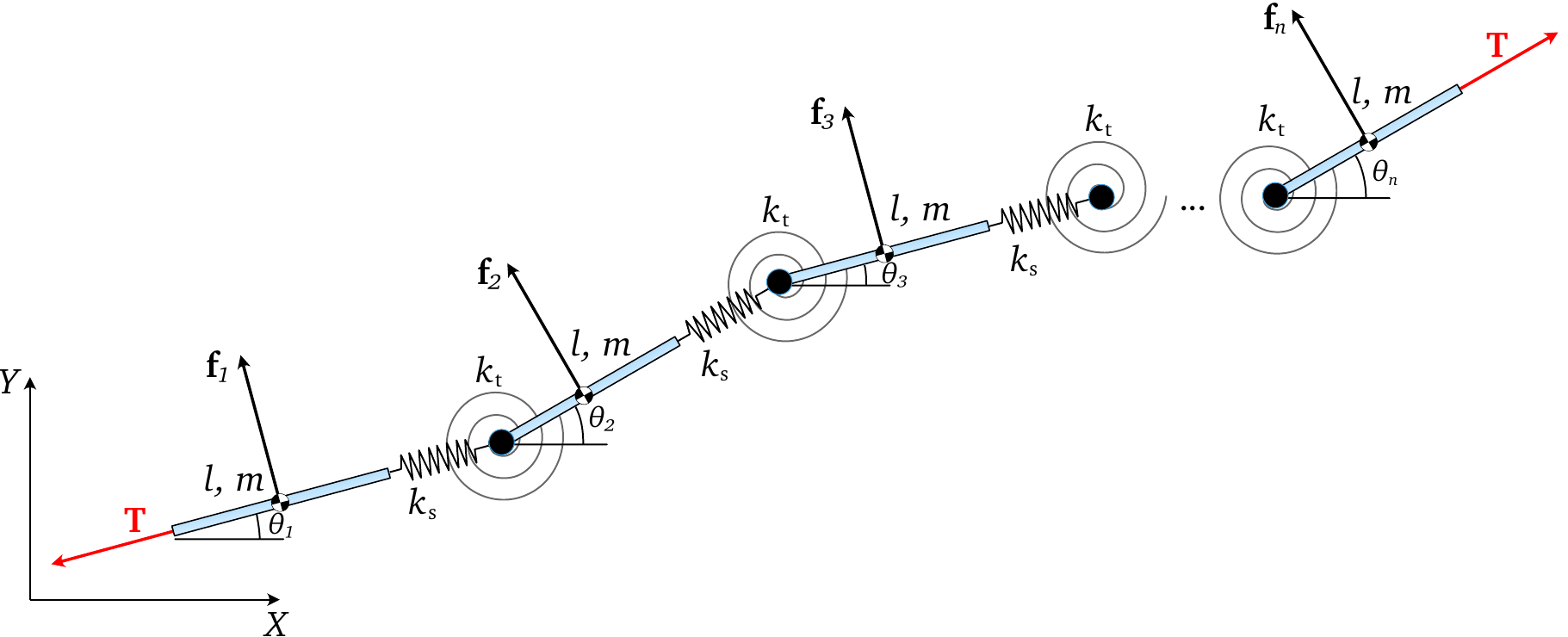}
    \caption{Final FE model generalization including the addition of rectilinear spring elements and boundary tension forces.}
    \label{fig:tensile01}
\end{figure*}

To include both bending stiffness \emph{and} tensile, linear stiffness of the material, the lightsail FE model was further generalized by the addition of rectilinear springs with stiffness constant ${k_\mathrm{s} = (n-1) \frac{E \, h \, W}{L}}$ and initial/rest length $l_\mathrm{s}$. Further, to simulate a \emph{pre-tensioned} sail, boundary tension was applied at both ends of the discretized lightsail. This new model is depicted in Fig.~\ref{fig:tensile01}. The reader is referred to the Supplementary Material for a validation of this model against linear wave theory.

The formulation of this new model is now considerably increased in complexity since, to keep track of the rectilinear spring deformation, an additional set of generalized coordinates ${\mathbf{z} = [z_1, \, \ldots, \, z_{n-1}]^\mathrm{T}}$ needs to be included along with the previous set of coordinates $x_1$, $y_1$, and ${\boldsymbol{\theta} = [\theta_1, \, \ldots, \, \theta_{n}]^\mathrm{T}}$. This makes for a full vector of generalized coordinates ${\mathbf{q}_{\mathrm{TnT}} = [x_1, \, y_1, \, \boldsymbol{\theta}, \, \mathbf{z}]^\mathrm{T}}$, which contains ${2n+1}$ equations instead of the previous ${n+2}$. Given the presence of rectilinear springs, the Lagrangian of the system was also modified with another potential energy term:
\begin{equation}
    \label{eqn:TnTLagrangian}
    \begin{aligned}
        \mathcal{L}_\mathrm{TnT} = \frac{1}{2} m &\sum_{j=1}^n \left( \Dot{x}_j^2 + \Dot{y}_j^2 \right)
        + \frac{1}{2} I_\mathrm{G} \sum_{j=1}^n \Dot{\theta}^2 \\
        &- \frac{1}{2} k_\mathrm{t} \sum_{j=2}^n \left( \theta_j - \theta_{j-1} \right)^2
        - \frac{1}{2} k_\mathrm{s} \sum_{j=1}^{n-1} z_j^2.
    \end{aligned}
\end{equation}
The constraint equations were also revised; to relate the coordinates of each element's center of mass to the first element's coordinate ($x_1$, $y_1$), the length of the springs, (${l_\mathrm{s} + z_k}$), now needed to be accounted for in addition to the length of each rigid element:
\begin{equation}
    \label{eqn:consX02}
    \begin{aligned}
        x_j = x_1(t) &+ \left( \frac{l}{2}+l_\mathrm{s}+z_1 \right) \cos{\theta_1} \\
        &+ \sum_{i=2}^{j-1} \Big\{ \left( l+l_\mathrm{s}+z_i \right) \cos{\theta_i} \Big\} + \frac{l}{2} \cos{\theta_j};
    \end{aligned}
\end{equation}
\vspace{-8 pt}
\begin{equation}
    \label{eqn:consY02}
    \begin{aligned}
        y_j = y_1(t) &+ \left( \frac{l}{2}+l_\mathrm{s}+z_1 \right) \sin{\theta_1} \\
        &+ \sum_{i=2}^{j-1} \Big\{ \left( l+l_\mathrm{s}+z_i \right) \sin{\theta_i} \Big\} + \frac{l}{2} \sin{\theta_j}.
    \end{aligned}
\end{equation}
Given the inclusion of boundary tension, the generalized force components were also modified:
\begin{equation}
    \label{eqn:genFTnT}
    \begin{aligned}
        Q_{j} = &- \mathrm{T}(\cos{\theta_1}, \sin{\theta_1}, 0) \cdot \frac{\partial \mathbf{r}_{T_1}}{\partial q_j}
        +\sum_{i=1}^{n} \mathbf{f}_i \cdot \frac{\partial \mathbf{r}_i}{\partial q_j} \\
        &+ \mathrm{T}(\cos{\theta_n}, \sin{\theta_n}, 0)\cdot \frac{\partial \mathbf{r}_{T_n}}{\partial q_j},
    \end{aligned}
\end{equation}
where
\begin{align*}
    \mathbf{r}_{T_1} &= \left( x_1 - \frac{l}{2} \cos{\theta_1}, \, y_1 - \frac{l}{2} \sin{\theta_1}, \, 0 \right), \\[0.5ex]
    \mathbf{r}_{T_n} &= \left( x_n + \frac{l}{2} \cos{\theta_n}, \, y_n + \frac{l}{2} \sin{\theta_n}, \, 0 \right), \\[0.5ex]
    \mathbf{T}_1 &= T(- \cos{\theta_1}, \, -\sin{\theta_1}, \, 0) \text{, and} \\[0.5ex]
    \mathbf{T}_n &= T(\cos{\theta_n}, \, \sin{\theta_n}, \, 0);
\end{align*}
with $\mathbf{f}_i$ and $\mathbf{r}_i$ as in the previous models.

Applying the Euler-Lagrange variational equation using equations (\ref{eqn:TnTLagrangian}) through to (\ref{eqn:genFTnT}) and simplifying yields a system of coupled ODEs of size ${2n+1}$: \\ \\
--- \quad{} $x_1$ equation of motion for the TnT model
\begin{equation}
    \label{eqn:x1TnT}
    \begin{aligned}
        - 2 n \Ddot{x}_1 &+ \sum_{j=1}^n A^{xy}_j \sin{\theta_j} \Ddot{\theta}_j
        - \sum_{j=1}^{n-1} B^{xy}_j \cos{\theta_j} \Ddot{z}_j \, = \\
        &- \sum_{j=1}^n A^{xy}_j \cos{\theta_j} \dot{\theta}_j^2
        - 2 \sum_{j=1}^{n-1} B^{xy}_j \sin{\theta_j} \dot{z}_j \dot{\theta}_j \\
        &+ \frac{2}{m} f_0 \sum_{j=1}^n \sin{\theta_j} \cos^2{\theta_j}
        + \frac{2}{m} T \left( \cos{\theta_1} - \cos{\theta_n} \right);
    \end{aligned}
\end{equation}
--- \quad $y_1$ equation of motion for the TnT model
\begin{equation}
    \label{eqn:y1TnT}
    \begin{aligned}
        2 n \Ddot{y}_1 + &\sum_{j=1}^n A^{xy}_j \cos{\theta_j} \Ddot{\theta}_j
        + \sum_{j=1}^{n-1} B^{xy}_j \sin{\theta_j} \Ddot{z}_j \, = \\
        &\sum_{j=1}^n A^{xy}_j \sin{\theta_j} \dot{\theta}_j^2
        - 2 \sum_{j=1}^{n-1} B^{xy}_j \cos{\theta_j} \dot{z}_j \dot{\theta}_j \\
        &+ \frac{2}{m} f_0 \sum_{j=1}^n \cos^3{\theta_j}
        + \frac{2}{m} T \left( \sin{\theta_n} - \sin{\theta_1} \right);
    \end{aligned}
\end{equation}
--- \quad $z_k$ equations of motion  for the TnT model where $k = 1, \, 2, \, 3, \, \ldots, \, n-1$
\begin{equation}
    \label{eqn:zkTnT}
    \begin{aligned}
        2 (n&-k) \cos{\theta_k} \Ddot{x}_1 + 2 (n-k) \sin{\theta_k} \Ddot{y}_1 \\
        &+ \sum_{j=1}^n A^{z}_{kj} \sin{\left( \theta_k - \theta_j \right)} \Ddot{\theta}_j
        + \sum_{j=1}^{n-1} B^{z}_{kj} \cos{\left( \theta_k - \theta_j \right)} \Ddot{z}_j \, = \\
        &\phantom{+ \,} \sum_{j=1}^n A^{z}_{kj} \cos{\left( \theta_k - \theta_j \right)} \dot{\theta}_j^2 \\
        & - 2 \sum_{j=1}^{n-1} B^{z}_{kj} \sin{\left( \theta_k - \theta_j \right)} \dot{z}_j \dot{\theta}_j \\
        &+ \frac{2}{m} f_0 \sum_{j=k+1}^n \left( \sin{\theta_k} \cos^3{\theta_j} - \cos{\theta_k} \sin{\theta_j}
        \cos^2{\theta_j} \right) \\
        & +\frac{2}{m} T \left( \sin{\theta_n} \sin{\theta_k} + \cos{\theta_n} \cos{\theta_k} \right)
        - \frac{2}{m} k_\mathrm{s} z_k;
    \end{aligned}
\end{equation}
--- \quad $\theta_1$ equation of motion  for the TnT model
\begin{equation}
    \label{eqn:th1TnT}
    \begin{aligned}
        &- 6 (n-1) \left( l+2l_\mathrm{s}+2z_1 \right) \sin{\theta_1} \Ddot{x}_1
        + 6 \left(n-1\right) \left( l+2l_\mathrm{s}+2z_1 \right) \cos{\theta_1} \Ddot{y}_1 \\
        &+ \sum_{j=1}^n A^{\theta_1}_j \cos{\left( \theta_1 - \theta_j \right)} \Ddot{\theta}_j
        - \sum_{j=1}^{n-1} B^{\theta_1}_j \sin{\left( \theta_1 - \theta_j \right)} \Ddot{z}_j \, = \\
        &- \sum_{j=1}^n A^{\theta_1}_j \sin{\left( \theta_1 - \theta_j \right)} \dot{\theta}_j^2
        - 2 \sum_{j=1}^{n-1} B^{\theta_1}_j \cos{\left( \theta_1 - \theta_j \right)} \dot{z}_j \dot{\theta}_j \\
        & + \frac{12}{m} \left( \frac{l}{2}+l_\mathrm{s}+z_1 \right) f_0 \sum_{j=2}^n \left( \sin{\theta_1} \sin{\theta_j} \cos^2{\theta_j} + \cos{\theta_1} \cos^3{\theta_j} \right) \\
        &+ \frac{12}{m} \left( \frac{l}{2} + l_\mathrm{s} + z_k \right) T
        \left( \sin{\theta_n} \cos{\theta_1} - \sin{\theta_1} \cos{\theta_n} \right)
        + \frac{12}{m} k_\mathrm{t} \left( \theta_2 - \theta_1 \right);
    \end{aligned}
\end{equation}
--- \quad $\theta_k$ equations of motion  for the TnT model where $k = 2, \, 3, \, 4, \, \ldots, \, n-1$
\begin{equation}
    \label{eqn:thkTnT}
    \begin{aligned}
        &- 6 \left[ (2n-2k+1)l + 2(n-k)l_\mathrm{s} + 2(n-k)z_k \right] \sin{\theta_k} \Ddot{x}_1 \\
        &+ 6 \left[ (2n-2k+1)l + 2(n-k)l_\mathrm{s} + 2(n-k)z_k \right] \cos{\theta_k} \Ddot{y}_1 \\
        &+ \sum_{j=1}^n A^{\theta_k}_{kj} \cos{\left( \theta_k - \theta_j \right)} \Ddot{\theta}_j
        - \sum_{j=1}^{n-1} B^{\theta_k}_{kj} \sin{\left( \theta_k - \theta_j \right)} \Ddot{z}_j \, = \\
        &- \sum_{j=1}^n A^{\theta_k}_{kj} \sin{\left( \theta_k - \theta_j \right)} \dot{\theta}_j^2
        - 2 \sum_{j=1}^{n-1} B^{\theta_k}_{kj} \cos{\left( \theta_k - \theta_j \right)} \dot{z}_j \dot{\theta}_j \\
        & +\frac{12}{m} \frac{l}{2} f_0 \left( \sin^2{\theta_k} \cos^2{\theta_k} + \cos^4{\theta_k} \right) \\
        & +\frac{12}{m} \left( l+l_\mathrm{s}+z_k \right) f_0 \sum_{j=k+1}^n \left( \sin{\theta_k} \sin{\theta_j} \cos^2{\theta_j} + \cos{\theta_k} \cos^3{\theta_j} \right) \\
        &+ \frac{12}{m} \left( l+l_\mathrm{s}+z_k \right) T \left( \sin{\theta_n} \cos{\theta_k} - \sin{\theta_k} \cos{\theta_n} \right) \\
        & -\frac{12}{m} k_\mathrm{t} \left( \theta_k - \theta_{k-1} \right)
        + \frac{12}{m} k_\mathrm{t} \left( \theta_{k+1} - \theta_k \right);
    \end{aligned}
\end{equation}
--- \quad $\theta_n$ equation of motion  for the TnT model
\begin{equation}
    \label{eqn:thnTnT}
    \begin{aligned}
        - &6 l \sin{\theta_n} \Ddot{x}_1 + 6 l \cos{\theta_n} \Ddot{y}_1 \\
        & + \sum_{j=1}^n A^{\theta_n}_{nj} \cos{\left( \theta_n - \theta_j \right)} \Ddot{\theta}_j 
        - \sum_{j=1}^{n-1} B^{\theta_n}_{nj} \sin{\left( \theta_n - \theta_j \right)} \Ddot{z}_j \, = \\
        &- \sum_{j=1}^n A^{\theta_n}_{nj} \sin{\left( \theta_n - \theta_j \right)} \dot{\theta}_j^2
        - 2 \sum_{j=1}^{n-1} B^{\theta_n}_{nj} \cos{\left( \theta_n - \theta_j \right)} \dot{z}_j \dot{\theta}_j \\
        &+ \frac{12}{m} \frac{l}{2} f_0 \left( \sin^2{\theta_n} \cos^2{\theta_n} + \cos^4{\theta_n} \right)
        - \frac{12}{m} k_\mathrm{t} \left( \theta_n - \theta_{n-1} \right);
    \end{aligned}
\end{equation}
with the new auxiliary coefficients
\footnotesize
\begin{equation*}
    A^{xy}_j =
    \begin{cases} 
        \left( n-1 \right) \left( l+2l_\mathrm{s}+2z_1 \right) &\text{for } j=1 \\
        (2n-2j+1)l + 2(n-j)l_\mathrm{s} + 2(n-j)z_1 &\text{for } j>1
    \end{cases}
\end{equation*}
\begin{equation*}
    B^{xy}_j = 2(n-j)
\end{equation*}
\begin{equation*}
    A^{z}_{kj} =
    \begin{cases} 
        \left( n-k \right) \left( l+2l_\mathrm{s}+2z_1 \right) &\text{for } j=1 \\
        2(n-k)l + 2(n-k)l_\mathrm{s} + 2(n-k)z_j &\text{for } 1<j \leq k \\
        (2n-2j+1)l + 2(n-j)l_\mathrm{s} + 2(n-j)z_j &\text{for } j>k
    \end{cases}
\end{equation*}
\begin{equation*}
    B^{z}_{kj} =
    \begin{cases} 
        2(n-k) &\text{for } j \leq k \\
        2(n-j) &\text{for } j>k 
    \end{cases}
\end{equation*}
\begin{equation*}
    A^{\theta_1}_{j} =
    \begin{cases} 
        3 (n-1) \left( l+2l_\mathrm{s}+2z_1 \right)^2 + l^2 &\text{for } j=1 \\
        3 \left( l+2l_\mathrm{s}+2z_1 \right) \left[ (2n-2j+1)l + 2(n-j)l_\mathrm{s} + 2(n-j)z_j \right] &\text{for } j>1
    \end{cases}
\end{equation*}
\begin{equation*}
    B^{\theta_1}_{j} = 6 \left( n-j \right) \left( l+2l_\mathrm{s}+2z_1 \right)    
\end{equation*}
\begin{equation*}
    A^{\theta_k}_{kj} =
    \begin{cases} 
        3 \left( l+2l_\mathrm{s}+2z_1 \right) \left[ (2n-2k+1)l + 2(n-k)l_\mathrm{s} + 2(n-k)z_k \right] &\text{for } j=1\\
        6 \left( l+l\mathrm{s}+z_j \right) \left[ (2n-2k+1)l + 2(n-k)l_\mathrm{s} + 2(n-k)z_k \right] &\text{for } 1<j<k \\
        6 \left( l+l_\mathrm{s}+z_k \right) \left[ 2(n-k)l + 2(n-k)l_\mathrm{s} + 2(n-k)z_k \right] + 4l^2
        &\text{for } j=k \\
        6 \left( l+l_\mathrm{s}+z_k \right) \left[ (2n-2j+1)l + 2(n-j)l_\mathrm{s} + 2(n-j)z_j \right] &\text{for } j>k 
    \end{cases}
\end{equation*}
\begin{equation*}
    B^{\theta_k}_{kj} =
    \begin{cases} 
        6 \left[ (2n-2k+1)l + 2(n-k)l_\mathrm{s} + 2(n-k)z_k \right] &\text{for } j< k \\
        6 \left[ 2(n-j)l + 2(n-j)l_\mathrm{s} + 2(n-j)z_k \right] &\text{for } j \geq k.
    \end{cases}
\end{equation*}
\normalsize

\subsection{Numerical Methods}
\label{subsec:numMethods}
Before pressing on with the numerical implementation of each model, note that each set of ODEs can be rewritten in the more compact matrix form
\begin{equation}
    \label{eqn:sys01}
    \mathbf{M}(\mathbf{q}_\mathrm{m}) \ddot{\mathbf{q}}_\mathrm{m} = \mathbf{f}(t, \mathbf{q}_\mathrm{m}, \dot{\mathbf{q}}_\mathrm{m}) 
\end{equation}
where the subscript ``$\mathrm{m}$" attached to the generalised coordinate vectors in the above denotes the respective model being studied be it the rigid, torsion, or TnT lightsail model. As per numerical fashion, the second-order system described by equation (\ref{eqn:sys01}) can readily be turned into a first-order system through the introduction of an auxiliary vector of unknowns ${\mathbf{u} = \dot{\mathbf{q}}_\mathrm{m}}$. The above system can thus be rewritten in the more numerically friendly form
\begin{equation}
    \label{eqn:sys02}
    \hat{\mathbf{M}}(\mathbf{x}) \dot{\mathbf{x}} = \hat{\mathbf{f}}(t, \mathbf{x}),
\end{equation}
where
\begin{equation*}
    \hat{\mathbf{M}}=\left[\begin{array}{cc}
        \mathbf{I} & \mathbf{0} \\
        \mathbf{0} & \mathbf{M}
    \end{array}\right]; \quad 
    \hat{\mathbf{f}}=\left[\begin{array}{l}
        \mathbf{u} \\
        \mathbf{f}
    \end{array}\right]; \quad 
    \mathbf{x}=\left[\begin{array}{l}
        \mathbf{q} \\
        \mathbf{u}
    \end{array}\right].
\end{equation*}
Given the dependency of the modified mass matrix $\hat{\mathbf{M}}$ upon the numerical vector of unknowns $\mathbf{x}$, the derivative of the unknown cannot be isolated in equation (\ref{eqn:sys02}). This renders the use of more conventional ODE solvers inadequate, and thus, to solve the numerical system of equations at hand, the set of first-order ODEs was instead treated as a set of differential-algebraic equations (DAEs). Opting for solvers better adapted to treat such DAEs, Mathematica's \verb|NDSolve| package was employed with the enforced residual method of simplification whereby the \verb|NDSolve| function first rewrites (\ref{eqn:sys02}) into the fully implicit form
\begin{equation}
    \label{eqn:sys03}
    \hat{\mathbf{F}}(t, \mathbf{x}, \dot{\mathbf{x}}) = \hat{\mathbf{M}}(\mathbf{x}) \dot{\mathbf{x}} 
    - \hat{\mathbf{f}}(t, \mathbf{x}) = \mathbf{0}
\end{equation}
before proceeding with the numerical computation of the solu-tion---see the Supplementary Material for a convergence study of this Lagrangian-based finite-element numerical lightsail approach using Mathematica's \verb|NDSolve|. Although all of the results found in this paper are an output of Mathematica, a number of the \verb|NDSolve| results were compared to the FORTRAN modified extended backward differentiation formulae (\verb|MEBDFV|) solver written by Abdulla and Cash of Imperial College, London (Department of Mathematics). This \verb|MEBDFV| solver was chosen because it deals precisely with systems of the form shown in (\ref{eqn:sys02}) with the mass matrix being dependant upon the vector of unknowns and because numerical stability of its solution has been demonstrated on multiple occasions for the simulations of discrete mechanical systems similar to the present problem \cite{fritzkowski2008tetherRigid,fritzkowski2009tetherTension,fritzkowski2011tetherTorsion}.

As its name implies, this \verb|MEBDFV| solver employs a modified version of the backward differentiation formulae whereby, if $\mathbf{x}_i$ is the solution of the unknown at the current time step, a superfuture value, $\mathbf{x}_{i+1}$ is computed to improve computational stability by refining the initial guess of the derivative of the unknown, $\dot{\mathbf{x}}_i$, thereby allowing for a more accurate Newton-Raphson solution to the nonlinear system (\ref{eqn:sys03}). For a concise description of the overall \verb|MEBDFV| scheme, the reader is directed to \cite{fritzkowski2009tetherTension}. The \verb|MEBDFV| solver also includes additional intricacies such as adaptive time stepping and Newton-Raphson scheme convergence rate estimates, but their detailed description goes beyond the scope of this present paper. For a thorough treatise of the inner workings of the \verb|MEBDFV| code, please refer to the papers written by Cash et al. on the subject \cite{cash1992,cash2000,abdulla2001,cash2003}. A Mathematica vs FORTRAN lightsail simulation output comparison test case can be found in the Supplementary Material.

\begin{figure*}[t]
    \centering
    \includegraphics[scale=0.75]{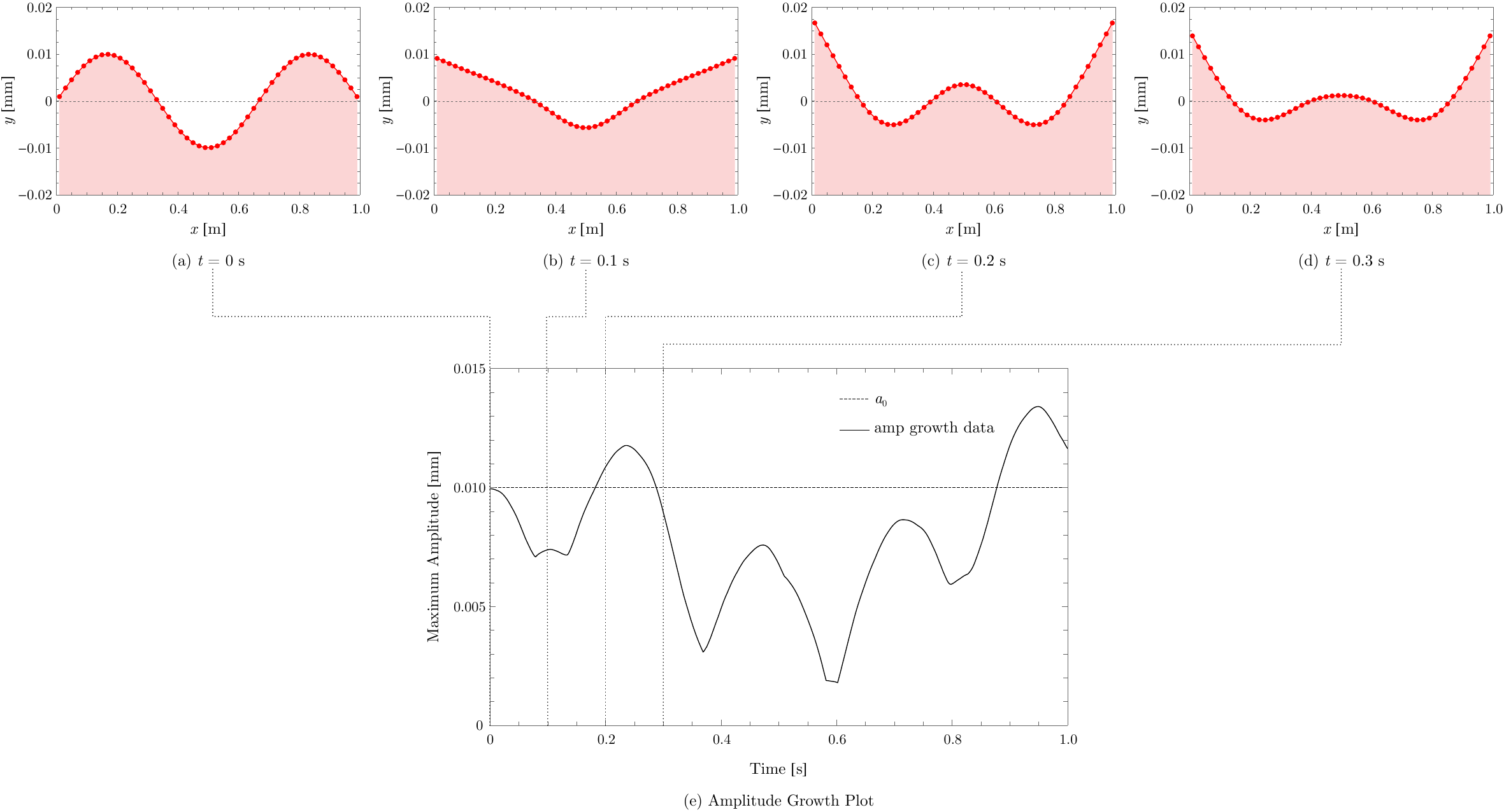}
    \caption{Torsion lightsail model \emph{stable oscillatory} sample run for mode 3/2. The lightsail shape snapshots (a)--(d) are visualized in a reference frame fixed to an ideally flat lightsail. Note the quasi-periodic evolution of the maximum perturbation amplitude.}
    \label{fig:TableauxTorsionStable}
\end{figure*}

While the authors recognize that a commercial FE software would have been able to numerically address this problem, the main goal of this study was to develop an analytic theory for lightsail stability under large radiation pressure. As such, it is important that the numerical method used to validate the theory should be derived from the same underlying physical model, and this would only be possible in a finite element model developed from the ground up. Further, the development of a FE model from first-principles allowed the authors full knowledge of the assumptions and limitations of each lightsail numerical model, which in turn granted the ability to isolate the influence of various effects upon lightsail shape stability such as bending stiffness, boundary tension, etc. The fact that bending stiffness was incorporated in this study is of note since conventional thin film theory usually dispenses with the resistance to bending deformation altogether. Finally, this first two-dimensional, rectangular model here studied both analytically and numerically was also chosen for inquiry as it could potentially act in future studies as an elementary building block for the construction of more general (i.e., non-rectangular) perturbed lightsails by coupling together, for example, multiple torsion or TnT lightsail models.


\section{Results and Discussion}
\label{sec:RandD}

\subsection{Torsion Model Results}
Given the rather large number of parameters required to perform a torsion model simulation, the theoretical expression (\ref{eqn:ECRMAX}) for the critical lightsail material modulus $E_\mathrm{cr_\mathrm{max}}$ was used as a guideline to reduce the parameter space to a more manageable size. As a result, only the material modulus, $E$, and the lightsail thickness, $h$, were varied during the torsion model simulations. In particular, $h$ was varied from $\SI{0.01}{\micro\meter}$ to $\SI{10}{\micro\meter}$ in jumps of powers of ten while the material modulus was varied over a few magnitudes from its theoretical critical value. Each simulation was run for a total of $\SI{1}{\second}$ and failure was considered whenever the perturbation amplitude doubled from its initial value. That is, at each numerical time increment, the maximum amplitude of the sail, $\mathrm{amp}_{\mathrm{max}}$ was computed by considering the vertical position of the CoM of all sail elements:
\begin{equation}
    \label{eqn:RandD01}
    \mathrm{amp}_{\mathrm{max}}(t) = \frac{\max{\{y_i(t)\}} - \min{\{y_i(t)\}}}{2}.
\end{equation}
Whenever $\mathrm{amp}_{\mathrm{max}}(t) \geq 2a_0$, the simulation was stopped and the time to failure, $\tau$, was set to $\tau = t$. If the amplitude of perturbations never doubled during the simulation, then the time to failure was set to the total runtime, $\tau = t_\mathrm{final}$. In total, for the torsion model, 900 lightsail simulations per mode number for each of modes $\nu = 1$ and $\nu = 3/2$ were run using 50 elements by varying the thickness, $h$,  from $\SI{0.01}{\micro\meter}$ to $\SI{10}{\micro\meter}$ and the elastic modulus, $E$, from $\SI{4.5e-3}{\giga\pascal}$ to $\SI{4.5e7}{\giga\pascal}$ with increments equally spaced on a logarithmic scale. Each data point was a 4-tuple recording the lightsail simulation thickness, elastic modulus, initial perturbation mode number, and time to failure $\left( h, \, E, \, \nu, \tau \right)$. Using 50 elements per simulation appeared justifiable because, despite the fact that asymptotic numerical convergence starts at about 100 sail elements (see Supplementary Material), the qualitative structural behavior of the sail was shown through convergence tests to remain the same independent of the number of elements used (i.e., a stable sail remained stable and an unstable sail remained unstable independent of the number of sail elements used; see Supplementary Material). All other simulation parameters were kept constant from simulation to simulation and their values are given below:
\begin{enumerate}
    \item Sail length, $L = \SI{1}{\meter}$
    \item Sail width, $W = \SI{1}{\meter}$
    \item Sail density, $\rho = \SI{1000}{\kilo \gram / \meter \cubed}$
    \item Incident laser intensity, $I_0 = \SI{10}{\giga \watt / \meter \squared}$
    \item Initial perturbation amplitude, $a_0 = \SI{0.01}{\milli \meter}$
    \item Perturbation mode number, $\nu = 1 \text{ or } 3/2$
    \item Number of elements, $n = 50$
    \item Total runtime, $t_\mathrm{final} = \SI{1}{\second}$ or to failure
\end{enumerate}

\begin{figure*}[t]
    \centering
    \includegraphics[scale=0.75]{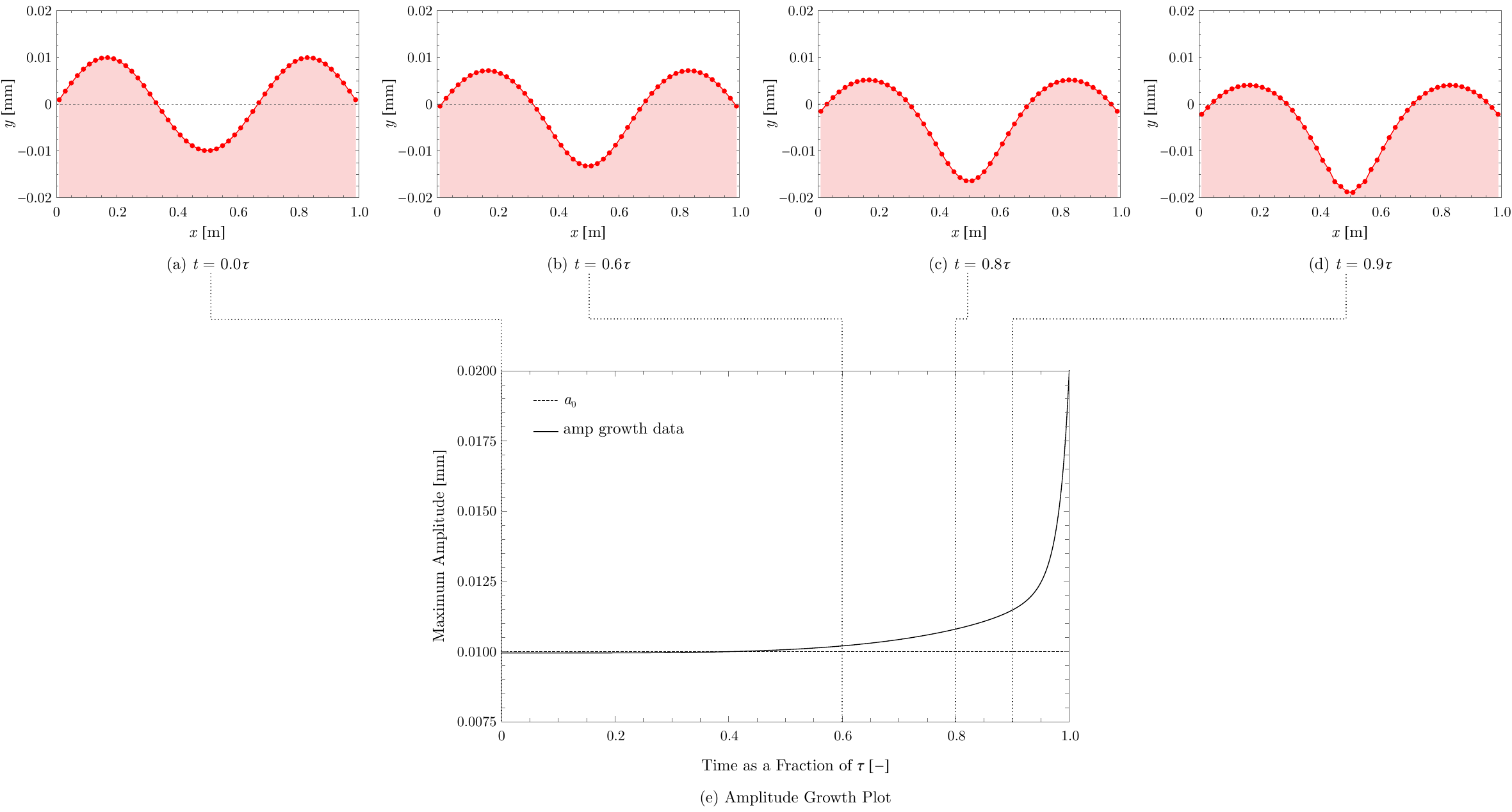}
    \caption{Torsion lightsail model \emph{potentially unstable} sample run for mode 3/2. The lightsail shape snapshots (a)--(d) are visualized in a reference frame fixed to an ideally flat lightsail. The perturbation amplitude grew in size exponentially with the largest growth occurring at perturbation mode trough.}
    \label{fig:TableauxTorsionUnstable}
\end{figure*}

\begin{figure*}[!htp]
    \centering
    \includegraphics[scale=1.0]{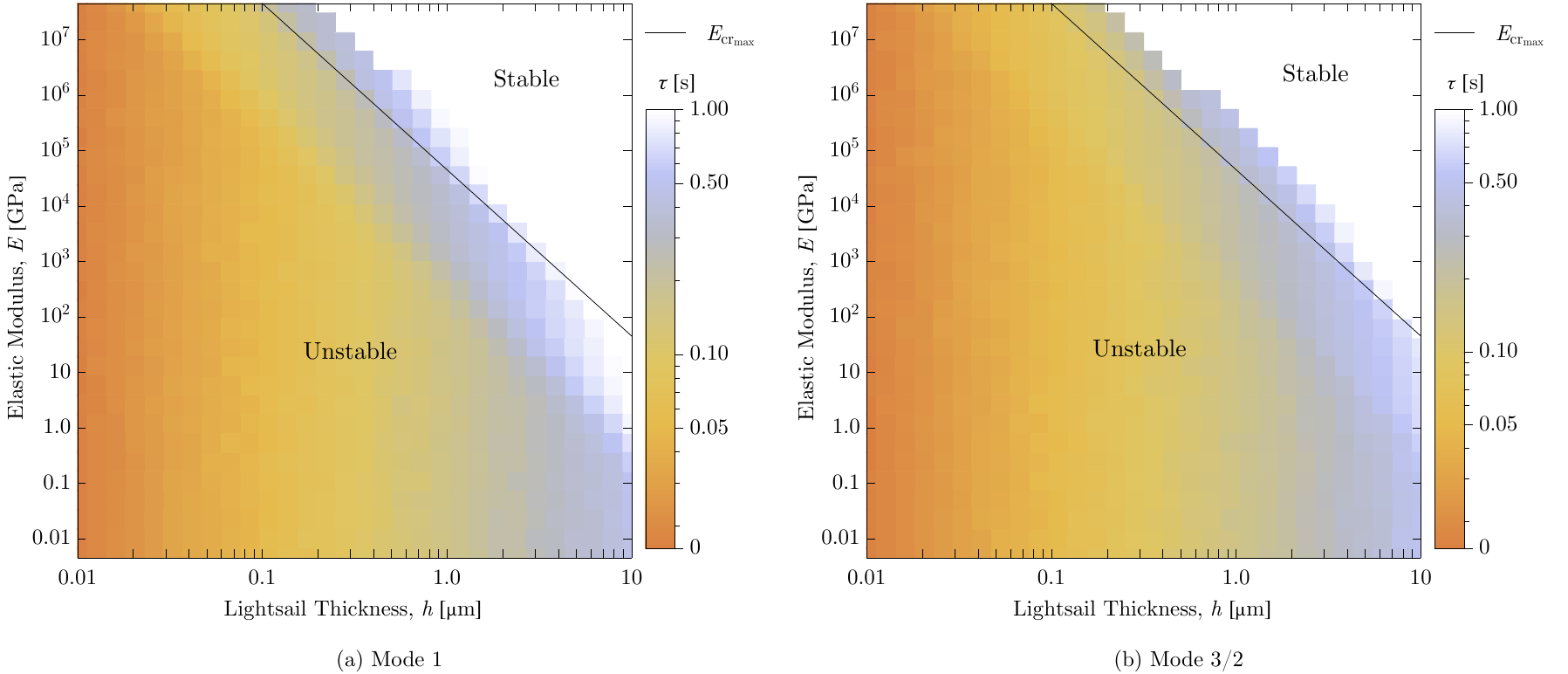}
    \caption{Stability map for the torsion lightsail model; (a)~for perturbation mode 1; (b)~for perturbation mode 3/2. Each map consists of $900$ ($30 \times 30$) simulations.}
    \label{fig:EMaps01Vertical}
\end{figure*}

Figure~\ref{fig:TableauxTorsionStable}e depicts the time evolution of the maximum amplitude of the $(h = \SI{10}{\micro \meter}$, $E = \SI{1480}{\giga \pascal}$, $\nu = 3/2$, \allowbreak $\tau = \SI{1}{\second})$ lightsail torsion model run where $E_\mathrm{cr_{max}} = \SI{45}{\giga \pascal}$. Given the quasi-periodic nature of the maximum amplitude value, the lightsail remained stable, that is, its initial perturbation amplitude never doubled during the runtime. Figures~\ref{fig:TableauxTorsionStable}a--d shows four snapshots of the lightsail's shape each recorded at different simulation times, which visually further reinforce the fact that the lightsail shape dynamics for this particular set of torsion model parameters is quasi-periodic.\footnote{In order to help better appreciate the lightsail shape dynamics, all numerical lightsail snapshot figures found in this paper were visualized in a reference frame fixed to the body of an ideally flat lightsail accelerating at $g_0 = \frac{2 I_0}{c \, \rho \, h}$.} Such lightsail configurations were termed \emph{stable oscillatory}. Figure~\ref{fig:TableauxTorsionUnstable}e, on the other hand, depicts the time evolution of the maximum amplitude for a torsion model run where $\left( h = \SI{10}{\micro \meter}, \, E = \SI{0.108}{\giga \pascal}, \, \nu = 3/2, \, \, \tau = \SI{0.541}{\second} \right)$. Notice that the sail perturbation amplitude grew without bound as indicated by the exponential behavior of the maximum amplitude, which effectively blew up as we approached the end of the simulation. Snapshots of the lightsail shape were taken at different times for this simulation as well and of note are both the absence of periodic behavior and the fact that the lightsail appears to have failed at a trough (see Figs.~\ref{fig:TableauxTorsionUnstable}a--d). Simulations such as these, which did not exhibit stable oscillatory behavior but whose initial perturbation amplitude doubled some time before the runtime ended, were termed \emph{potentially unstable}.

A summary of $1800$ simulations ($900$ per perturbation mode) in the form of stability plots is found in Fig.~\ref{fig:EMaps01Vertical} where Fig.~\ref{fig:EMaps01Vertical}a and Fig.~\ref{fig:EMaps01Vertical}b contain the data points for perturbation mode 1 and 3/2, respectively. Onto each plot was added the theoretical stability threshold predicted by the analytical equation (\ref{eqn:ECRMAX}). Of most remarkable note is that, for both perturbation modes, nearly all stable oscillatory sample points (i.e., all points with a $\tau$ value of $\SI{1}{\second}$) fall \emph{above} the theoretical curve and vice versa for potentially unstable sample points where, the further away from the theoretical curve the unstable data point was, the more unstable the simulation was (i.e., the faster the doubling of initial perturbation amplitude occurred). This distribution of stable/unstable data points indicates agreement between theory and numerical simulations: if the elastic modulus of a given sample run stood a few orders of magnitude above the critical value predicted by theory, that particular lightsail configuration behaved in a quasi-periodic, stable manner, otherwise the configuration displayed unstable behavior. Of further note is the fact that qualitative and quantitative differences between the two initial perturbation modes appear relatively minor, indicating that the agreement between theory and computation is independent of initial perturbation mode.

\vspace{0cm}

\subsection{Tensile Model Results}
\label{subsec:TnTResults}
\begin{figure*}[t]
    \centering
    \includegraphics[scale=0.75]{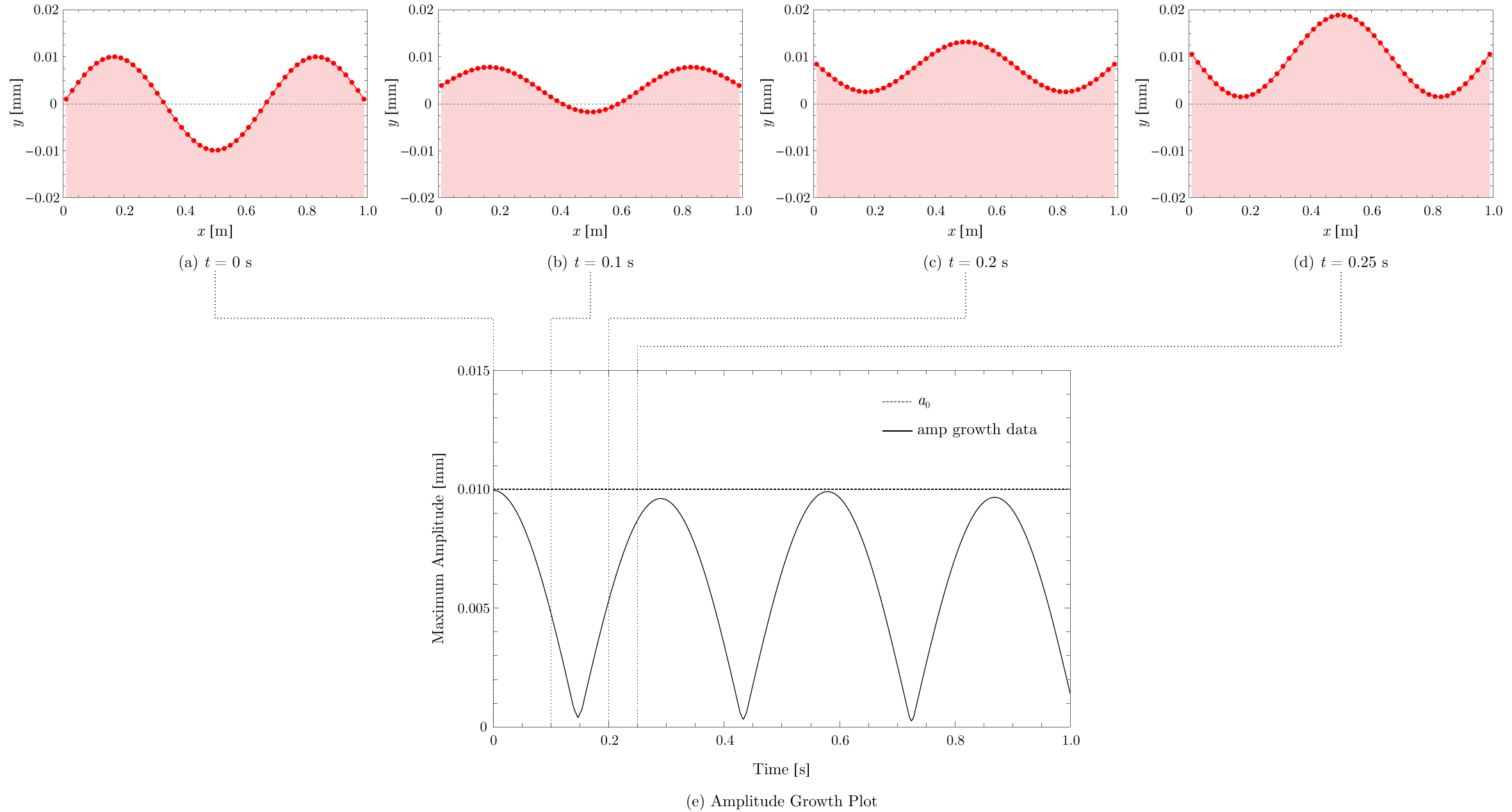}
    \caption{TnT lightsail model \emph{stable oscillatory} sample run for mode 3/2. The lightsail shape snapshots (a)--(d) are visualized in a reference frame fixed to an ideally flat lightsail. Note the quasi-periodic evolution of the maximum perturbation amplitude.}
    \label{fig:TableauxTensionStable}
\end{figure*}

\begin{figure*}[t]
    \includegraphics[scale=0.75]{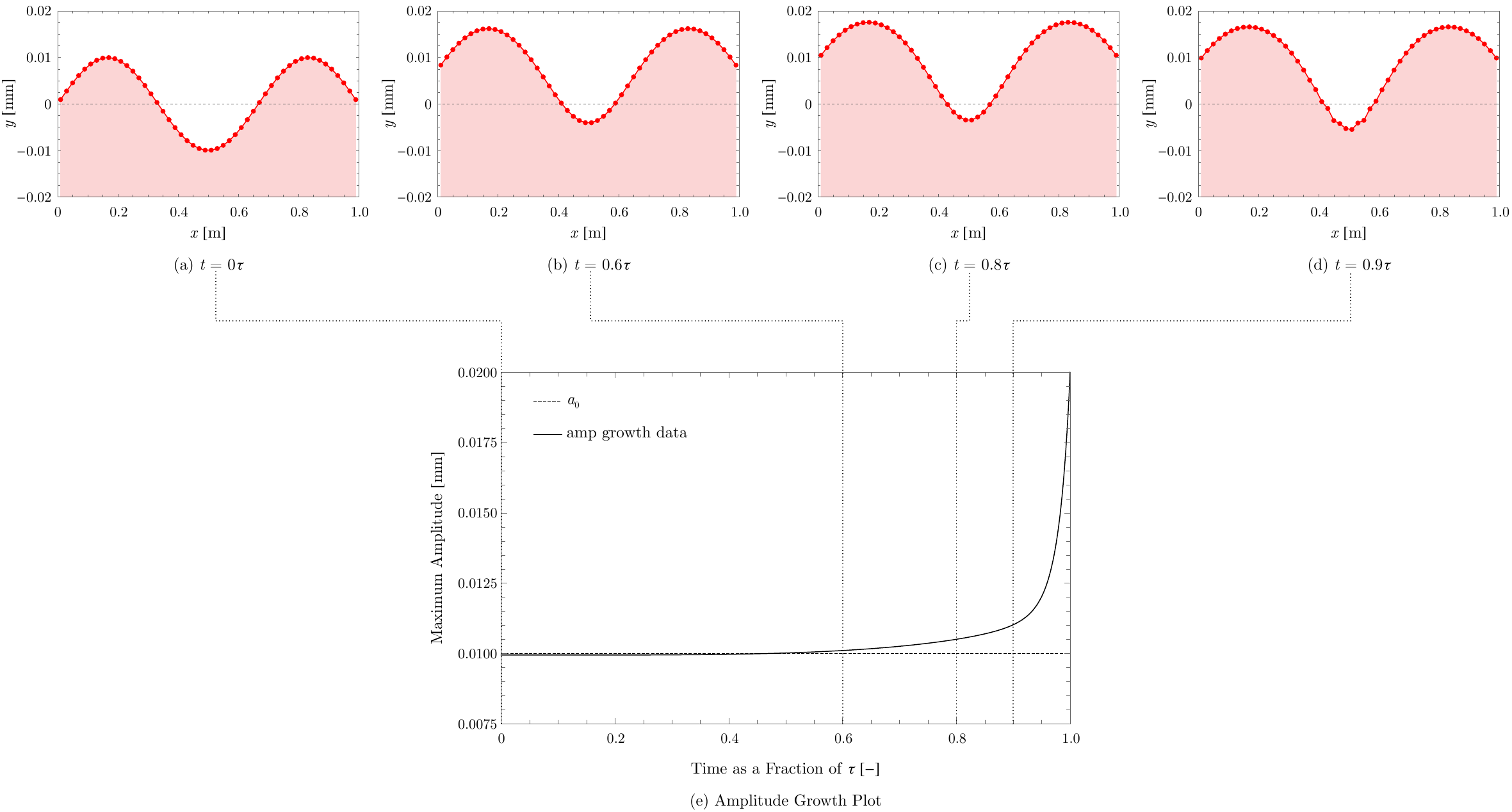}
    \caption{TnT lightsail model \emph{potentially unstable} sample run for mode 3/2. The lightsail shape snapshots (a)--(d) are visualized in a reference frame fixed to an ideally flat lightsail. The perturbation amplitude grew in size exponentially with the largest growth occurring at the perturbation mode trough.}
    \label{fig:TableauxTensionUnstable}
\end{figure*}

As with the torsion model, the expression (\ref{eqn:TCRMAX}) for $T_\mathrm{cr_{max}}$ as derived from theory was used to make the parameter space of the TnT model more tractable. Consequently, only the boundary tension, $T$, and the lightsail thickness, $h$, were varied. A total of 900 lightsail runs for each of modes $\nu = 1$ and $\nu = 3/2$ were performed using the TnT model by varying the lightsail thickness, $h$, from $\SI{0.01}{\micro \meter}$ to $\SI{10}{\micro \meter}$ and the boundary tension magnitude, $T$, from $\SI{3.34e-7}{\newton / \meter}$ to $\SI{3.34}{\newton / \meter}$.\footnote{As a reminder to the reader, recall that all values of the boundary tension magnitude, $T$, are given on a per unit width basis.} Each simulation or data point once more recorded the lightsail thickness, boundary tension magnitude, initial perturbation mode, and time to failure $\left( h, \, T, \, \nu, \, \tau \right)$. Note also that, in order to solely investigate the influence of membrane action or tensile stiffness upon  the lightsail structural stability, the bending stiffness of the lightsail model was manually tuned down to $0$ by setting $k_{\mathrm{t}} = 0$ (although, because of the dependency of bending stiffness upon $1/h^3$, the presence or absence of torsion springs had no significant influence given the range of the thickness values studied). Once more, each simulation was run for a total of $\SI{1}{second}$ using 50 elements and failure was considered to have occurred when the light-sail perturbation amplitude doubled. The simulation properties were as listed below:
\begin{enumerate}
    \item Sail length, $L = \SI{1}{\meter}$
    \item Sail width, $W = \SI{1}{\meter}$
    \item Sail density, $\rho = \SI{1000}{\kilo \gram / \meter \cubed}$
    \item Material Elastic Modulus, $E = \SI{5}{\giga \pascal}$
    \item Incident laser intensity, $I_0 = \SI{10}{\giga \watt / \meter \squared}$
    \item Initial perturbation amplitude, $a_0 = \SI{0.01}{\milli \meter}$
    \item Perturbation mode number, $\nu = 1 \text{ or } 3/2$
    \item Number of elements, $n = 50$
    \item Total runtime, $t_\mathrm{final} = \SI{1}{\second}$ or  to failure
\end{enumerate}
Figure~\ref{fig:TableauxTensionStable}e shows the time evolution of the maximum amplitude of the lightsail TnT run with values $\Big( h = \SI{10}{\micro \meter}$, $T = \SI{1.29e-2}{\newton / \meter}$, $\nu = 3/2$, $\tau = \SI{1}{\second} \Big)$ where $T_\mathrm{cr_\mathrm{max}} = \SI{3.34e-4}{\newton / \meter}$. As in the bending stiffness case, the simulation of a lightsail whose boundary tension magnitude was at least an order of magnitude above that of its critical tension value displayed stable oscillatory motion. Conversely, decreasing the boundary tension value by an order of magnitude or more below the critical value always made the lightsail model display potentially unstable behavior. See, for example, the $\Big( h = \SI{10}{\micro \meter}$, $T = \SI{3.09e-6}{\newton / \meter}$, $\nu = 3/2$, $\tau = \SI{0.494}{\second} \Big)$ lightsail TnT configuration whose uncontrolled perturbation growth is shown in Fig.~\ref{fig:TableauxTensionUnstable}.

Figure~\ref{fig:TMapsVertical} depicts the stability maps for perturbation modes 1 and 3/2 for the TnT model simulations. As with the torsion model, plotting the theoretical stability curve onto the stability maps indicates agreement between theory and simulations: nearly all stable data points lie above the theoretical stability curve and almost all unstable data points lie below the theoretical curve, with the data points farthest below the theoretical curve being more unstable than the data points closest to the theoretical curve. Noteworthy is also the fact that, once more, there appears to be little qualitative and quantitative difference between the different perturbation mode stability maps.

\subsection{Engineering Implications}
\label{subsec:EngImp}
\begin{figure*}[!htp]
    \centering
    \includegraphics[scale=1.0]{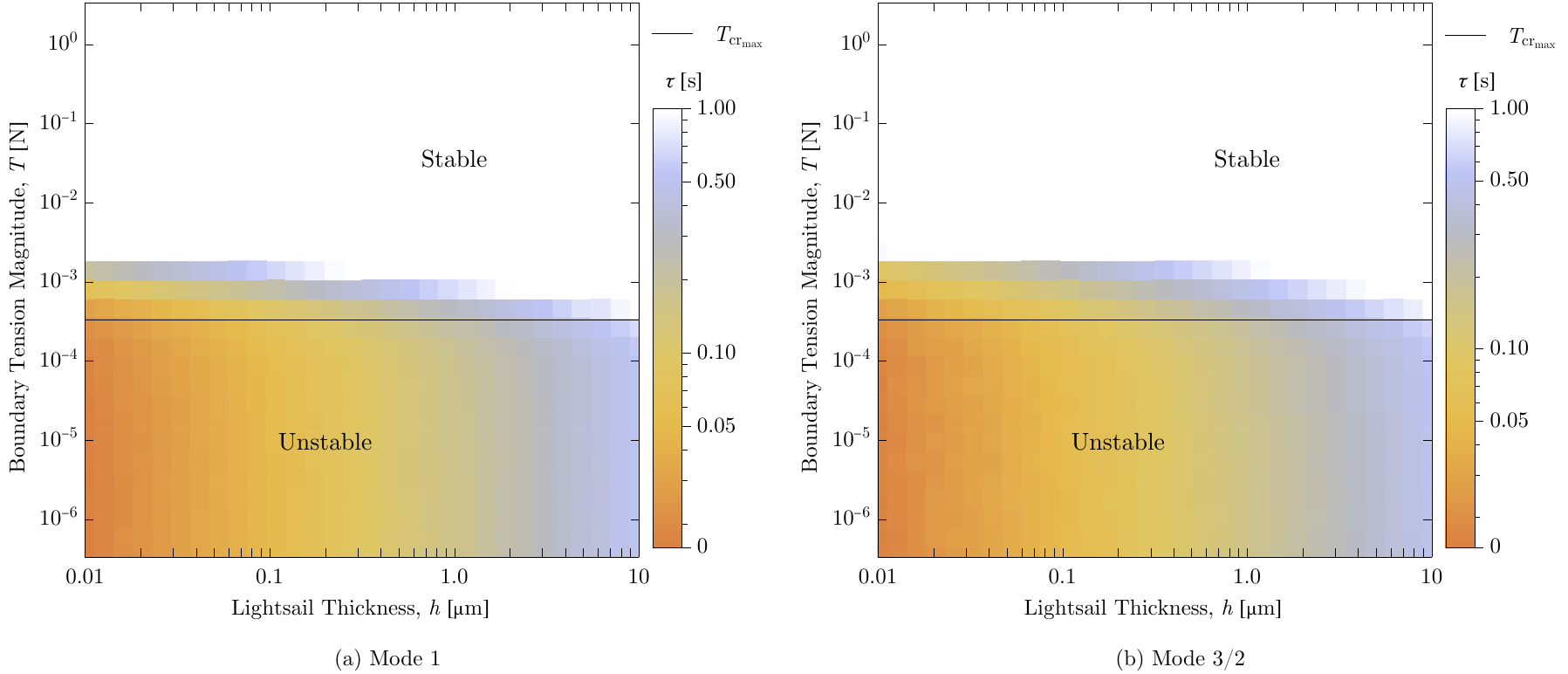}
    \caption{Stability map for the TnT lightsail model; (a)~for perturbation mode 1; (b)~for perturbation mode 3/2.}
    \label{fig:TMapsVertical}
\end{figure*}

The results encapsulated in the stability maps (Figs.~\ref{fig:EMaps01Vertical} and \ref{fig:TMapsVertical}) showed good agreement between theory and numerical computations. It should be noted, however, that stability is not strictly guaranteed when either the material modulus, $E$, or the boundary tension magnitude, $T$, are greater than their theoretical critical maximum value; typically, it is necessary to exceed the critical values by an order of magnitude to ensure that the lightsail shape remains stable. The inclusion of a safety factor in equations (\ref{eqn:ECRMAX}) and (\ref{eqn:TCRMAX}) should, in principle allow for the construction of a structurally stable lightsail configuration.

Upon closer inspection of the torsion model stability maps depicted in Fig.~\ref{fig:EMaps01Vertical}, it should be remarked that the material moduli required to generate a stable sail configuration are impractically large ($E_\mathrm{cr_{max}} = \SI{45 000}{\giga \pascal}$ for a 1-$\upmu$m-thick sail with a mode 3/2 and $\SI{0.01}{\milli \meter}$ amplitude perturbation under a laser flux intensity of $\SI{10}{\giga \watt / \meter \squared}$, compared to the elastic modulus of diamond, $E \approx \SI{1000}{\giga \pascal}$). This is believed to be in large part due to the dependence of the required material strength on the inverse of the thickness \emph{cubed}, $E \propto 1/h^3$, a dependence emerging from moment-curvature relation (\ref{eqn:MomCurv}), where the second moment of area of the lightsail is $I = Wh^3/12$. Consequently, if the lightsail is to be kept thin for practical accelerations to be reached, the inclusion of boundary tension into the design of the lightsail would be necessary in order to achieve structural stability since the boundary tension thresholds for stability appear feasible. For example, for the same 1-$\upmu$m-thick sail with a $\SI{0.01}{\milli \meter}$ amplitude perturbation, $T_\mathrm{cr_{max}} = \SI{0.334}{\milli \newton}$ per unit width for a laser flux intensity of $\SI{10}{\giga \watt / \meter \squared}$, which can be achieved by inflating the lightsail (much akin to a balloon) using a gas pressure of $\SI{1.4}{\milli \pascal}$ or by spinning the lightsail about the laser beam axis at a rate of $\SI{1.6}{\radian / \second}$---although, in the spinning case, the critical tension would be achieved only at the periphery of the lightsail. Ultimately, how the boundary tension is added into the lightsail concept is left at the discretion of the design engineer, and other ideas like the Cassenti and Cassenti proposed boundary ring \cite{cassenticassenti2020} and the inverse cat eye lightsail optical metasurface concept of Siegel et al. \cite{siegel2019selfstabilizing}, etc. have been proposed in the literature. Not only do the boundary tension magnitudes required for structural stability appear readily achievable, but of important note is \emph{the absence of material strength and lightsail thickness in the functional expression of $T_\mathrm{cr_{max}}$}. The use of boundary tension would then effectively leave the design space of material strength and lightsail thickness entirely open and thus allows the conceptual engineer to, for example, base their choice of material on optical properties alone without having to worry about structural stability implications.

Further, a closer look at equations (\ref{eqn:ECRMAX}) and (\ref{eqn:TCRMAX}) reveals that both the critical material modulus and boundary tension magnitude are directly proportional to the perturbation amplitude, $a_0$. Given this direct correspondence between the critical stability values and the amplitude of the perturbations, it is an imperative that good quality control be undertaken during the manufacturing and packaging/handling processes of the lightsail. As already discussed, the previously quoted $T_\mathrm{cr_{max}} = \SI{0.334}{\milli \newton}$ per unit width was taken for a perturbation amplitude of $\SI{0.01}{\milli \meter}$. If the amplitude of perturbation is increased such that $a_0 = \SI{10}{\centi \meter}$ instead while maintaining all other parameters constant, the critical maximum value becomes $T_\mathrm{cr_{max}} = \SI{3.34}{\newton}$ per unit width. We see here a substantial increase in the threshold value and the reader is reminded that, as per numerical simulations, stability is only guaranteed when the boundary tension or material modulus are at least an order of magnitude above their respective thresholds.

\subsection{Lightsail Model Limitations}

It remains of note that the analytical and numerical models and results studied thus far hold certain limitations. The laser beam intensity was here taken as uniform, and the non-uniformity of a more realistic laser beam intensity distribution like that of a Gaussian beam may drive further instability. Further, and perhaps of greater importance, is the fact that no lightsail model included the presence of dissipation. This is especially important given the stable oscillatory behavior of the torsion and TnT models when above their respective $E_\mathrm{cr_{max}}$ and $T_\mathrm{cr_{max}}$ values. In particular, it was noted during the generation of the TnT model stability maps that further increasing the boundary tension beyond its critical value increased the frequency of oscillations of the lightsail. Increasing frequency of oscillation in the presence of dissipation effects can potentially increase the internal heat generation to a point where the lightsail may melt, even in the absence of any radiation absorption. Finally, each sail element was modeled as perfectly reflective. As noted in the introductory section, the material optical properties can potentially be used to help enforce lightsail stability. In order to improve the physical accuracy of the lightsail model it will be thus necessary in the future to include: a realistic, non-uniform laser beam intensity distribution, heat dissipation effects, and material optical properties.


\section{Conclusion}
\label{sec:conlcusion}

The large directed-energy loads required for laser-driven propulsion pose the problem of the stability of the lightsail shape. Given the unavoidable presence of perturbations along the reflective sail surface, it is important to determine whether the sail will retain its shape or crumple under large photon pressures. An analytical theory of stability was first proposed wherein critical values for the material modulus, $E_\mathrm{cr}$, and for boundary tension, $T_\mathrm{cr}$, were derived using a moment balance, involving key lightsail parameters such as thickness and laser intensity. The analytical model was compared against various numerical simulations of the dynamics of an accelerating lightsail, and agreement was found between theory and numerical computations. In the absence of boundary tension, both theory and numerical simulations predict the necessity of unrealistically large elastic moduli to guarantee the lightsail's stability. Increasing the sail thickness can potentially generate stable sail configurations, but to the detriment of spacecraft acceleration. With the addition of boundary tension, it was found from both theory and numerical simulations that the lightsail can be kept under a stable oscillatory state through the application of relatively small tensile loads, even for very thin (sub-micron) lightsails. The tension required to maintain a stable shape could be readily provided via a sail inflation, spinning of the sail, or other tensioning mechanics. The parameter maps provided by the analytical theory and verified through numerical simulation could potentially provide the mission designer with a convenient method to engineer a lightsail capable of sustaining the large directed-laser pressures needed for high-acceleration spacecraft applications.


\section*{Acknowledgments}
The authors acknowledge the assistance of Hansen Liu in the early stages of this work. The authors would like to thank Philip Lubin and Geoffrey Landis for stimulating discussions. DCS would also like to thank Zhuo Fan Bao, John Kokkalis, Monika Azmanska, and Emmanuel Duplay for their constructive suggestions on this paper. This work was supported by the Natural Sciences and Engineering Research Council of Canada (NSERC) Discovery Grant “Dynamic Materials Testing for \allowbreak Ultrahigh-Speed Spaceflight”, the McGill Summer Undergraduate Research in Engineering program, and Fonds de Recherche du Québec Nature et Technologies (FRQNT).



\appendix
\section{Continuous Lightsail Model---Energy Method}
\label{sec:appendix A}

To help solidify the results obtained from equations (\ref{eqn:ECR}) and (\ref{eqn:ECRMAX}), the quasi-static analysis of the critical point between structural stability and instability of the lightsail is considered here using a different, energy-based approach. 

The principle of virtual work stipulates that a system can only be in equilibrium if the variation of its internal energy equals the virtual work applied to the system:
\begin{equation}
    \label{eqn:VW01}
    \delta U = \delta \mathcal{W}
\end{equation}
where all variations (denoted using the operator symbol $\delta$) are performed in accordance with the kinematic constraints of the system. From a stability perspective, equation (\ref{eqn:VW01}) considers the tipping point or threshold for a stable equilibrium. If, in practice, the change in internal energy is greater than the applied work ($\Delta U > \Delta \mathcal{W}$), then the system will be stable about its current equilibrium point. If the change in internal energy is smaller than the applied work ($\Delta U < \Delta \mathcal{W}$), the system will be unstable and move away from its equilibrium point.

Ignoring the effects of tension, the internal energy of the continuous beam lightsail model is its internal bending strain energy. Assuming the deformations to be linear-elastic, on a per element basis, the internal bending strain energy is
\begin{equation}
    \label{eqn:VW02}
    \mathrm{d} U = \frac{1}{2} \, \mathcal{M}_\mathrm{b} \, \mathrm{d}\theta = \frac{1}{2} \, \mathcal{M}_\mathrm{b} \, \kappa \, \mathrm{d}s.
\end{equation}
Integrating the above result while employing the moment-curvature relation and a first order small angle approximation, the total internal strain energy becomes
\begin{equation}
    \label{eqn:VW04}
    U = \int_0^L \frac{E \mkern 2mu I}{2} \, \left( \frac{\mathrm{d}^2w}{\mathrm{d}x^2} \right)^2 \, \mathrm{d}x.
\end{equation}
The first variation of the lightsail's internal bending energy can thus be found by applying the functional derivative:
\begin{equation}
    \label{eqn:VW05}
    \delta U = \delta \int_0^L \frac{E \mkern 2mu I}{2} \, \left( \frac{\mathrm{d}^2w}{\mathrm{d}x^2} \right)^2 \, \mathrm{d}x = \int_0^L E \mkern 2mu I \, \frac{\mathrm{d}^2w}{\mathrm{d}x^2} \, \delta
    \left( \frac{\mathrm{d}^2w}{\mathrm{d}x^2} \right) \, \mathrm{d}x.
\end{equation}
After performing integration by parts twice, $\delta U$ can be expressed only in terms of the first variation of the vertical deformation \footnote{The boundary term $E \mkern 2mu I\, \frac{\mathrm{d}^2w}{\mathrm{d}x^2}  \, \delta \left(\frac{\mathrm{d}w}{\mathrm{d}x}\right) \Big|_0^L$ is neglected because the curvature of the lightsail is zero at the endpoints by construction.}
\begin{equation}
    \label{eqn:VW06}
    \delta U = - E \mkern 2mu I \, \frac{\mathrm{d}^3w}{\mathrm{d}x^3} \, \delta w \Big|_0^L +  \int_0^L E \mkern 2mu I \, \frac{\mathrm{d}^4w}{\mathrm{d}x^4} \, \delta w \, \mathrm{d}x.
\end{equation}

In the absence of tension, the virtual work applied to the lightsail is composed of the vertical and horizontal components of the radiation pressure loads and of the d'Alembert body force,
\begin{equation}
    \label{eqn:VW07}
    \begin{aligned}
        \delta W = &\int_{0}^{L} \frac{2 I_0}{c} \, W \, \cos{\theta} \sin{\theta} \, \delta u \, \mathrm{d}x \\
        &+ \int_{0}^{L} \frac{2 I_0}{c} \, W \,  \cos^2{\theta} \, \delta w \, \mathrm{d}x \\
        &- \int_{0}^{L} \rho \, h \, g_0 \, W \, \delta w \, \frac{\mathrm{d}x}{\cos{\theta}}.
    \end{aligned}
\end{equation}
Once again employing small angle approximations, the vertical radiation pressure and d'Alembert body force virtual work contributions cancel out and the total virtual work simplifies to
\begin{equation}
    \label{eqn:VW08}
    \delta \mathcal{W} = \int_{0}^{L} \frac{2 I_0}{c} \, W \, \frac{\mathrm{d}w}{\mathrm{d}x} \, \delta u \, \mathrm{d}x.
\end{equation}
The virtual displacement in the horizontal direction, $\delta u$, present in equation (\ref{eqn:VW08}) can be related to the virtual displacement in the vertical direction, $\delta w$, by assuming the continuous beam lightsail model to be inextensible. Inextensibility can rightly be assumed in the case where the tension running through the sail before the application of loads is of such magnitude that any additional deformation will not induce the change of the body's total length. The inextensibility assumption can also be applicable in the case where the tensile stiffness of the lightsail is altogether ignored with the investigation efforts being solely focused on the bending stiffness of the lightsail as is presently done. Assuming an inextensible lightsail model, it can be shown that
\begin{equation}
    \label{eqn:VW09}
    \frac{\partial u}{\partial x} + \frac{1}{2} \, \left( \frac{\partial w}{\partial x} \right)^2 = 0 \implies u(x)
    = - \int_{0}^{x} \frac{1}{2} \, \left( \frac{\partial w}{\partial \xi} \right)^2 \, \mathrm{d}\xi,
\end{equation}
a standard result found in the literature \cite{Dowell2016InexBeam,novozhilov1999foundations,timoshenko2012stability}. From the above kinematic constraint, finding the first variation of $u(x)$ becomes a straightforward application of the functional derivative followed by an integration-by-parts:
\begin{equation}
    \label{eqn:VW10}
    \delta u(x) = - \frac{\mathrm{d}w}{\mathrm{d}\xi} \, \delta w \, \Big|_0^x + \int_{0}^{x} \frac{\mathrm{d}^2w}{\mathrm{d}\xi^2} \, \delta w \, \mathrm{d}\xi,
\end{equation}
where the partial derivative symbols have been replaced by a total derivative because the vertical displacement here spatially and temporally depends by construction only on $\xi$, which is here a dummy variable of integration for $x$. The boundary term in equation (\ref{eqn:VW10}) can be more explicitly written to yield
\begin{equation}
    \label{eqn:VW11}
    \delta u(x) = - \frac{\mathrm{d}w(x)}{\mathrm{d}x} \, \delta w(x)
    + \left[ \frac{\mathrm{d}w}{\mathrm{d}x} \, \delta w \, \right]_{x=0}
    + \int_{0}^{x} \frac{\mathrm{d}^2 w}{\mathrm{d}\xi^2} \, \delta w \, \mathrm{d}\xi,
\end{equation}
with the second boundary term, $\left[ \frac{\mathrm{d}w}{\mathrm{d}x} \, \delta w \, \right]_{x=0}$, not being dependent upon $x$. Substituting equation (\ref{eqn:VW11}) into equation (\ref{eqn:VW08}) and expanding the virtual work expression gives
\begin{equation}
    \label{eqn:VW12}
    \begin{aligned}
        \delta \mathcal{W} =  &-\int_{0}^{L} \frac{2 I_0}{c} \, W \, \left( \frac{\mathrm{d}w}{\mathrm{d}x} \right)^2 \, \delta w(x) \, \mathrm{d}x \\
        &+ \int_{0}^{L} \frac{2 I_0}{c} \, W \, \frac{\mathrm{d}w}{\mathrm{d}x} \,
        \left[ \frac{\mathrm{d}w}{\mathrm{d}x} \, \delta w \, \right]_{x=0} \mathrm{d}x \\
        &+ \int_{0}^{L} \frac{2 I_0}{c} \, W \,  \frac{\mathrm{d}w}{\mathrm{d}x} \,
        \left[ \int_{0}^{x} \frac{\mathrm{d}^2 w}{\mathrm{d}\xi^2} \, \delta w \, \mathrm{d}\xi \right] \, \mathrm{d}x.
    \end{aligned}
\end{equation}
The second integrals can be solved for to yield $0$ given that the term in brackets, $\left[ \frac{\mathrm{d}w}{\mathrm{d}x} \, \delta w \, \right]_{x=0}$, effectively acts as a constant in the integrand. The order of integration of the double spatial integral in equation (\ref{eqn:VW12}) can be interchanged by properly accounting for the limits of integration and by noting that the dummy variables of integration $x$ and $\xi$ can themselves also be interchanged:
\begin{equation}
    \label{eqn:VW13}
    \begin{aligned}
        \delta \mathcal{W} = &- \int_{0}^{L} \frac{2 I_0}{c} \, W \, \left( \frac{\mathrm{d}w}{\mathrm{d}x} \right)^2 \, \delta w(x) \, \mathrm{d}x \\
        &+ \int_{0}^{L} \left[ \int_x^L \frac{2 I_0}{c} \, W \, \frac{\mathrm{d}w}{\mathrm{d}\xi} \, \mathrm{d}\xi \right] \frac{\mathrm{d}^2 w}{\mathrm{d}x^2} \, \delta w \, \, \mathrm{d}x.
    \end{aligned}
\end{equation}
Considering the mode number, $\nu$, to take on only whole and halved integer values, the inner integral, $\left[ \int_x^L \frac{2 I_0}{c} \, W \, \frac{\mathrm{d}w}{\mathrm{d}\xi} \, \mathrm{d}\xi \right]$ evaluates to $-\frac{2 I_0}{c} \, W \, w(x)$ thereby simplifying equation (\ref{eqn:VW13}) to
\begin{equation}
    \label{eqn:VW14}
    \delta \mathcal{W} = - \int_{0}^{L} \frac{2 I_0}{c} \, W \,
    \left[ \left( \frac{\mathrm{d}w}{\mathrm{d}x} \right)^2 + w(x) \frac{\mathrm{d}^2 w}{\mathrm{d}x^2} \right] \,
    \delta w(x) \, \mathrm{d}x.
\end{equation}

Having found an expression for the virtual work depending only on the vertical virtual displacement, $\delta w$, the principle of virtual work may finally be used to investigate the question of structural stability. Substituting equations (\ref{eqn:VW04}) and (\ref{eqn:VW14}) into equation (\ref{eqn:VW01}) results in the following:
\begin{equation}
\label{eqn:VW15}
    \begin{aligned}
        &\delta U - \delta \mathcal{W} = 0 = - E \mkern 2mu I \, \frac{\mathrm{d}^3w}{\mathrm{d}x^3} \, \delta w \Big|_0^L \\
        &+ \int_{0}^{L} \left\{ E \mkern 2mu I \, \frac{\mathrm{d}^4w}{\mathrm{d}x^4}
        + \frac{2 I_0}{c} \, W \, \left[ \left( \frac{\mathrm{d}w}{\mathrm{d}x} \right)^2 + w(x) \frac{\mathrm{d}^2 w}{\mathrm{d}x^2} \right] \right\} \, \delta w(x) \, \mathrm{d}x. 
    \end{aligned}
\end{equation}
Given the arbitrary nature of $\delta w$, it is noted that for the second term in equation (\ref{eqn:VW15}) to be equal to zero, the entire quantity found inside the braces in the integrand must be equal to zero. Thus,
\begin{equation}
    \label{eqn:VW16}
    E \mkern 2mu I \, \frac{\mathrm{d}^4w}{\mathrm{d}x^4} + \frac{2 I_0}{c} \, W \,
    \left[ \left( \frac{\mathrm{d}w}{\mathrm{d}x} \right)^2 + w(x) \frac{\mathrm{d}^2 w}{\mathrm{d}x^2} \right] = 0,
\end{equation}
or,
\begin{equation}
    \label{eqn:VW17}
    E \mkern 2mu I \, \frac{\mathrm{d}^4w}{\mathrm{d}x^4} = - \frac{2 I_0}{c} \, W \,
    \left[ \left( \frac{\mathrm{d}w}{\mathrm{d}x} \right)^2 + w(x) \frac{\mathrm{d}^2 w}{\mathrm{d}x^2} \right].
\end{equation}
Equation (\ref{eqn:VW17}) is noted to look the same as the familiar Euler-Bernoulli (EB) beam theory result, $E \mkern 2mu I \, \frac{\mathrm{d}^4w}{\mathrm{d}x^4} = q(x)$ where $q(x)$ stands for the distributed loading applied onto the EB beam. The radiation pressure loading and d'Alembert body forces applied onto the lightsail are also noted to generate a RHS in equation (\ref{eqn:VW17}) whose dependency on the lightsail slope and curvature renders the differential equation nonlinear. Recalling that the initial deformation of the lightsail is, by construction, ${w = a_0 \sin \left ( \frac{2 \, \pi \, \nu \, x}{L} \right)}$, the critical material elastic modulus can be solved for in terms of the lightsail parameters by substituting in the appropriate expressions for $w$, $\frac{\mathrm{d}w}{\mathrm{d}x}$, $\frac{\mathrm{d}^2w}{\mathrm{d}x^2}$, and $\frac{\mathrm{d}^4w}{\mathrm{d}x^4}$ into equation (\ref{eqn:VW17}):
\begin{equation}
    \label{eqn:VW18}
    E_\mathrm{cr_{\delta \mathcal{W}}} = - 3 \left( \frac{2 I_0}{c} \right) \frac{a_0 L^2}{\pi^2 \, \nu^2 \, h^3} \,
    \frac{\cos{\left( \frac{4 \, \pi \, \nu \, x}{L} \right)}}{\sin{\left( \frac{2 \, \pi \, \nu \, x}{L} \right)}}.
\end{equation}
Looking at the crests and troughs of the lightsail, that is, looking at positions $x$ where $\sin{\left( \frac{2 \, \pi \, \nu \, x}{L} \right)} = \pm 1 \implies \cos{\left( \frac{4 \, \pi \, \nu \, x}{L} \right)} = \pm 1$, the virtual work approach to the lightsail stability problem predicts material strength extrema values of
\begin{equation}
    \label{eqn:VW19}
    E_\mathrm{cr_{{max}_{\delta \mathcal{W}}}}
    = 3 \left( \frac{2 I_0}{c} \right) \frac{a_0 L^2}{\pi^2 \, \nu^2 \, h^3},
\end{equation}
Comparing equations (\ref{eqn:ECRMAX}) and (\ref{eqn:VW19}) shows that the direct approach and the energy approach predict near identical results as to the critical material strength required for equilibrium:
\begin{align*}
    E_\mathrm{cr_{max}} &=  \frac{3}{2 } \left( \frac{2 I_0}{c} \right) \frac{a_0 L^2}{\pi^2 \, \nu^2 \, h^3},
    &E_\mathrm{cr_{{max}_{\delta \mathcal{W}}}}
    &= 3 \left( \frac{2 I_0}{c} \right) \frac{a_0 L^2}{\pi^2 \, \nu^2 \, h^3}.
\end{align*}

\begin{figure}[t!]
    \centering
    \includegraphics[scale=1.0]{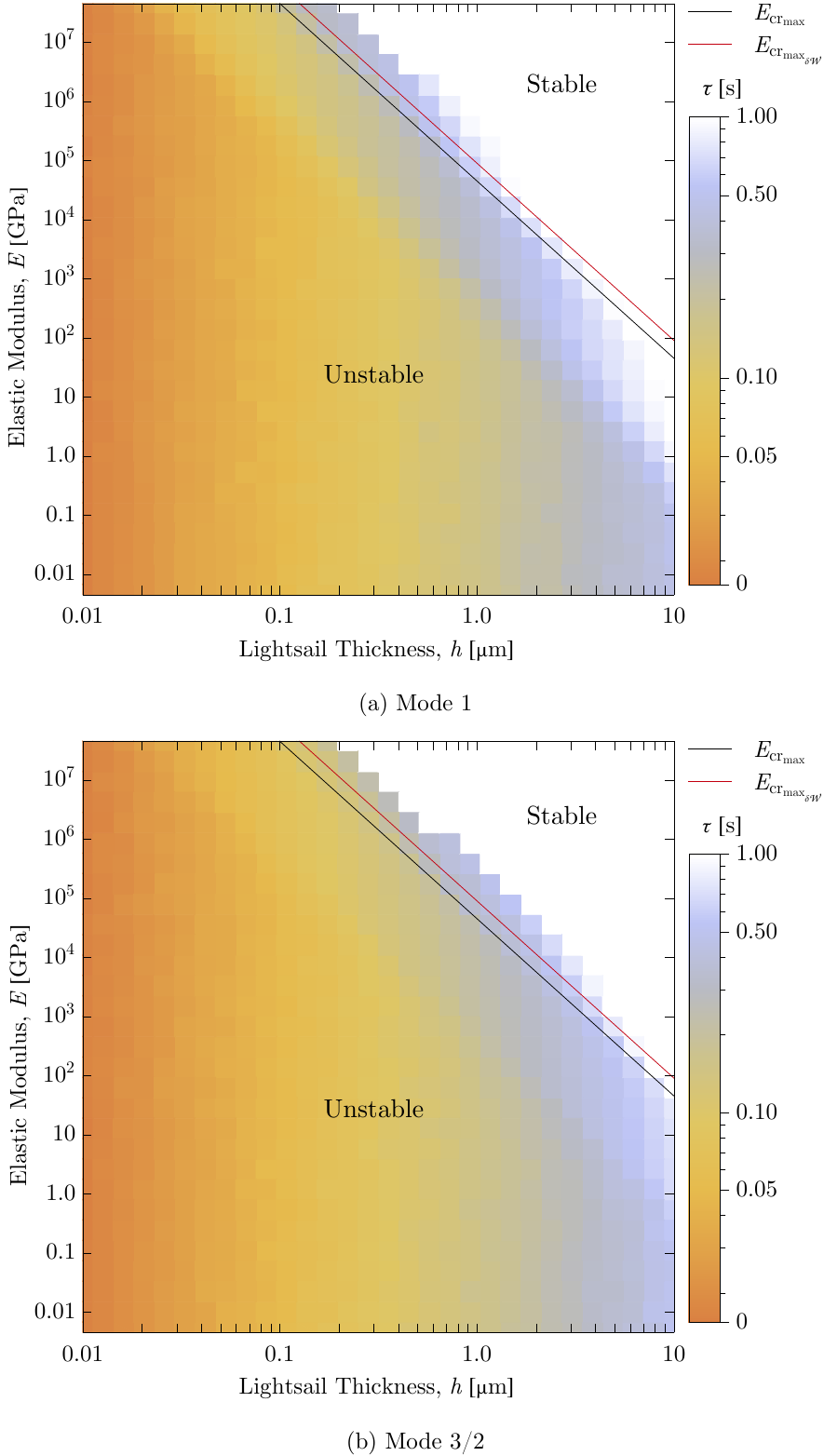}
    \caption{Stability map for the torsion lightsail model with both the moment approach (black) and energy approach (red) theoretical boundaries; (a) for perturbation mode 1; (b) for perturbation mode 3/2.}
    \label{fig:EMaps02Vertical}
\end{figure}
These two equations only differ by a factor of $1/2$. Figure~\ref{fig:EMaps02Vertical} plots these two theoretical results on the same mode $1$ and mode $3/2$ stability maps previously encountered in Section~\ref{sec:RandD}. After laying out both of the theoretical curves onto the numerical maps, there appears to be little difference between the direct method and the energy method in terms of qualitatively predicting the lightsail stability boundary: no matter the analytical expression chosen, if the material strength of the lightsail is an order of magnitude or so above that of the theoretical maximum critical value, then the lightsail configuration ends up being stable and vice versa. The factor of $1/2$ difference between equations (\ref{eqn:ECRMAX}) and (\ref{eqn:VW19}) is ultimately inconsequential since, as previously remarked in the Engineering Implications Section \ref{subsec:EngImp}, both equations require the addition of a safety factor to ensure that the lightsail configuration is stable as stability is guaranteed only when the material modulus of the sail is an order of magnitude greater than either $E_\mathrm{cr_{max}}$ or $E_\mathrm{cr_{{max}_{\delta \mathcal{W}}}}$. This strong similarity between the two analytical approaches helps further verify the agreement between theory and numerical computations found in this paper with respect to the lightsail structural stability problem.


\bibliographystyle{elsarticle-num}
\bibliography{MAIN}





\end{document}


\begin{frontmatter}
\title{\textbf{Supplementary Material}:\\Structural Stability of a Lightsail for Laser-Driven Interstellar Flight}

\author[inst1]{Dan-Cornelius Savu \texorpdfstring{\fnref{DCS}}{}}
\author[inst1]{Andrew J. Higgins \texorpdfstring{\fnref{AJH}}{}}

\affiliation[inst1]{organization={Department of Mechanical Engineering, McGill University}, 
            addressline={817 Sherbrooke St. W.}, 
            city={Montreal},
            state={Quebec},
            postcode={H3A 0C3}, 
            country={Canada}}

\fntext[DCS]{Undergraduate Research Assistant, Department of Mechanical Engineering, 817 Sherbrooke St. W. Email address: \href{mailto:x@x.com}{\underline{dan-cornelius.savu@mail.mcgill.ca}}}
\fntext[AJH]{Professor, Department of Mechanical Engineering, 817 Sherbrooke St. W. Email address: \href{mailto:x@x.com}{\underline{andrew.higgins@mcgill.ca}}}
\end{frontmatter}


\noindent This Supplementary Material documents the methods used to validate the various lightsail numerical models. In particular, this material contains the Euler-Bernoulli beam validation case for the torsion lightsail model (Section~\ref{sec:EBVAL}) and the harmonic string validation case for the tension and torsion (TnT) model (Section~\ref{sec:GSVAL}). A study of the convergence of both lightsail models is described in Section~\ref{sec:CONVS}. Section~\ref{sec:FORTRAN} contains a Mathematica vs FORTRAN lightsail simulation output comparison test case.


\section{Euler-Bernoulli Beam Validation}
\label{sec:EBVAL}
\begin{figure}[H]
    \centering
    \includegraphics[scale=0.7]{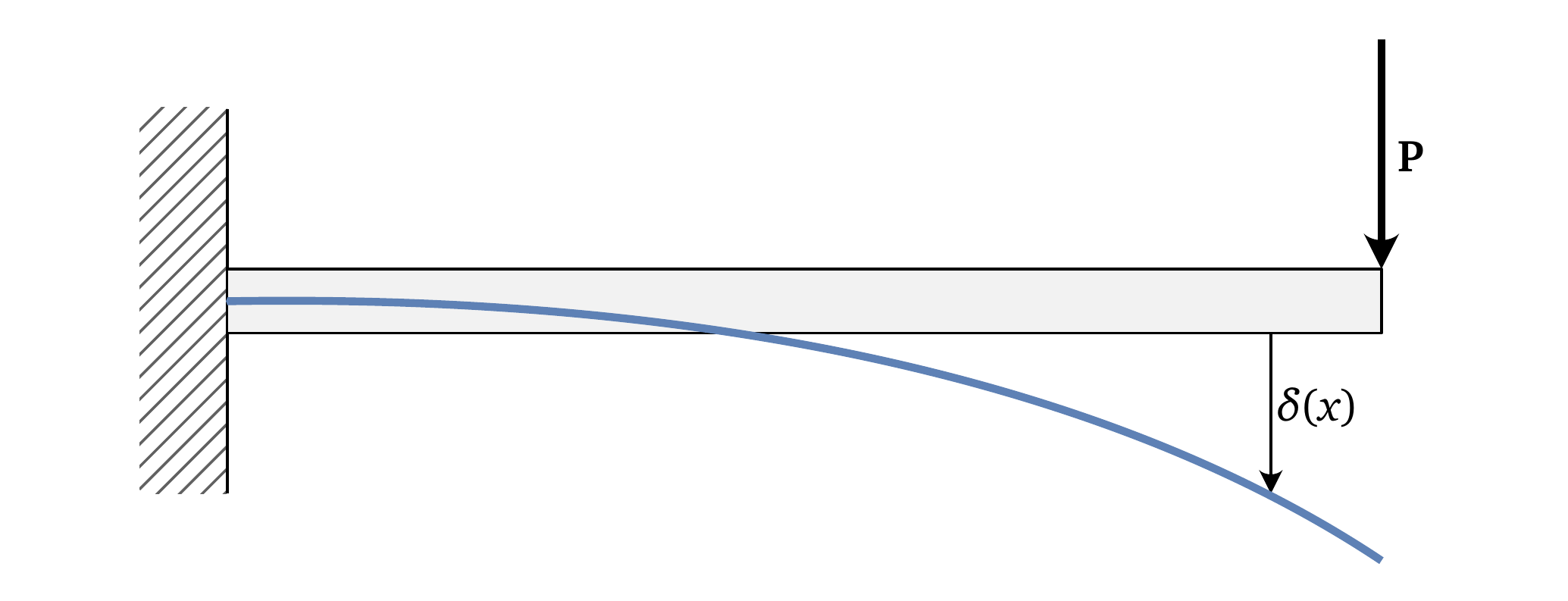}
    \caption{The analytical Euler-Bernoulli model.}
    \label{fig:EB01}
\end{figure}

\begin{figure}[H]
    \centering
    \includegraphics[scale=0.9]{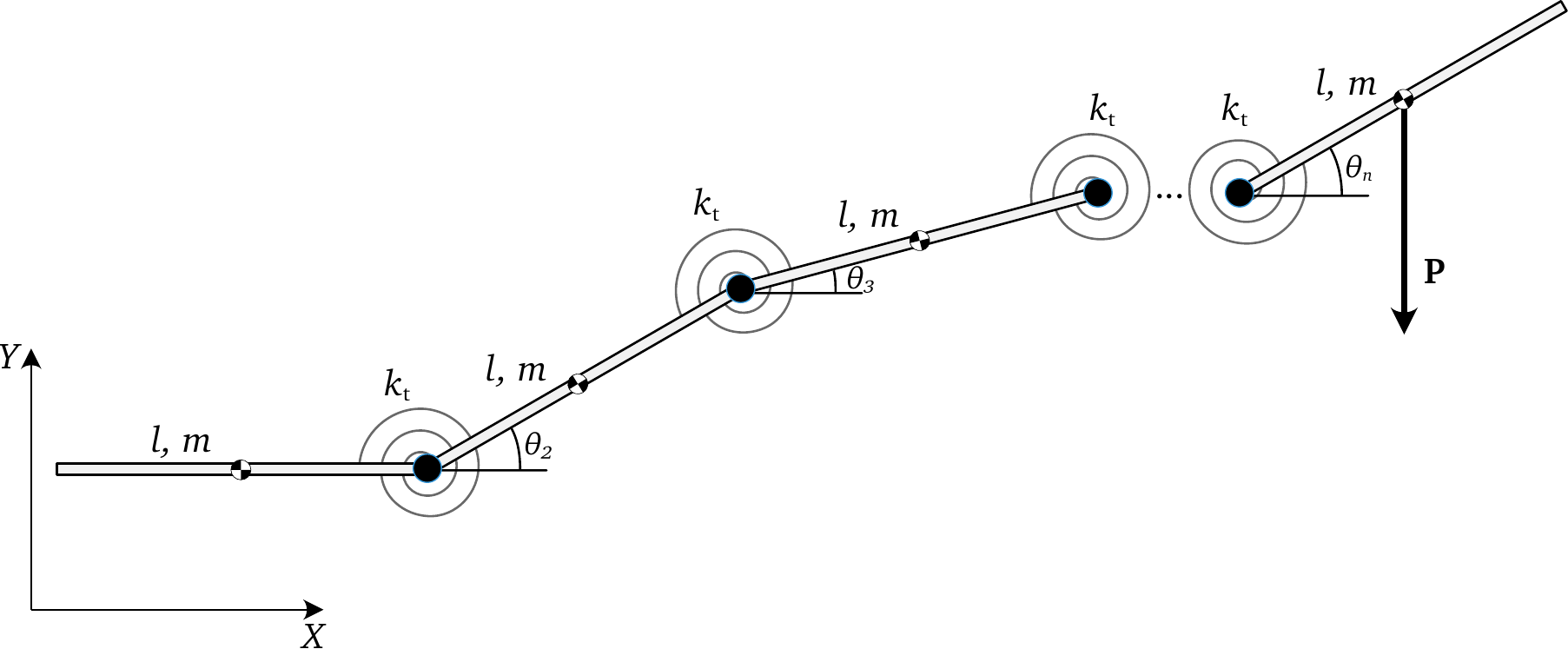}
    \caption{The numerical Euler-Bernoulli model. Note that, for the sake of visual clarity, the angular dashpots are absent in this figure.}
    \label{fig:EB02}
\end{figure}

Given that the torsion lightsail model was constructed specifically to account for material bending stiffness in the elastic regime, a relevant validation case here is that of the classic Euler-Bernoulli cantilever beam. The beam, clamped at its left end, sustains an applied concentrated load $\mathbf{P}$ at its right end (see Fig.~\ref{fig:EB01}). The Euler-Bernoulli theory predicts, for small deformations, the deflection
\begin{equation}
    \label{eqn:EB01}
    \delta(x) = \frac{P_0 \, x^{2}}{6 E I} (x-3 L).
\end{equation}
The setup for the numerical simulation is similar to the torsion lightsail model with only a few minor exceptions to account for the applied forces and boundary conditions. Since the left end is clamped, the coordinates of the center of mass of the first element is set to ($x_1=\frac{l}{2}$, $y_1=0$) and the angular position of the first element is held constant as $\theta_1(t)=0$. This, in turn, reduces the size of the system of equations from $n+2$ to $n-1$. Finally, the forces imparted to each sail element by the photon pressure $\mathbf{f}_i$ were set to $\mathbf{0}$ and replaced by a static concentrated force, $\mathbf{P}=(0, \, -P_0, \, 0)$ applied at the center of mass of the last element ($x_n$, $y_n$). This new FE model can be seen in Fig.~\ref{fig:EB02}.

\begin{figure}[H]
    \centering
    \includegraphics[scale=0.45]{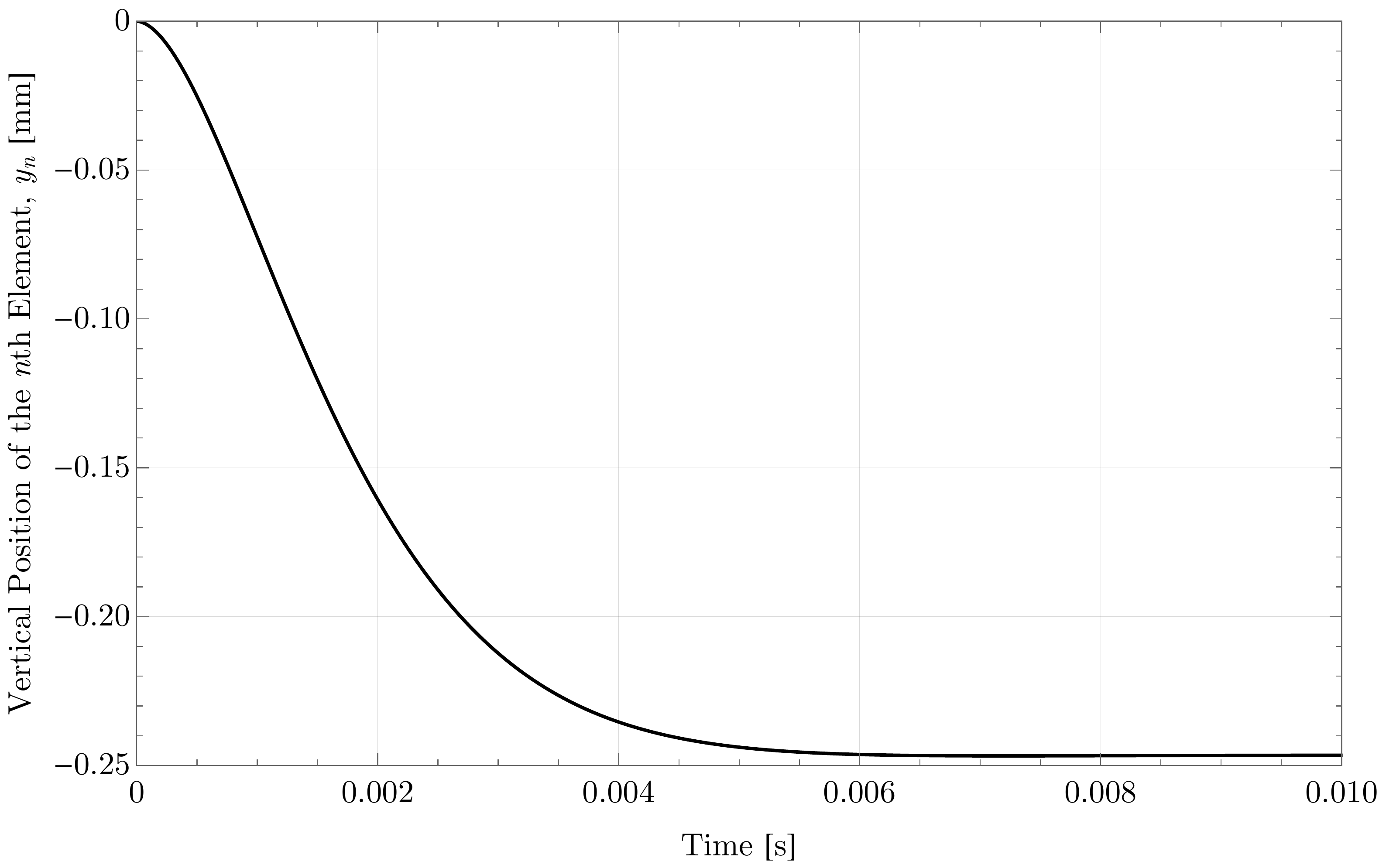}
    \caption{The vertical position of the $n$th numerical element as a function of time for the chosen dampening coefficient of $c_\mathrm{d} = \SI{e7}{\newton \meter \second}$. The plateau that starts to form towards the end of the simulation runtime indicates an overdamped response of the system to the step force input.}
    \label{fig:EB03}
\end{figure}

To properly simulate \emph{static} loading while keeping the computational demands of the validation to a minimum, it was decided not to make the transverse load a function of time. Instead, the applied load was kept as a simple step input while the numerical model was modified to include (linear) angular dashpots at its hinges to help dampen the response of the system following the application of the load. The addition of these angular dashpots further modifies the numerical model's EOMs ever so slightly due to the presence of dissipative moments. These dampening moments can be modeled using the Rayleigh dissipation function
\begin{equation}
    \label{eqn:EB02}
    R = \frac{1}{2} c_\mathrm{d} \sum_{j=2}^n \left( \dot{\theta}_j - \dot{\theta}_{j-1} \right)^2.
\end{equation}
The generalized dissipative forces that need to be accounted for are
\begin{equation}
    \label{eqn:EB03}
    Q_j^\mathrm{d} = - \frac{\partial R}{\partial \dot{\theta}_j},
\end{equation}
whose final form is
\begin{equation}
    \label{eqn:EB04}
    Q_j^\mathrm{d} = 
        \begin{cases} 
        - c_\mathrm{d} \left( \dot{\theta}_j - \dot{\theta}_{j-1} \right)
        + c_\mathrm{d} \left( \dot{\theta}_{j+1} - \dot{\theta}_j \right) &\text{for } j = 2 \, , \, 3,  \, ..., \, n-1, \\[6pt]
        - c_\mathrm{d} \left( \dot{\theta}_j - \dot{\theta}_{j-1} \right) &\text{for } j = n.
    \end{cases}
\end{equation}
In the absence of such dissipative forces the numerical model would output periodic to quasi-periodic behavior given the step load input and would never settle to a static deformation curve within computationally reasonable time. To find the magnitude of the dissipative forces required to ensure a static numerical response, the material and geometrical simulation parameters first had to be set. The simulation parameters for the beam validation are:
\begin{enumerate}
    \item Beam length, $ L =\SI{1}{\meter}$
    \item Beam thickness, $h = \SI{0.2}{\meter}$
    \item Beam width, $W = \SI{0.1}{\meter}$
    \item Beam density (steel), $\rho = \SI{7800}{\kilo \gram/\meter \cubed}$
    \item Beam modulus (steel), $E = \SI{200}{\giga \pascal}$
    \item Transverse load magnitude, $P_0 = \SI{10 000}{\newton}$
    \item Total runtime, $t_\mathrm{final} = \SI{10}{\milli \second}$
\end{enumerate}
For the simulation parameters above, the ideal Rayleigh dampening constant value was $c_\mathrm{d} = \SI{e7}{\newton \meter \second}$. For this $c_\mathrm{d}$ value the numerical model quickly settled to a static deflection curve within the allowed runtime without overshooting the static value as seen in Fig.~\ref{fig:EB03}. Ideal here means that reducing the dampening constant value by more than one order of magnitude will cause the numerical system to overshoot its static curve before settling, that is, reducing the value of the ideal $c_\mathrm{d} = \SI{e7}{\newton \meter \second}$ by more than an order of magnitude will cause the numerical system to respond like an underdamped system.

\begin{figure}[H]
    \centering
    \includegraphics[scale=0.45]{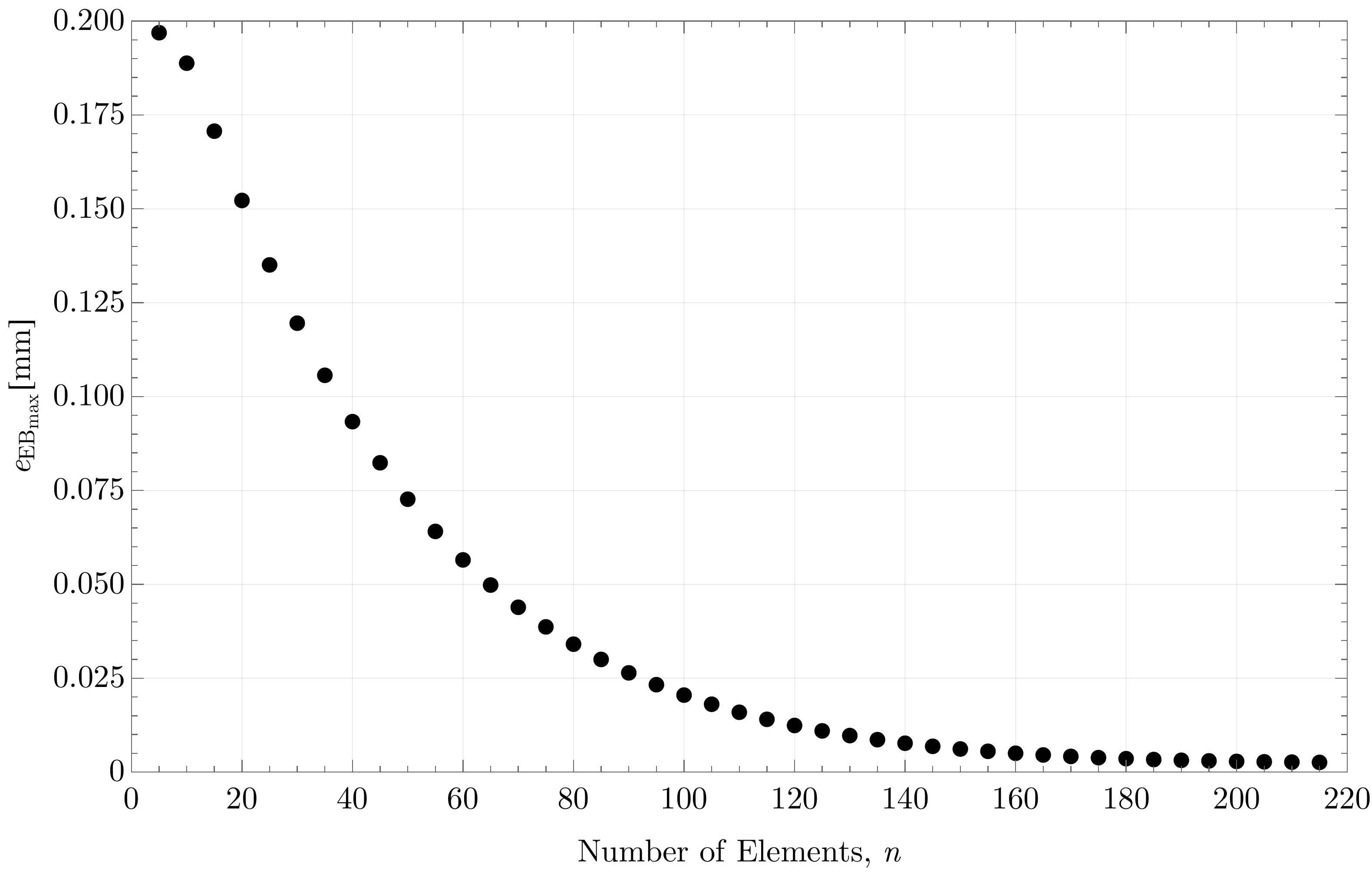}
    \caption{The Euler-Bernoulli validation convergence results.}
    \label{fig:EB04}
\end{figure}

\begin{figure}[H]
    \centering
    \includegraphics[scale=0.45]{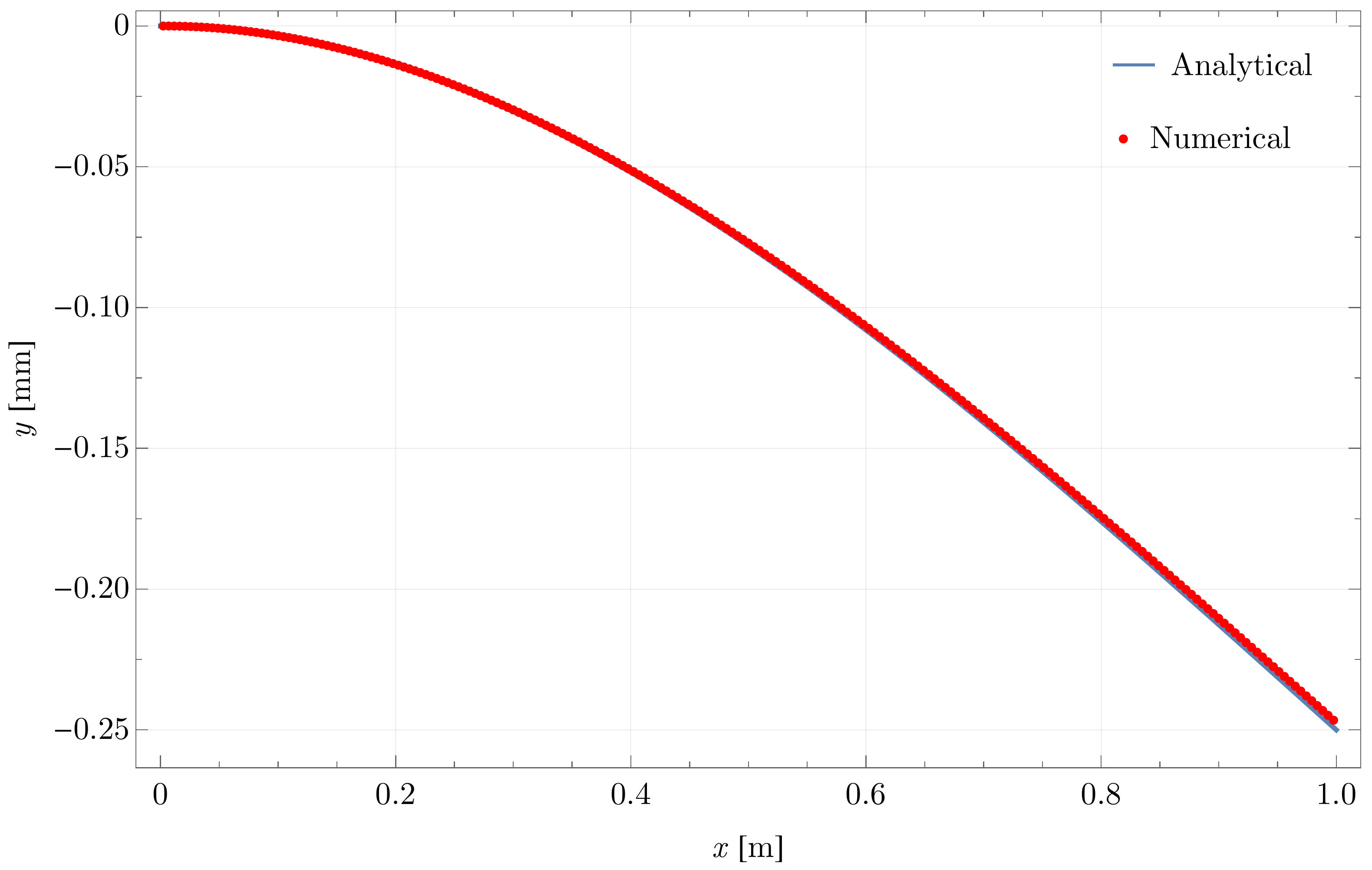}
    \caption{The Euler-Bernoulli beam analytical (blue) and numerical (red) deflection results at $n = 215$. The red dots represent the center of mass of each numerical element. Note how the numerical model sits on top of the analytical curve.}
    \label{fig:EB05}
\end{figure}

To validate the torsion model, the Euler-Bernoulli beam simulations were run for different numbers of elements and the maximum error between the analytical deflection, $\delta(x)$, and the numerical beam deflection at end of runtime was computed for each run by taking the absolute difference between a given element's vertical coordinate and its theoretical deflection:
\begin{equation}
    \label{eqn:EB05}
    e_{\mathrm{EB}_\mathrm{max}} = \mathrm{max} 
    \left\{ \left| \delta(x_j(t_{\mathrm{final}})) - y_j(t_{\mathrm{final}}) \right| \right\}
    \text{ for } i=1, \,2 \,, \,3,\, ...,\, n.
\end{equation}
Figure~\ref{fig:EB04} plots the maximum absolute error vs the number of elements. The number of elements was increased from $n = 5$ to $n = 215$ in increments of 5. Note that, after about $n = 150$, the absolute maximum error seems to start plateauing with the final maximum error at $n = 215$ being $e_{\mathrm{EB}_\mathrm{max}} = \SI{0.00257}{\milli \meter}$. Figure~\ref{fig:EB05} shows the deflection curves of both the numerical and analytical model for the $n = 215$ simulation run where it is noted that the two curves are nearly identical. This relatively small error between theory and numerical computations instills confidence into the ability of the torsion model to correctly describe material bending stiffness.

\section{Harmonic String Validation}
\label{sec:GSVAL}

Given its inclusion of rectilinear springs and boundary tension, a pertinent scenario to validate the TnT model for transient problems against is that of a vibrating string's harmonics. From Newtonian first principles, it is known that the fundamental frequency, $f_0$, of a string under constant tension, $T$, and kept fixed at both ends can be expressed as
\begin{equation}
    \label{eqn:GG01}
    f_0 = \frac{1}{2} \sqrt{\frac{T}{M L}}
\end{equation}
where $L$ is the working length, that is, the stretched length of the string and $M$ is its total mass. To help make the validation as general as can be, as set of geometric and material simulation parameters were first set. From these starting parameters, a theoretical fundamental frequency for the string could be computed using equation (\ref{eqn:GG01}). This theoretical value was then compared to the fundamental frequency of the simulated spring measured via the use of a discrete Fourier transform (DFT). The simulation parameters chosen for the validation test were the following\footnote{Note that the cross-sectional area of the numerical model and second moment of area of this particular string are, respectively, $A_\mathrm{c} = \frac{\pi d^2}{4}$ and $I = \frac{\pi}{4}\left(\frac{d}{2}\right)^2$.}\textsuperscript{,}\footnote{The reader might recognize these parameters as the ones defining a string whose fundamental frequency is G3 in scientific pitch notation.}:
\begin{enumerate}
    \item String length, $L = \SI{0.6477}{\meter}$
    \item String gauge (diameter), $d = \SI{0.430}{\milli \meter}$
    \item String density (tempered high-carbon steel), $\rho = \SI{7800}{\kilo \gram/\meter \cubed}$
    \item String modulus (tempered high-carbon steel), $E = \SI{210}{\giga \pascal}$
    \item Applied boundary tension magnitude, $T = \SI{72.485}{\newton}$
\end{enumerate}

In adapting the TnT model to the taut string scenario, the following issue is encountered: in order to keep the right end of the TnT model fixed, it is not enough to simply set $y_n=\mathrm{constant}$. This holonomic constraint would, in turn, eliminate the connection between the last TnT element and the remainder of the string. It would be as though the hinge connecting the last extensible element to the other elements was deleted at the start of the simulation. Ideally, one would want the vertical acceleration component of the last element to be zero throughout the simulation, that is, ideally, $\ddot{y}_n(t) = 0$. In reality, however, $\ddot{y}_n=f\left(\mathbf{q},\dot{\mathbf{q}}\right)$, and, in fact, $\ddot{y}_n$ is non-linear in $\mathbf{\dot{q}}$ and thus Lagrangian multipliers cannot be used to account for the generalized force(s) needed to maintain the $\ddot{y}_n(t) = 0$ constraint. Fortunately, maintaining the first string element fixed, that is, setting ($x_1=\mathrm{constant}$, $y_1=\mathrm{constant}$) and ensuring that the two applied boundary tension forces remain co-linear to their respective element, that is, enforcing $\mathbf{T}_1 = T(-\cos\theta_1, \,-\sin\theta_1, \,0)$ and $\mathbf{T}_n = T(\cos\theta_n, \,\sin\theta_n, \,0)$ appears to keep the end element from straying too much from its starting position.

Because the objective of this procedure was to validate the fundamental frequency of the simulated string against that of a physical string, the numerical string was initially deformed in a half sine wave fashion where $y(x) =  a_0 \sin \left ( \frac{ \, \pi \, \, x}{L}  \right )$ with $a_0 = \SI{10}{\milli \meter}$. An initial amplitude of $\SI{10}{\milli \meter}$ was used instead of the  $\SI{0.01}{\milli \meter}$ used throughout the main paper for the lightsail stability maps in order to ensure that the initial mode shape of the simulated spring was of significant scale as opposed to a perturbation. Further, to simulate pre-tension, the initial elongation of each rectilinear spring was set such that they each exerted a total force equal in magnitude to the boundary tension, that is, initially $z_i = T/k_\mathrm{s}$ for $i=1,\,2,\,3\,...,\,n-1$.

\begin{figure}[H]
    \centering
    \includegraphics[scale=0.62]{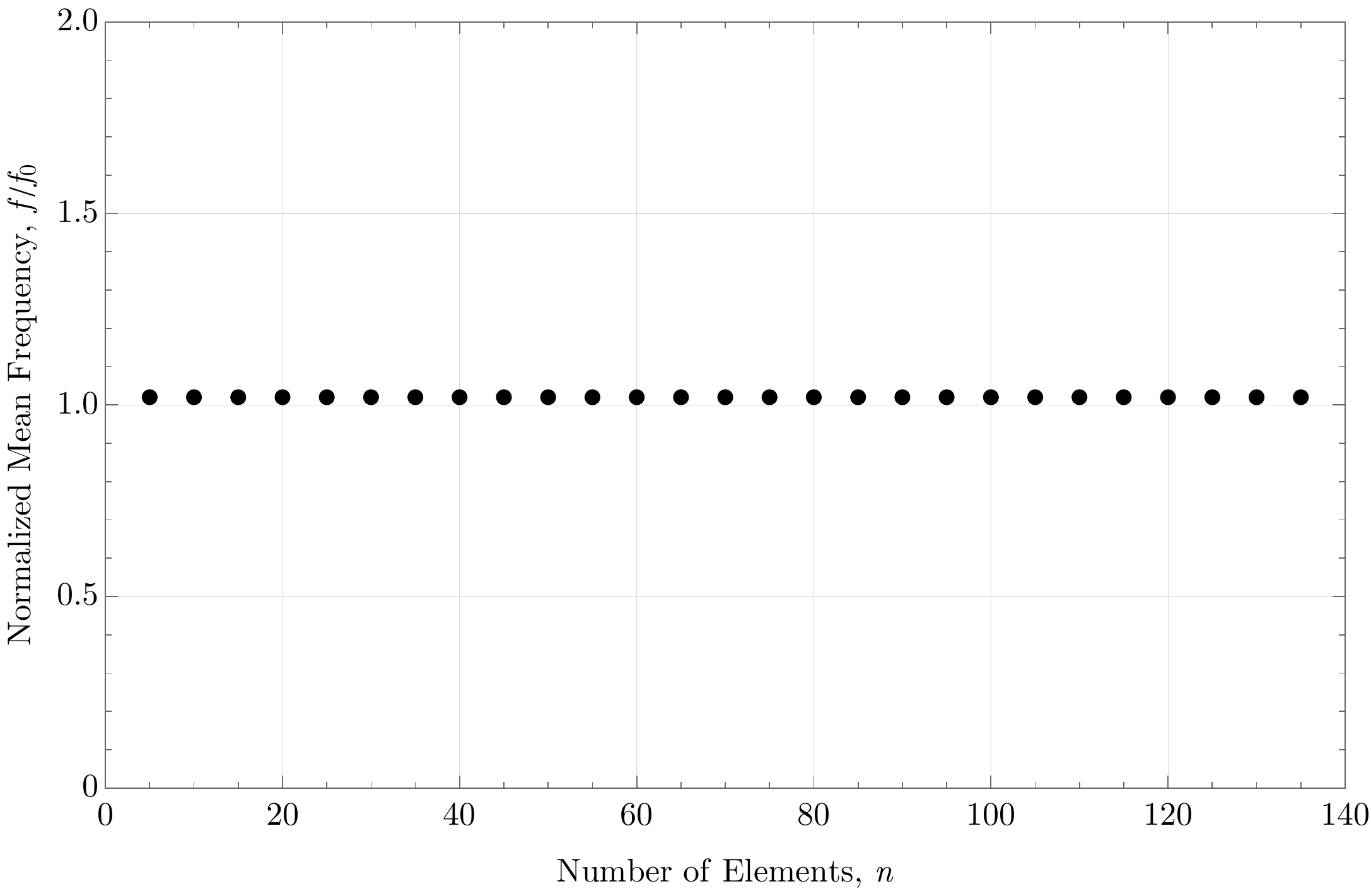}
    \caption{Validation results of the numerical model plotting the convergence of the mean fundamental frequency of the model normalized by the theoretical frequency.}
    \label{fig:GG02}
\end{figure}

\begin{figure}[H]
    \centering
    \includegraphics[scale=0.62]{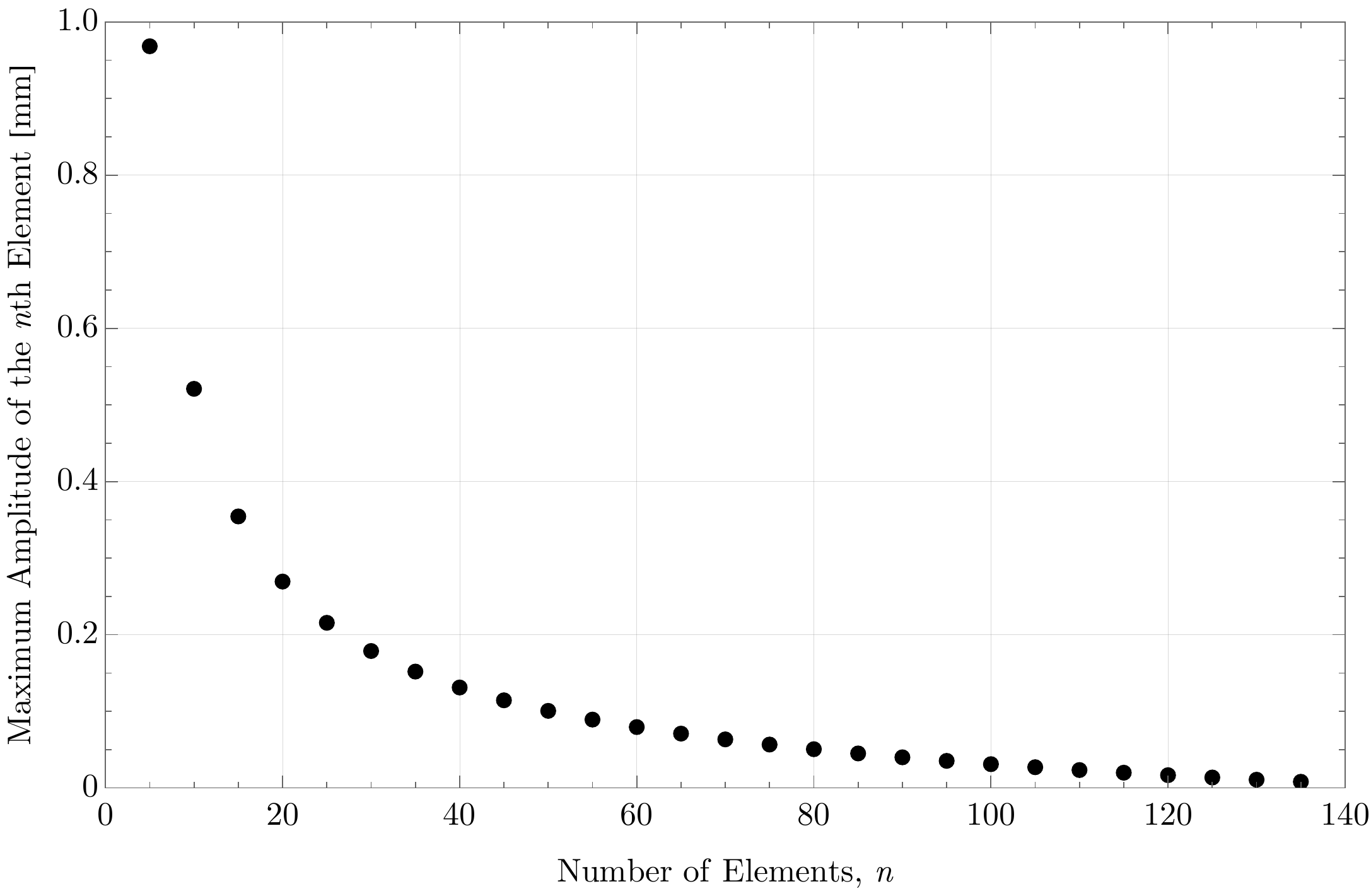}
    \caption{Maximum amplitude of the CoM of the $n$th element.}
    \label{fig:GG03}
\end{figure}

Figures~\ref{fig:GG02} and~\ref{fig:GG03} display the validation results which were gathered by increasing the number of elements used to simulate the string from $n = 5$ to $n = 135$ in increments of 5. To properly measure the initially deformed string's fundamental frequency, the frequency of vertical oscillations of each string element with the exception of the ends was measured via a discrete Fourier transform. The mean fundamental frequency of the total elements is plotted in Fig.~\ref{fig:GG02}. Note that, as per theory, each element ended up oscillating with the same frequency and thus the standard deviation from the mean is always zero, hence the absence of error bars in Fig.~\ref{fig:GG02}. The mean fundamental frequency of the string appears to have settled, from the very start of the convergence test at $n=5$, to that of $\SI{199.96}{\hertz}$ which is less than $\SI{3}{\hertz}$ away from the target fundamental frequency of $\SI{196.00}{\hertz}$. The discrepancy is suspected to be due at least in part to the DFT scheme and perhaps in part also due to the ``wobbling'' of the $n$th element (i.e., the final element, which should remain fixed). Figure~\ref{fig:GG03} portrays the maximum amplitude of the vertical oscillations of the $n$th element. Note that the wobbling of the $n$th element steadily tends to disappear as the number of elements is increased.

\section{Numerical Convergence}
\label{sec:CONVS}

Following the validations undertaken for the torsion model and for the TnT model, the task of proving numerical convergence for both numerical models was undertaken. To demonstrate convergence, the time to doubling of amplitude, $\tau$, was recorded as the number of elements used to construct each of the lightsail models was progressively increased. To be precise, the number of elements was gradually increased from 10 all the way to 150 in increments of 5, and for each run the time when uncontrolled perturbation growth occurred, $\tau$, was recorded with uncontrolled perturbation growth being defined to occur when the maximum lightsail shape amplitude equals or exceeds double its initial perturbation amplitude, $a_0$. The maximum lightsail shape amplitude was computed as follows:
\begin{equation}
    \label{eqn:CONV01}
    \mathrm{amp}_\mathrm{max}(t) = \frac{\max\{y_i(t)\}-\min\{y_i(t)\}}{2}
\end{equation}
where $y_i(t)$, as already encountered, stands for the vertical position of the center of mass of the $i$th lightsail element.

For each numerical model, the simulation properties of an exact potentially unstable sample case from the corresponding stability maps were used to probe convergence. The chosen convergence simulation properties were
\begin{itemize}
    \item for the torsion model
    \begin{enumerate}
        \item Sail length, $L = \SI{1}{\meter}$
        \item Sail width, $W = \SI{1}{\meter}$
        \item Sail thickness, $h = \SI{10}{\micro\meter}$
        \item Sail density, $\rho = \SI{1000}{\kilo \gram/\meter \cubed}$
        \item Material Elastic Modulus, $E = \SI{0.527}{\giga \pascal}$
        \item Incident laser intensity, $I_0 = \SI{10}{\giga \watt/\meter \squared}$
        \item Initial perturbation amplitude, $a_0 = \SI{0.01}{\milli \meter}$
        \item Perturbation mode number, $\nu = 3/2$
        \item Total runtime, $t_\mathrm{final} = \SI{1}{\second}$ or  to doubling of initial perturbation amplitude
    \end{enumerate}
\end{itemize}
\begin{itemize}
    \item for the TnT model
    \begin{enumerate}
        \item Sail length, $L = \SI{1}{\meter}$
        \item Sail width, $W = \SI{1}{\meter}$
        \item Sail thickness, $h = \SI{10}{\micro\meter}$
        \item Sail density, $\rho = \SI{1000}{\kilo \gram/\meter \cubed}$
        \item Material Elastic Modulus, $E = \SI{5}{\giga \pascal}$
        \item Boundary Tension Magnitude, $T = \SI{3.09e-6}{\newton/\meter}$
        \item Incident laser intensity, $I_0 = \SI{10}{\giga \watt/\meter \squared}$
        \item Initial perturbation amplitude, $a_0 = \SI{0.01}{\milli \meter}$
        \item Perturbation mode number, $\nu = 3/2$
        \item Total runtime, $t_\mathrm{final} = \SI{1}{\second}$ or  to doubling of initial perturbation amplitude
    \end{enumerate}
\end{itemize}

\begin{figure}[H]
    \centering
    \includegraphics[scale=0.45]{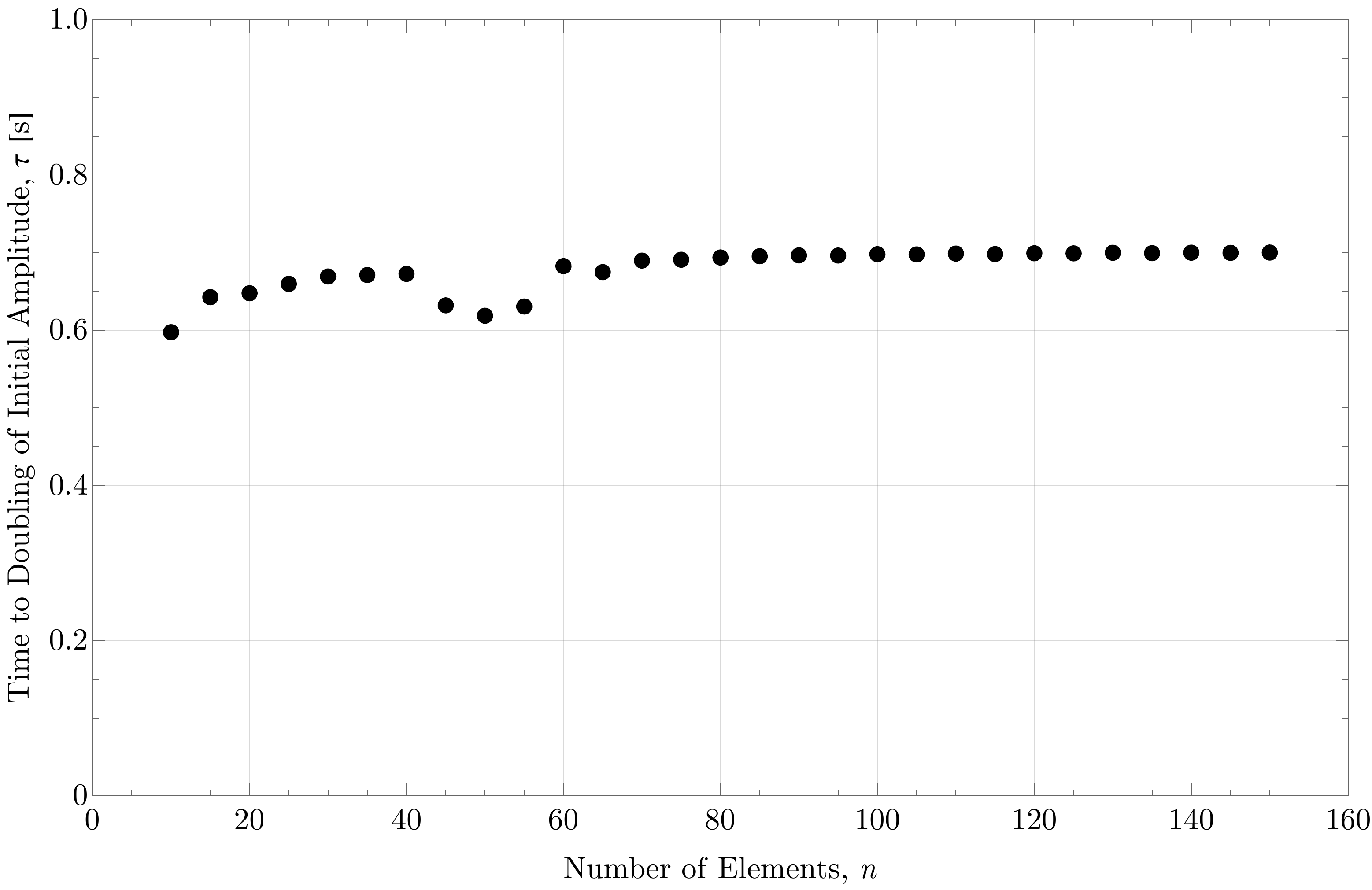}
    \caption{Torsion lightsail numerical model convergence plot for the time to doubling of initial amplitude, $\tau$.}
    \label{fig:CONV01}
\end{figure}

\begin{figure}[H]
    \centering
    \includegraphics[scale=0.45]{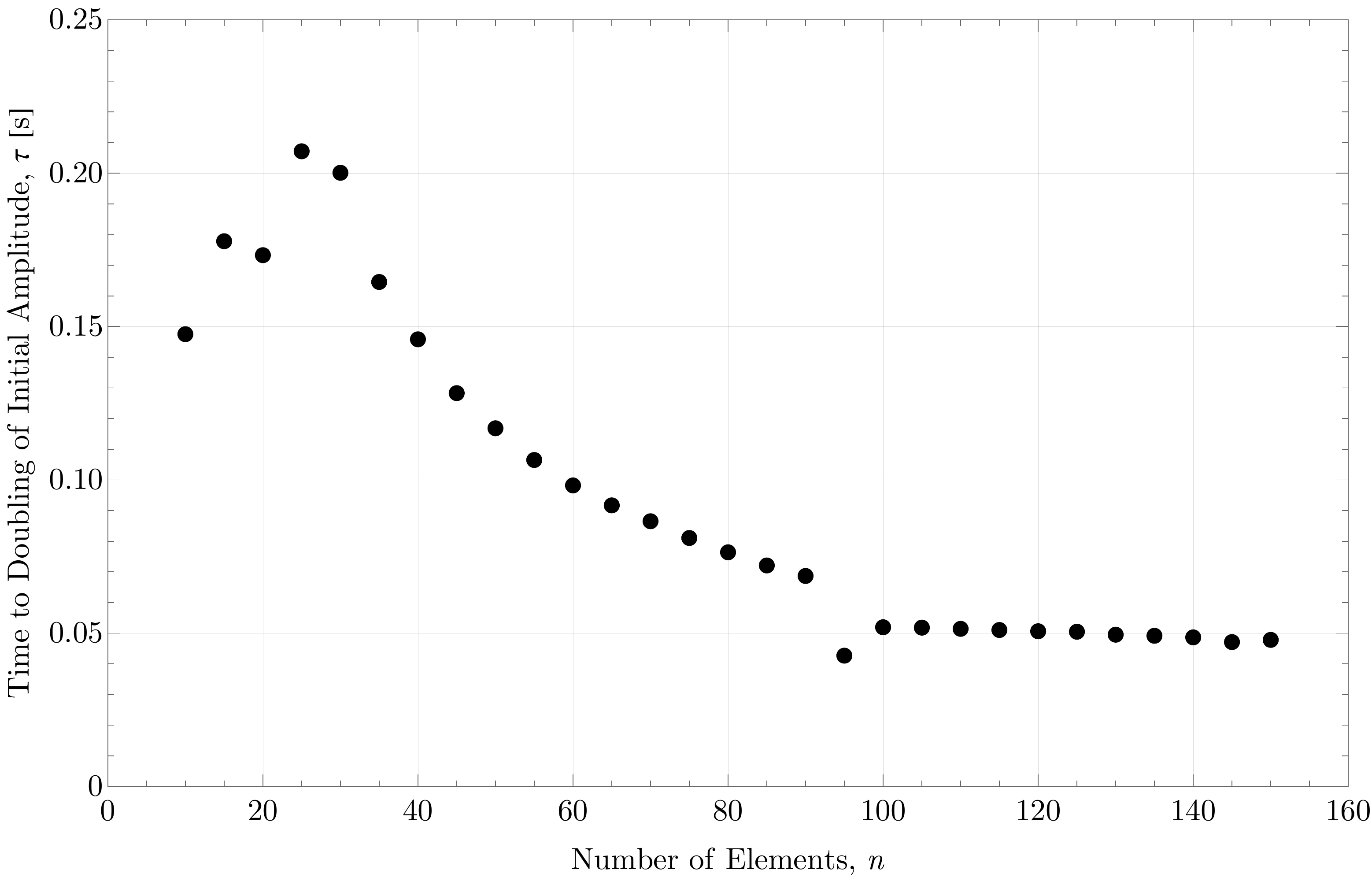}
    \caption{TnT lightsail numerical model convergence plot for the time to doubling of initial amplitude, $\tau$.}
    \label{fig:CONV02}
\end{figure}

Figure~\ref{fig:CONV01} plots the convergence data for the torsion model. The torsion model appears to converge to the $\tau \approx \SI{0.7}{\second}$ value following the $n = 70$ simulation case. The convergence study results for the TnT model are shown in Fig.~\ref{fig:CONV02} where it can be seen that, for the chosen lightsail configuration, the time to doubling of initial perturbation amplitude steadily converges to $\tau \approx \SI{0.05}{\second}$ with the asymptotic behavior clearly starting at around $n=100$ elements. The lightsail simulation runs performed for the construction of both stability maps found in the Results and Discussion section are noted to have used a lesser number of elements with $n$ being set to 50 elements for the sake of numerical efficiency. Given the convergence results, it may be suspected that the number of elements used in the stability maps --- especially for the stability maps employing the TnT model --- was too low. However, while increasing the number of elements does affect the results quantitatively, \emph{the number of elements appears not to affect the qualitative stability results}. That is, a stable lightsail configuration remains stable no matter the number of elements employed and likewise for an unstable lightsail configuration.

\section{Mathematica \large \texttt{NDSolve} \normalsize vs FORTRAN \large \texttt{MEBDFV} \normalsize Comparison}
\label{sec:FORTRAN}

All results shown in the above two validation sections, in the numerical convergence section, and in the main paper were a consequence of numerically integrating a discrete lightsail model or slight modification thereof using the \verb|NDSolve| solver package of Mathematica. To further gain confidence into the validity of the numerical models, a number of lightsail configurations where also simulated using the FORTRAN modified backward differentiation formulae (\verb|MEBDFV|) solver of Abdulla and Cash from Imperial College, London (Department of Mathematics), a FORTRAN solver written to deal with the kind of numerical systems encountered in this study (see Section~3.4 of the main paper for a more complete discussion of the inner workings of the \verb|MEBDFV| solver). This section of the Supplementary Material will focus on the comparison between the Mathematica \verb|NDSolve| and FORTRAN \verb|MEBDFV| outputs for a representative lightsail configuration of the mathematically and numerically more complex TnT lightsail model. The TnT lightsail model configuration in question is the $\big( h = \SI{10}{\micro \meter}$, $T = \SI{1.29e-2}{\newton / \meter}$, $\nu = 3/2$, $\tau = \SI{1}{\second} \big)$ simulation run whose behavior was visually displayed in a set of temporal snapshots in the main paper. For the sake of completion and convenience, all other parameters relevant to the lightsail configuration studied here are restated below from Section~4.2 of the main paper:
\begin{enumerate}
    \item Sail length, $L = \SI{1}{\meter}$
    \item Sail width, $W = \SI{1}{\meter}$
    \item Sail density, $\rho = \SI{1000}{\kilo \gram / \meter \cubed}$
    \item Material Elastic Modulus, $E = \SI{5}{\giga \pascal}$
    \item Incident laser intensity, $I_0 = \SI{10}{\giga \watt / \meter \squared}$
    \item Initial perturbation amplitude, $a_0 = \SI{0.01}{\milli \meter}$
    \item Perturbation mode number, $\nu = 3/2$
    \item Number of elements, $n = 50$
    \item Total runtime, $t_\mathrm{final} = \SI{1}{\second}$ or  to failure
\end{enumerate}

\begin{figure}
    \centering
    \begin{subfigure}{\textwidth}
        \hspace{2.75cm}
        \includegraphics[scale=0.44]{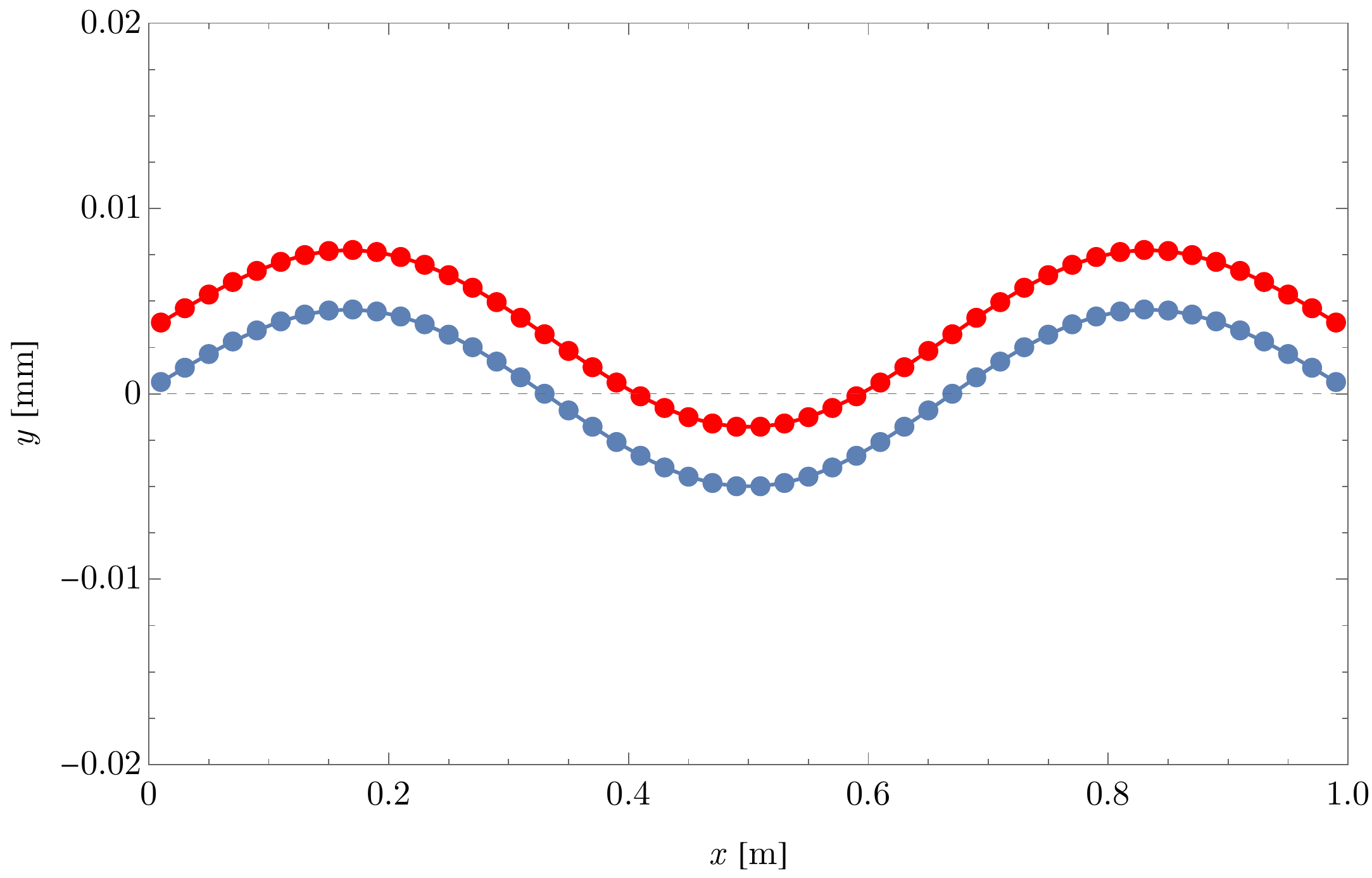}
        \caption{$t=\SI{0.1}{\second}$}
        \label{subfig:FORTRAN01}
    \end{subfigure}
    \hfill
    \vspace{0cm}
    \begin{subfigure}{\textwidth}
        \hspace{2.75cm}
        \includegraphics[scale=0.44]{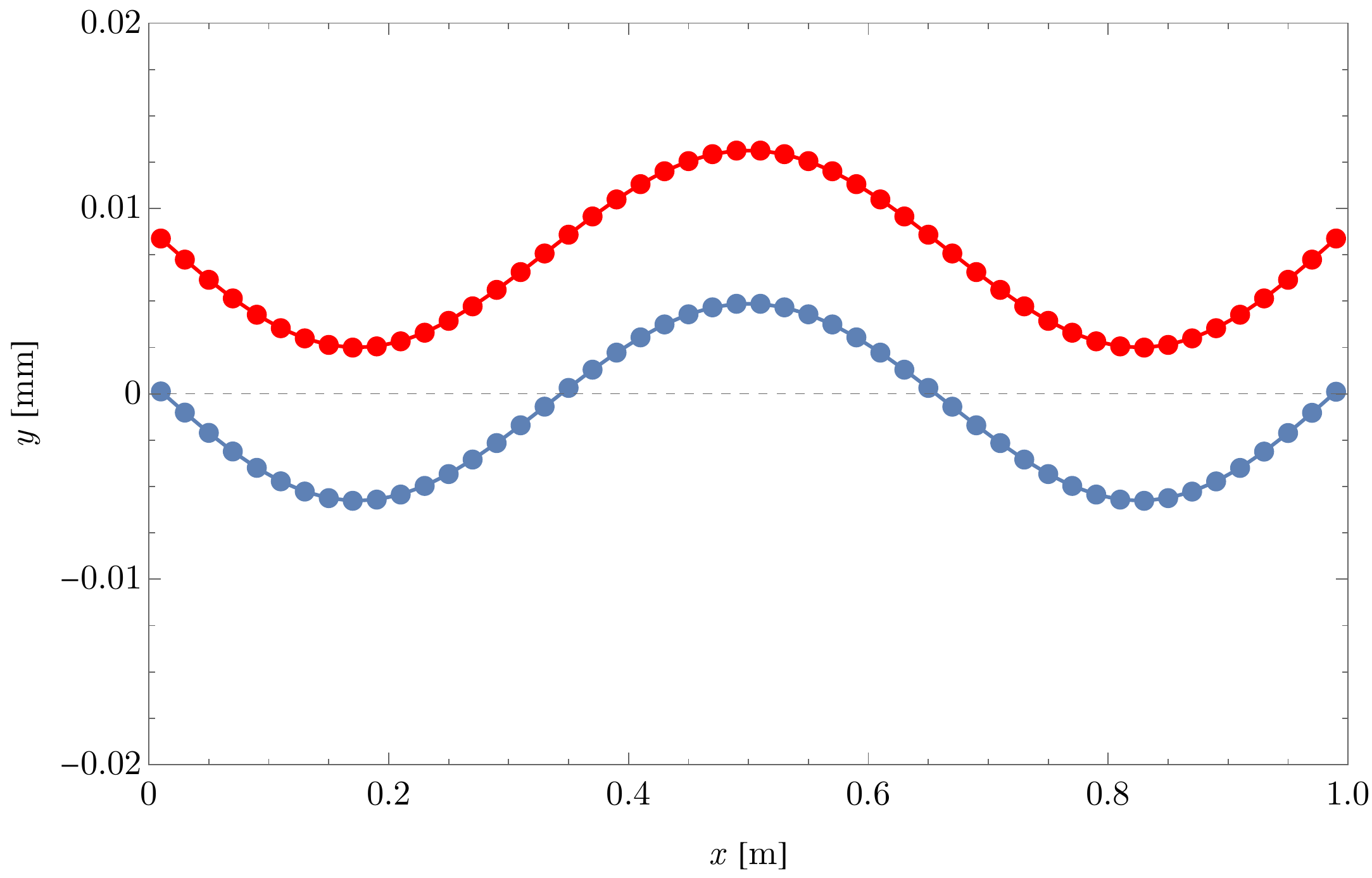}
        \caption{$t=\SI{0.2}{\second}$}
        \label{subfig:FORTRAN02}
    \end{subfigure}
    \hfill
    \vspace{0cm}
    \begin{subfigure}{\textwidth}
        \hspace{2.75cm}
        \includegraphics[scale=0.44]{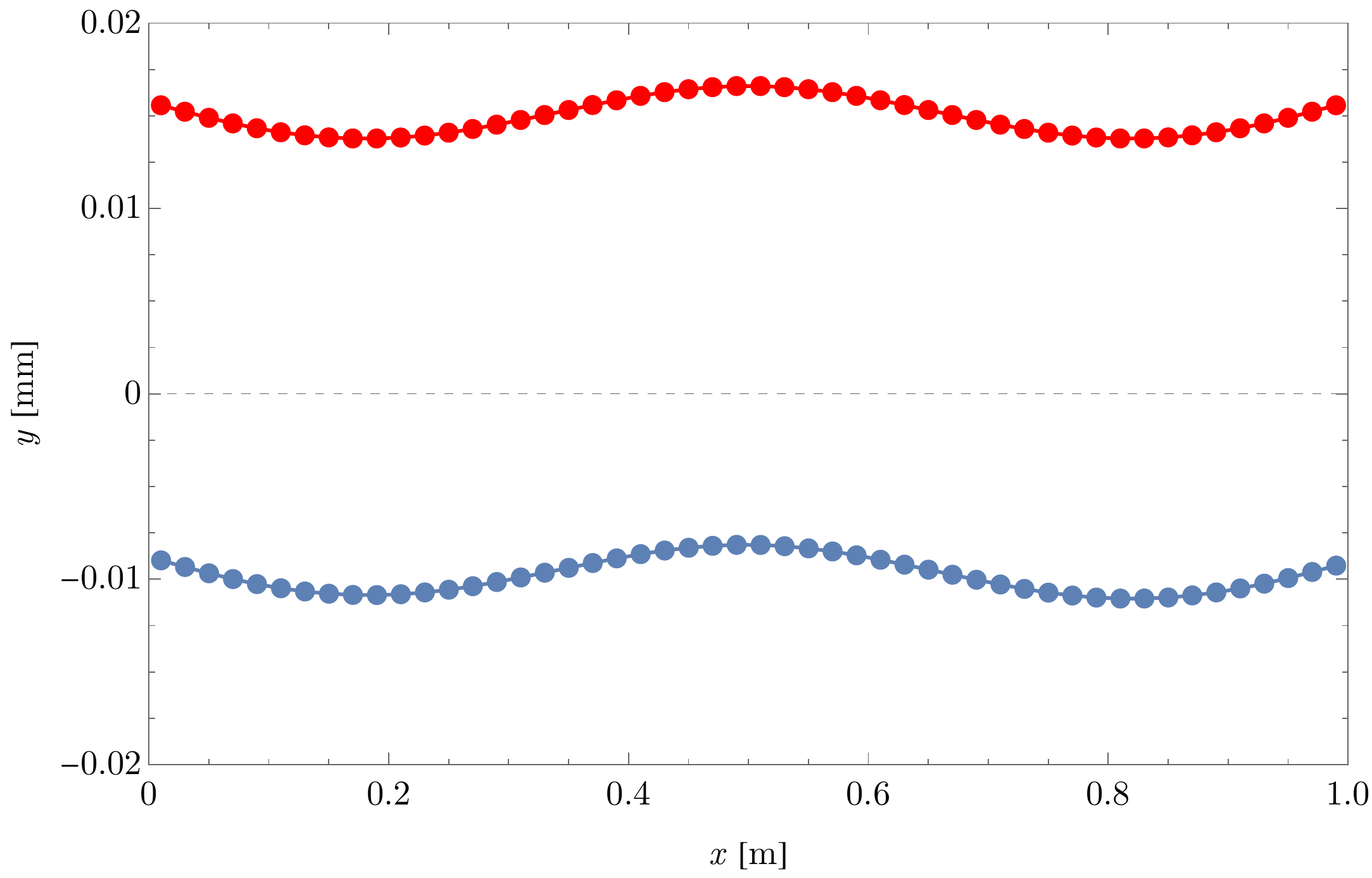}
        \caption{$t=\SI{1}{\second}$}
        \label{subfig:FORTRAN03}
    \end{subfigure}
    \caption{Comparison between the Mathematica \texttt{NDSolve} (red) and FORTRAN \texttt{MEBDFV} (blue) outputs for the same lightsail configuration test case. The dashed line represents the position of a flat (i.e., unperturbed) lightsail. The lightsail shape snapshots are visualized in a reference frame fixed to an ideally flat lightsail.}
    \label{fig:FORTRANComp}
\end{figure}

Figures~\ref{fig:FORTRANComp}a--c show the Mathematica lightsail output (red) and the FORTRAN lightsail output (blue) at three different snapshots in time as seen from a reference frame attached to a flat lightsail---the flat lightsail reference frame was used to help better appreciate the perturbed shape of the lightsail. At each instant in time there was no visually appreciable difference in lightsail shape between the Mathematica and FORTRAN outputs with the greatest percentage difference between the two solvers in lightsail shape maximum amplitude being of $2.4\%$ and occurring at the end of the simulation time, $t=\SI{1}{\second}$. While there was some appreciable difference between the two solver outputs in absolute vertical position of the sail with the Mathematica output being consistently ahead of the FORTRAN output, the greatest difference in vertical position was of $\SI{0.025}{\milli \meter}$ and also occurred at the end of the simulation, that is, at time $t=\SI{1}{\second}$. Considering the fact that, given the $\SI{10}{\giga \watt}$ laser intensity, both lightsails traveled a total of $\SI{3335}{\meter}$ during the $\SI{1}{\second}$ runtime, this difference in absolute position between the two solvers is here considered acceptable. Of note is that both solvers displayed stable oscillatory lightsail behavior, and it is important to emphasize that the Mathematica vs FORTRAN comparison test case shown in this section is a representative of all the comparison tests made between the two solvers: Barring the presence of minor differences in absolute positioning of the lightsail, the evolution of the lightsail shape throughout simulation runtime was identical between the Mathematica \verb|NDSolve| and the FORTRAN \verb|MEBDFV| outputs, that is, whenever a Mathematica lightsail simulation displayed stable oscillatory behavior, the FORTRAN lightsail simulation also displayed stable oscillatory behavior, and whenever a Mathematica lightsail simulation displayed potentially unstable behavior so did the FORTRAN lightsail simulation.

